\documentclass[journal=jacsat,manuscript=article]{achemso}

\usepackage[version=3]{mhchem} % Formula subscripts using \ce{}
\usepackage{xcolor}
\usepackage{soul,color}
\usepackage[T1]{fontenc}
\usepackage[nolist]{acronym}
\usepackage[font=normalsize,labelfont=bf]{caption}
\usepackage{subcaption}

\usepackage{tcolorbox}
\tcbuselibrary{breakable}

\usepackage{listings}
\lstdefinestyle{xyzstyle}{
    language=XYZ,
    basicstyle=\ttfamily\small,
    breaklines=true,
    frame=single,
    numbers=none,
    showstringspaces=false,
    tabsize=2
}
\author{Lena T. T. Nguyen}
\affiliation[SDU]
{Department of Physics, Chemistry and Pharmacy, University of Southern Denmark, Campusvej~55, DK--5230 Odense M, Denmark}
\author{Ernst D. Larsson}
\affiliation[SDU]
{Department of Physics, Chemistry and Pharmacy, University of Southern Denmark, Campusvej~55, DK--5230 Odense M, Denmark}
\author{Kajsa M. F. Niklasson}
\affiliation[SDU]
{Department of Physics, Chemistry and Pharmacy, University of Southern Denmark, Campusvej~55, DK--5230 Odense M, Denmark}
\affiliation[SDU]
{Department of Physics, Chemistry and Pharmacy, University of Southern Denmark, Campusvej~55, DK--5230 Odense M, Denmark}
\author{Erna K. Wieduwilt}
\affiliation[Essen]
{Physical Chemistry, Faculty of Chemistry, Center of Medical Biotechnology (ZMB) and Centre for Water and Environmental Research (ZWU), University of Duisburg-Essen, Universitätsstraße 5, Essen 45141, Germany}
\author{Erik D. Hedegård}
\affiliation[SDU]
{Department of Physics, Chemistry and Pharmacy, University of Southern Denmark, Campusvej~55, DK--5230 Odense M, Denmark}
\email{erdh@sdu.dk}

\title[]{Efficient Calculation of Absorption Spectra of Platinum
Complexes Used as Luminescent Probes for Cancer Detection}
\abbreviations{IR,NMR,UV}

\begin{acronym}[MPC]
\acro{BJ}{Becke-Johnson}
\acro{CPCM}{conductor-like polarizable continuum model}
\acro{CSS}{closed shell-singlet}
\acro{CT}{charge-transfer}
\acro{DFT}{density functional theory}
\acro{FWHM}{full-width at half-maximum}
\acro{G4}{G-quadruplex}
\acro{MLCT}{metal to ligand charge transfer}
\acro{MAE}{mean absolute error}
\acro{MSE}{mean signed error}
\acro{PACT}{photoactivated chemotherapy}
\acro{PDB}{protein data bank}
\acro{PDT}{photodynamic therapy}
\acro{RI}{the resolution of the identity}
\acro{RMSD}{root mean square deviation}
\acro{RSF}{range-separated functional}
\acro{SOC}{spin-orbit coupling}
\acro{TDA}{Tamm-Dancoff approximation}
\acro{TD-DFT}{time-dependent density functional theory}
\acro{QM}{quantum mechanics}
\end{acronym}

\begin{document}

%% graphical table of contents. It is given here in some journals
\begin{tocentry}

Insert TOC

\end{tocentry}

%% if the target journal does not use an abstract.
%%%%%%%%%%%%%%%%%%%%%%%%%%%%%%%%%%%%%%%%%%%%%%%%%%%%%%%%%%%%%%%%%%%%%
\begin{abstract}
Despite major advances in oncology, many chemotherapeutic agents still cause severe side effects that reduce quality of life, motivating new approaches for early detection and targeted elimination of cancer cells. Luminescent transition metal complexes are promising biomolecular probes as their photo-physical properties are dependent on the surrounding environment. This makes it possible to differentiate between different environments and as a result, allows for identification of abnormalities in DNA.

However,  reliable computational protocols to predict optical properties of transition metal intercalators are limited, making accurate absorption spectra calculations essential for screening candidates.

Here, we benchmark methods for computing UV–Vis spectra of a Pt(II) pincer complex. The complex is studied both in isolation and intercalated in a small DNA model, representing probes designed to target DNA-associated molecular abnormalities.
We find that the Tamm–Dancoff approximation (TDA) and the resolution of identity (RI) approximations provide a significant increase in speed for TD-DFT with only a modest loss of accuracy. Since geometry optimization is often the dominant cost, PBEh-3c emerges as an efficient alternative to conventional DFT, introducing errors comparable to those from TDA. Tight-binding methods (GFN-xTB) offer further acceleration, but yield larger deviations in structures and UV–Vis spectra; thus, unless extensive optimization is required, PBEh-3c provides the best balance between accuracy and efficiency. The largest source of uncertainty stems from the exchange–correlation functional used in the TD-DFT calculation, where we obtain good results with PBE0 based on PBEh-3c structures.
\end{abstract}

\section{Introduction}

According to the World Health Organization (WHO)\cite{whoCancer}, cancer accounted for almost 10 million deaths in 2020\cite{iarcCancerToday}, and it is one of the most significant threats to global health \cite{ferlay2020gco}. Cancer treatment with transition metal complexes as chemotherapeutic agents started in the 1970s with the development of cisplatin\cite{ROSENBERG1969}. Subsequently, various types of platinum complexes have been approved for cancer treatment globally \cite{Imran2018}. 

Despite advances in oncology in the last decades, many of the associated side effects of chemotherapeutic agents -- some of which reduce life quality significantly -- still persist to this day.\cite{Schirrmacher2018, Oun2018}  The side effects occur since chemotherapeutic agents possess high toxicity, but have low tumor specificity, resulting in damage to healthy tissue.\cite{Schirrmacher2018}  New methods for early detection and targeted elimination of cancer cells are therefore crucial \cite{Oun2018}. To reduce side effects, treatments utilizing light, such as \ac{PDT} and \ac{PACT}, have emerged. By exploiting light as a non-invasive method to regulate the activity of the chemotherapeutic drug, \ac{PDT} and \ac{PACT} can reduce undesired off-target side reactions and increase target specificity. \cite{Nkune2025,AlonsodeCastro2018,GurruchagaPereda2019}  

\begin{figure}[htb!]
\includegraphics[width=\textwidth]{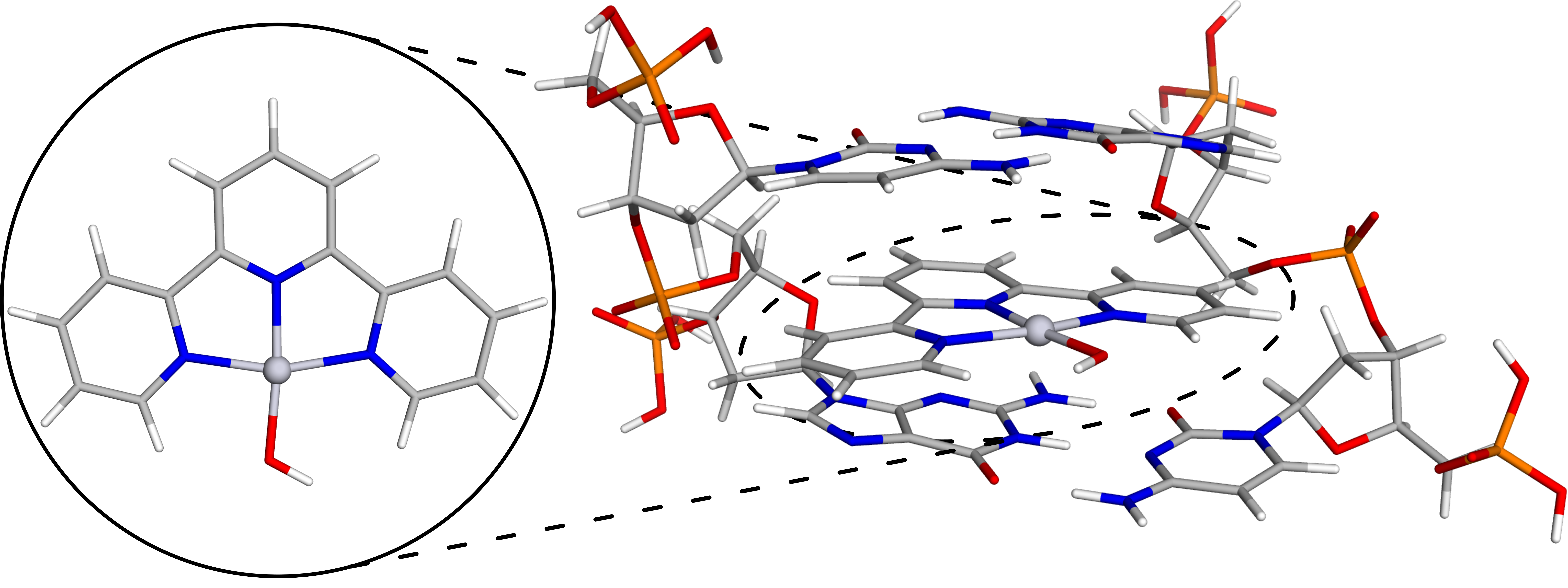}
\caption{The isolated Pt(II) model complex (\ce{[Pt(OH)(terpy)]+}) and the Pt(II) model intercalated in a double helical DNA fragment investigated in this paper.  For both systems, the oxygen atoms are red, the phosphor atoms are orange, the nitrogen atoms are blue, the hydrogen atoms are white, and the carbon and platinum atoms are gray.}
\label{fig:system_and_zoom}
\end{figure}

In addition to regulating the activity of anti-cancer agents, light can also induce luminescence.\cite{Ma2013,baggaley2014,berrones-reyes2021} Luminescence is exploited in optical microscopy to visualize and detect biological processes such as tumor formation.\cite{Ma2013,baggaley2014,berrones-reyes2021} Since cancer is associated with specific molecular abnormalities in DNA\cite{Salmaninejad2021} (e.g.~G-quadruplex structures and mismatched DNA), these abnormalities can serve as cancer-associated targets.\cite{Fung2016} Transition metal complexes are highly versatile synthetic platforms and target abnormalities in DNA.\cite{Imberti2019, Imran2018} Moreover, they often have long luminescence lifetimes\cite {Ma2013,baggaley2014} -- properties that make transition metal complexes promising as probes in optical microscopy. Many of the complexes currently investigated as probes have a planar structure that allows them to \textit{intercalate} between DNA base pairs, as shown in Figure \ref{fig:system_and_zoom} for a Pt(II) complex with a pincer ligand -- an early known class of transition metal intercalators.\cite{Peyratout1995, Liu2011}. In addition to their ability to intercalate, their photo-physical properties change depending on the surrounding environment. This behavior enables differentiation between different environments, which again allows identification of abnormalities in DNA.\cite{Fung2016} In some cases, it is even possible that the probes themselves show significant anti-cancer activity.\cite{Zou2014}

Computational methods have become an integral part of drug discovery.\cite{darocha2024,sun2025} However, for luminescent transition metal biomarkers, we are still far from establishing efficient and accurate protocols. At the very least, it is important to know the absorption properties of a potential transition metal biomarker. This requires a method based on \ac{QM}\cite{scoditti2021, Creutzberg2020, Creutzberg2023, Hedegaard2022, Freitag2021, zhao2013-1, Zhao2013-2, Shaili2019, Salassa2009, Mackay2009, Farrer2010, Westendorf2011, Westendorf2012} that can target excited states. Further, the heavy atoms generally require relativistic methods\cite{Saue2011}, and if it is also intercalated in a DNA segment, the efficiency of the  \ac{QM} method can become a concern. Establishing an adequate protocol that is computationally feasible without too large sacrifices in accuracy is therefore crucial to allow investigations on real systems.  

In this paper, we take the first steps to establish a protocol for calculating absorption spectra of intercalated pincer ligands used as anti-cancer drugs and probes. We will use 2,2$'$:6$'$,2$''$-terpyridine platinum(II) hydroxide -- \ce{[Pt(OH)(terpy)]+} in short -- as a model. It will be used in both an isolated form and intercalated in the small DNA duplex (see Figure \ref{fig:system_and_zoom}), formed from the tetranucleotide 5$'$-d(CGCG)-3$'$. This model was chosen for its structural simplicity and because it shares key structural features with complexes known to exhibit anti-cancer activity.\cite{Zou2014,baggaley2014,Fung2016} Further, UV-Vis spectra for both the intercalated and isolated complex have been obtained for \ce{[Pt(OH)(terpy)]+}\cite{Peyratout1995} and X-ray structures of very similar Pt(II) complexes intercalated in the same tetranucleotide exist, validating the mode of intercalation.\cite{Wang1978} 

We first investigate different methods to obtain the underlying structure of both isolated and intercalated complexes, which is a prerequisite for accurate spectra calculations. This is an obvious step for savings of computational resources since fast \ac{DFT}-based methods for structure optimisations\cite{Grimme2015, Grimme2017, Bannwarth2019} have in recent years been developed. However, their accuracy for intercalated transition metal complexes has, to the best of our knowledge, not been investigated. While the underlying structure is one important factor for the calculated UV-Vis spectrum, another important factor is the approximations used to calculate the excited states and the associated oscillator strengths. While we stay within a \ac{TD-DFT} formulation, we investigate the effect of \ac{SOC}, the \ac{TDA}, and  \ac{RI}. The choice of basis sets and exchange-correlation functional was also investigated. Since we do not include dynamics in this study, we will, in most cases, use DFT as a reference against the more approximate methods rather than an experimental reference. We will only occasionally discuss the performance with respect to the experimental spectra, e.g., in the question of which DFT functional that is most likely to perform best.  

\section{Computational Details}

The \ce{[Pt(OH)(terpy)]+} complex (see Figure \ref{fig:system_and_zoom}) was built in the Maestro 14.3\cite{maestro14.3} software. The complex was then optimized as a closed-shell singlet state in ORCA 6.0.1\cite{Neese2020}, using the methods HF-3c\cite{Sure2013}, PBEh-3c \cite{Grimme2015}, PBE\cite{Perdew1996}/def2-SVP\cite{Weigend2005}, PBE0\cite{Perdew1996}/def2-SVP, and B3LYP\cite{Becke1993, Lee1988, Vosko1980, Stephens1994}/def2-SVP. For the def2 basis sets, an effective core potential is included for platinum. For the last three DFT methods, we also carried out calculations with the scalar-relativistic X2C Hamiltonian\cite{Kutzelnigg2005, Liu2006, Ilia2007} and the corresponding x2c-SVPall\cite{Franzke2020} basis set. Henceforth, when the x2c-SVPall basis set is mentioned, it will be implied that the scalar-relativistic X2C Hamiltonian was used. All \ac{DFT} structure optimizations included Grimme's D3 dispersion correction\cite{Grimme2016} with \ac{BJ} damping\cite{Schrder2015, Becke2005, Becke2005.2}. In ORCA, the keyword tightopt was used along with the \ac{CPCM}\cite{Marenich2009} representing a water solvent for all geometry optimizations.

%\textcolor{blue}{A second system with explicit solvation was also made for the \ce{[Pt(OH)(terpy)]+} complex using the geometry optimized with PBE/def2-SVP. In PACKMOL\cite{Martnez2009}, the optimized \ce{[Pt(OH)(terpy)]+} complex was solvated in a sphere of 20 Å containing 1125 water molecules. In Maestro, the OPLS$\_$2005 force field was used through the minimization tool to optimize the solvent while the solute was frozen. Afterwards, with the complex selected, a 6 Å expansion was made. This selection was then used to make an excerpt from the 20 Å sphere (See Figure S1). Next, the excerpt, with the solute frozen, was optimized in ORCA using GFN-FF and the keyword opt.}

Additionally, a B-DNA (the most common conformation of DNA\cite{Hartmann1996}) fragment with two guanine and cytosine base pairs (5'-CG-3') was built in Maestro (see Figure \ref{fig:system_and_zoom}). The \ce{[Pt(OH)(terpy)]+} complex was intercalated manually in the B-DNA model between the guanine and cytosine base pairs using Maestro and then optimized with PBEh-3c. On the four terminal phosphate groups, hydrogen atoms were added to reduce the negative charge from -5 to -1 (see Figure S1). Two additional structures were optimized: One without added hydrogen atoms (charge -5) and one with hydrogen atoms added to all of the phosphate groups (charge +1). A visual comparison of the UV-Vis spectra obtained with each of these underlying structures showed that they did not differ significantly (see Figure S2).Thus, only the structure with a charge of -1 was considered for the remaining spectra calculations. Afterwards, the PBEh-3c-optimized structure was used for the geometry optimizations of the intercalated complex with PBE/def2-SVP, PBE0/def2-SVP, GFN1-xTB \cite{Grimme2017}, and GFN2-xTB. \cite{Bannwarth2019}

The UV-Vis spectra discussed in the Results and Discussion section were calculated with PBE0/x2c-SVPall with the exception of the part discussing the effect of using the range-separated functional, LC-PBE. These calculations were performed with the default value (0.47)  for the range-separation parameter, $\omega$. We additionally investigated different $\omega$ values ranging from 0.20 to 0.40 in intervals of 0.05. All calculations with changed $\omega$ were performed with \ac{TDA} and the \ac{RI}; both \ac{TDA} and the \ac{RI} were generally disabled in all other calculations, unless explicitly stated. 
All \ac{TD-DFT} calculations included \ac{CPCM} to represent water.

The effect of \ac{TDA} and \ac{SOC} where tested on  geometries obtained with PBE/def2-SVP for the isolated and intercalated complex, respectively. This was done by comparing calculations that 1) included or disabled \ac{SOC}\cite{deSouza2019} (with TDA disabled) and 2) included or disabled \ac{TDA} (with \ac{SOC} included). We further investigated the use of the effect of employing \ac{RI} as well as the use of larger basis sets (PBE0/def2-TZVP\cite{Weigend2005} and  PBE0/def2-TZVPD\cite{Weigend2005, Rappoport2010, andrae1990a}). Upon visual inspection, the  effect of both the \ac{RI} and expanding the basis set was estimated to be rather small, and these results are therefore given in the SI (Figures S3 and S4, respectively). The \ac{TD-DFT} calculations were done with  24 states when the \ac{SOC} was not included and 79 states when the \ac{SOC} was included.

In addition to visual comparison, we also quantified differences between the obtained spectra. As a method for comparison, the \ac{MAE} and \ac{MSE} were chosen. The objective of the comparison is to determine if it is possible to use a method of lower accuracy, but higher computational efficiency. The formulae are ${\sum^{n}_{i=1}|y_i-x_i|}{n^{-1}}$ and ${\sum^{n}_{i=1}(y_i-x_i)}{n^{-1}}$ for the \ac{MAE} and \ac{MSE}, respectively.
The values were calculated for both the energies and the intensities. To determine the values for the energy, the energies obtained from the \ac{SOC} corrected stick spectrum were used. Ideally, the \ac{MAE} and \ac{MSE} values for the energy would have been determined by comparing each state from one method with the corresponding state in another method. It was not possible to accurately determine the corresponding state, as access to calculate the overlap could not be obtained through ORCA. Instead, the states were compared in order of appearance on the spectrum.

For the intensities, the values were determined from the intensities of the smeared spectra. In all cases, intensities are compared starting from 2.5 eV for the isolated complex and from 2.0 eV for the intercalated complex. The only exception is the comparison between range-separated and non-range-separated functionals for the intercalated complex. Here, we begin from 2.5 eV. The smeared spectra were obtained by convoluting the stick spectra with Gaussian functions for each state, using a \ac{FWHM} of 0.3 eV.

For the structural comparisons between the obtained geometries, the \ac{RMSD} was chosen as a method for comparison. The formula for \ac{RMSD} is given by $\sqrt{{\bigl(\sum_{i=1}^{n}(x_j-x_i)^2+(y_j-y_i)^2+(z_j-z_i)^2}\bigl){n^{-1}}}$ where \textit{x}, \textit{y}, and \textit{z} are the cartesian coordinates of an atom in the system. The value was calculated using Maestro. For the intercalated complex, an additional \ac{RMSD}-value was calculated; a ``local'' value that only encompasses the isolated complex and the base pairs (i.e. excluding the DNA backbone).

\section{Results and discussion}

We first discuss the method employed for the underlying structure and its impact on the UV-Vis spectra from a subsequent \ac{TD-DFT} calculation. Next, we investigate the approximations to the electronic structure in greater detail. 

\subsection{Structures}
 We have here chosen PBE/def2-SVP as the reference, i.e.,  each optimized structure was superimposed on the PBE/def2-SVP structure. These overlays will be used for a qualitative assessment by visually comparing the structural differences in addition to the calculated  \ac{RMSD}-values.

We first discuss the \ac{RMSD}-values for the isolated complex compiled in Table \ref{tab:isolated-intercalated-RMSD-values}. The values obtained for the PBEh-3c method are comparable or only slightly larger than the \ac{RMSD}-values between the different functionals, while the difference can almost not be discerned visually based on the overlays (see Figures S5--S7).
\begin{table}[htb!]
\centering
\caption{Overview of obtained \ac{RMSD}-values (Å) for the optimized geometries of the isolated complex compared to the geometry obtained with PBE/def2-SVP. For the intercalated complex, the values in parentheses are the ``local'' \ac{RMSD}-values, with only the platinum complex and the base pairs included.}
\begin{tabular}{l c  c}
\hline
\hline
                            \textbf{Method}  &  \textbf{Isolated Complex} & \textbf{Intercalated Complex} \\
 \hline
 \hline
 \multicolumn{1}{l}{\textbf{B3LYP/def2-SVP}} & 0.023                   &   Not calculated  \\
 
 \multicolumn{1}{l}{\textbf{PBE0/def2-SVP}}  & 0.027                      &  0.492 (0.308)  \\

 \multicolumn{1}{l}{\textbf{HF-3c}}         & 0.049                      &  Not calculated  \\
 \multicolumn{1}{l}{\textbf{PBEh-3c}}       & 0.088                      & 0.509 (0.239)     \\ 
 \multicolumn{1}{l}{\textbf{GFN1-xTB}}      &  Not calculated            & 0.968  (0.691)   \\
 \multicolumn{1}{l}{\textbf{GFN2-xTB}}      &  Not calculated             & 0.703 (0.459)    \\
\hline
\hline
\end{tabular}
\label{tab:isolated-intercalated-RMSD-values}
\end{table}
The most noticeable visual discrepancy between PBEh-3c and the reference is at the hydroxy ligand, suggesting that the semi-empirical methods can be used to increase computational efficiency during the geometry optimization. 

\begin{figure}[htb!]
    \centering
    \begin{subfigure}[b]{0.48\textwidth}
        \centering
        \includegraphics[width=\linewidth]{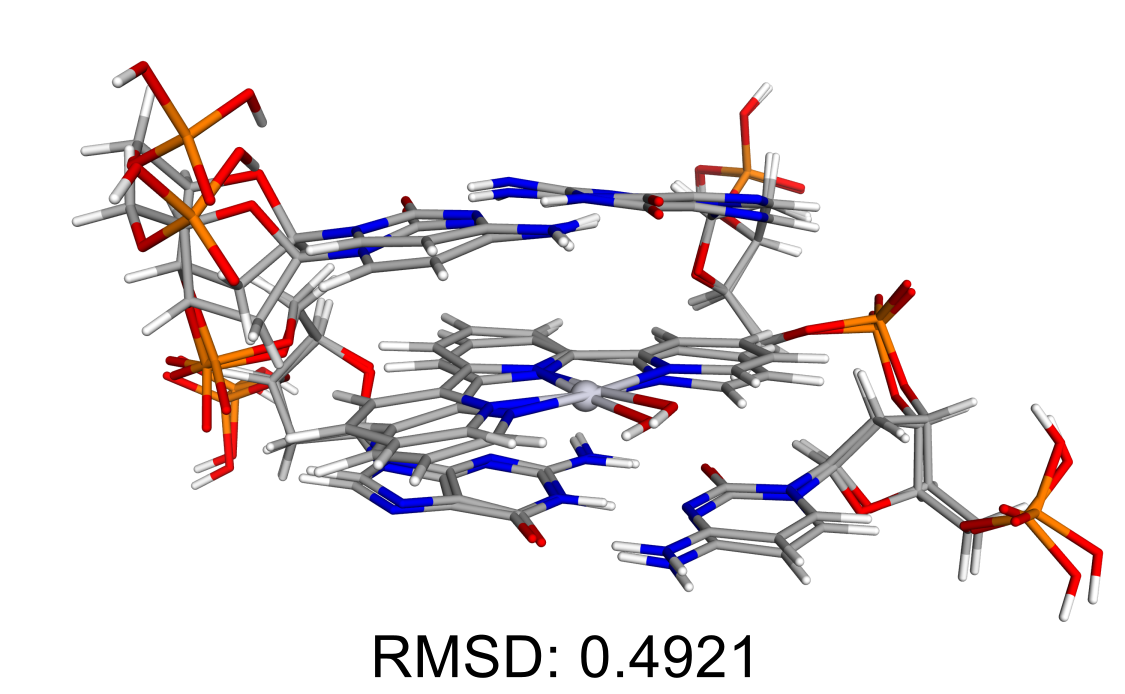}
        \caption{}
        \label{pbe0}
    \end{subfigure}
    \hfill
    \begin{subfigure}[b]{0.48\textwidth}
        \centering
        \includegraphics[width=\linewidth]{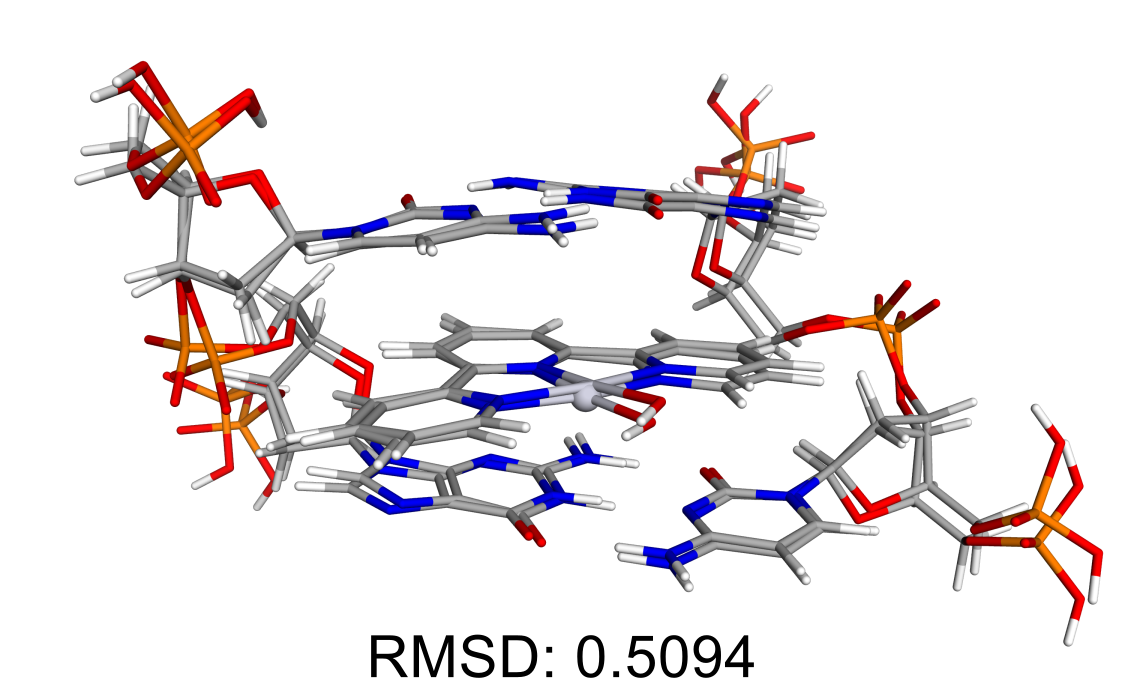}
        \caption{}
        \label{pbeh3c}
    \end{subfigure}
    \caption{The intercalated complex optimized with (a) PBE0/def2-SVP and (b) PBEh-3c, respectively, overlayed on the PBE/def2-SVP structure. All structures have a total charge of –1.}
    \label{fig:intercalated_structures_and_overlays}
\end{figure}

For the intercalated complex, the structures from PBE0/def2-SVP and PBEh-3c are shown in  Figure \ref{fig:intercalated_structures_and_overlays} and overlayed with the reference structure (PBE/def2-SVP). The structures optimized with PBE0/def2-SVP and PBEh-3c both have an \ac{RMSD}-value of 0.51 Å. Visual inspection of Figure \ref{pbeh3c} shows that the largest discrepancies occur at the sugar-phosphate backbone. We therefore calculated a ``local'' \ac{RMSD}-value, including only the base pairs and the platinum complex. This value is, as expected, significantly smaller (0.30 and 0.24 Å for PBE0 and PBEh-3c, respectively, cf.~Table \ref{tab:isolated-intercalated-RMSD-values}), although large enough to expect changes in the calculated UV-Vis spectrum.

\begin{figure}[htb!]
    \centering
    \begin{subfigure}[b]{0.48\textwidth}
    \centering
    \includegraphics[width=\linewidth]{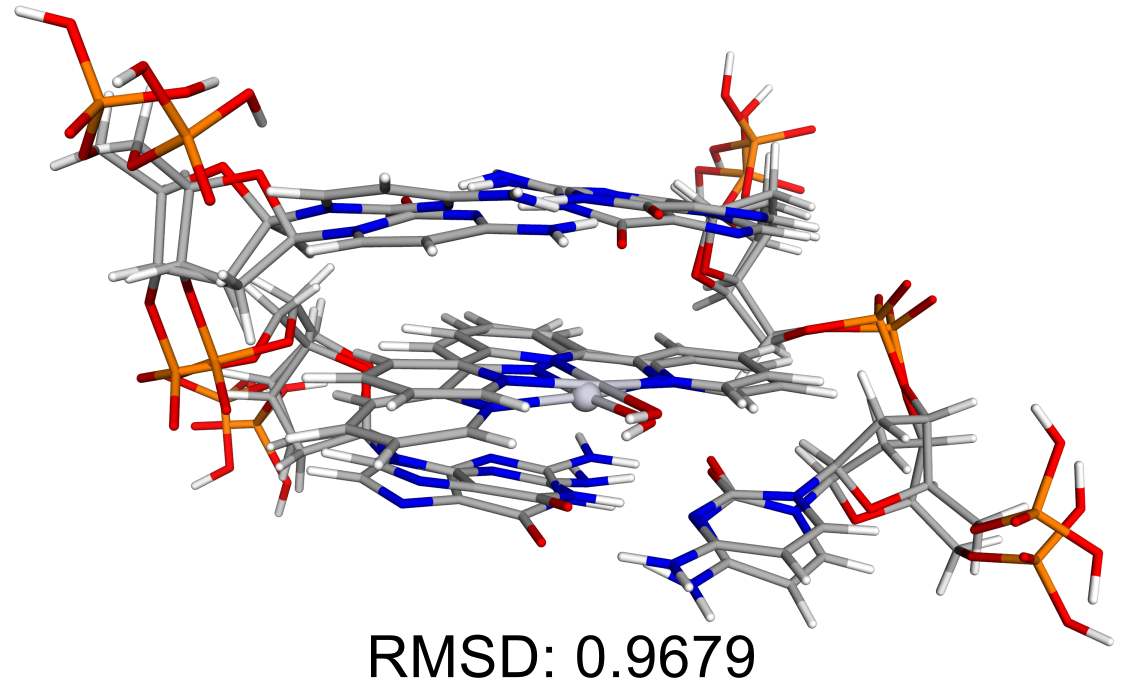}
        \caption{}
        \label{fig:RFS_xtb1}
     \end{subfigure}
     \begin{subfigure}[b]{0.48\textwidth}
    \centering
    \includegraphics[width=\linewidth]{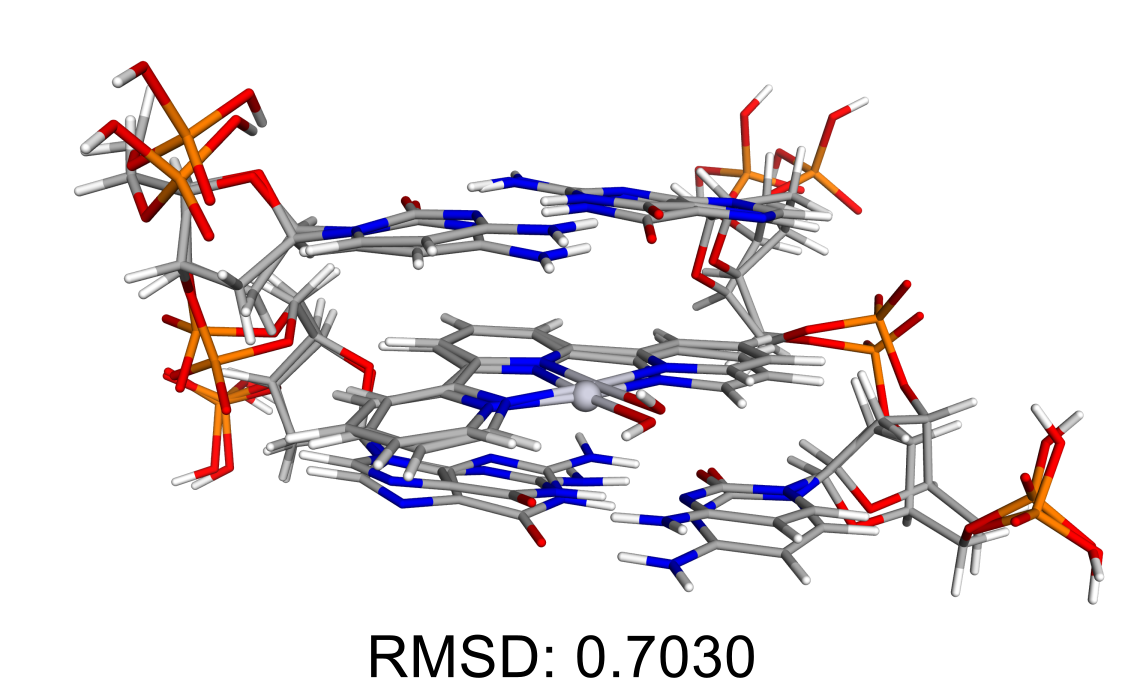}
        \caption{}
        \label{fig:RFS_xtb2}
     \end{subfigure}
    \caption{The intercalated complex optimized with (a) GFN1-xTB and (b) GFN2-xTB, respectively, overlayed on the PBE/def2-SVP structure. All structures have a total charge of –1.}
    \label{fig:RFS_overlays_intercalated}
\end{figure}

Similar comparisons were made for the tight-binding \ac{DFT} methods GFN-1xTB and GFN-2xTB.\cite{Bannwarth2020} The obtained geometries are shown in Figure \ref{fig:RFS_overlays_intercalated},  overlayed with the reference structure.
Here, the deviations are larger than the results obtained through PBEh-3c (RMSD of 0.97 Å and 0.70 Å for GFN1-xTB and GFN2-xTB, respectively). The ``local'' \ac{RMSD}-values remain relatively high (RMSD of 0.69 Å and 0.46 Å, see Table \ref{tab:isolated-intercalated-RMSD-values}).

\begin{figure}[htb!]
    \centering
    \begin{subfigure}[b]{0.48\textwidth}
        \centering
        \includegraphics[width=\textwidth]{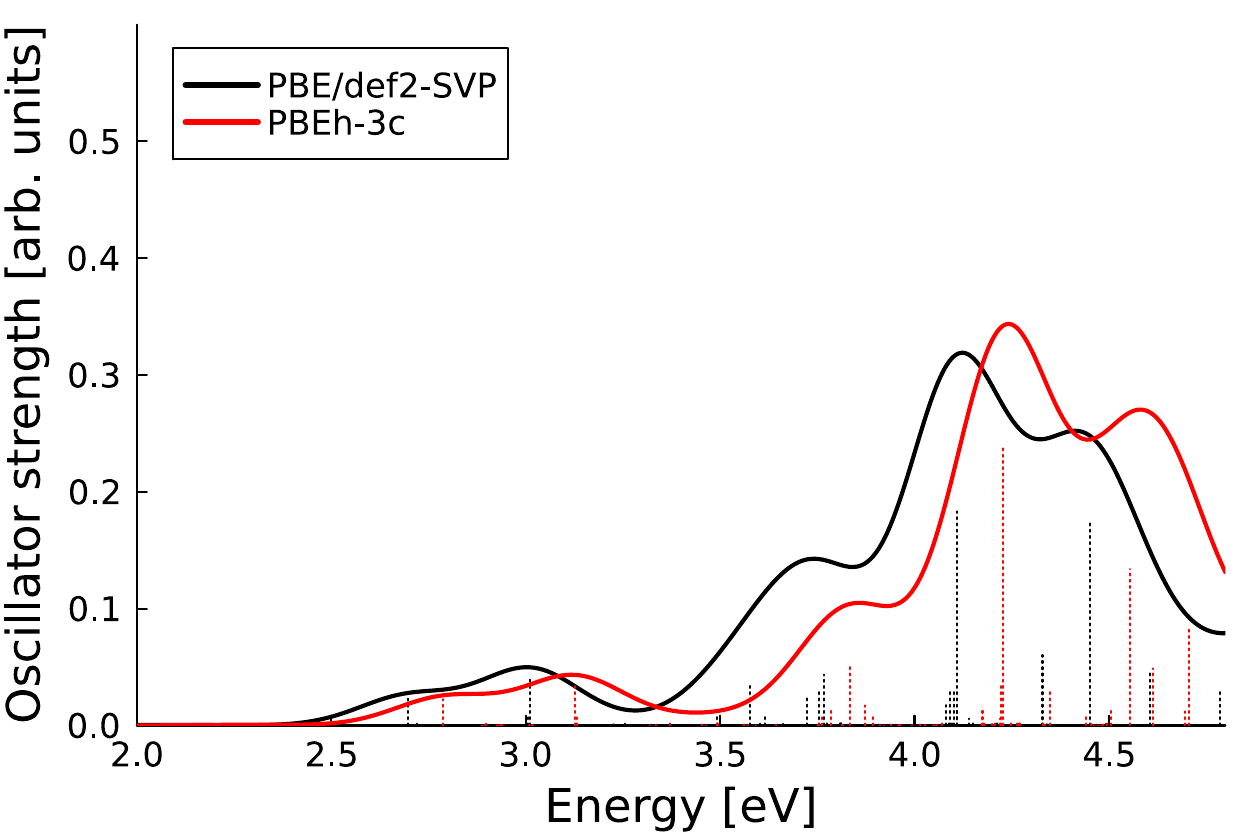}
        \caption{}
        \label{fig:iso_method_comparison_PBEh-3c_and_PBE}
    \end{subfigure}
    \hfill
    \begin{subfigure}[b]{0.48\textwidth}
        \centering
        \includegraphics[width=\textwidth]{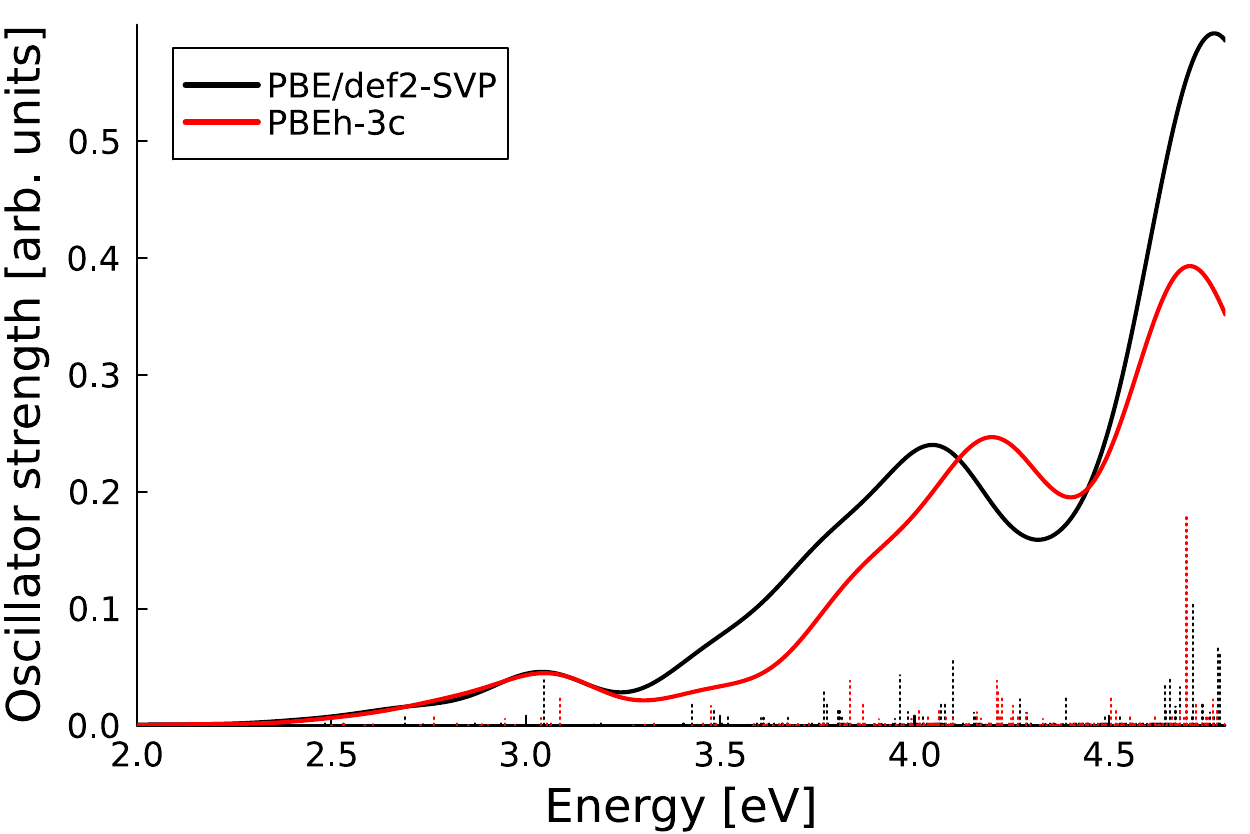}
        \caption{}
        \label{fig:inter_method_comparison_PBEh-3c_and_PBE}
    \end{subfigure}
    \caption{UV-Vis spectra calculated with PBE0/x2c-SVPall and \ac{SOC}, but without \ac{TDA} and the \ac{RI} for (a) the isolated complex and (b) the intercalated complex. The compared structures were optimized with PBE/def2-SVP (black) and PBEh-3c (red). A \ac{FWHM} of 0.3 eV was used to smear the spectra.}
    \label{fig:method_comparison_PBEh-3c_and_PBE}
\end{figure}

To investigate the effect of the underlying structure on the calculated spectra, we conducted \ac{TD-DFT} calculations on the geometries obtained from PBEh-3c and PBE/def2-SVP for both systems; the results are shown in Figure \ref{fig:iso_method_comparison_PBEh-3c_and_PBE} for the isolated complex. All \ac{MAE} and \ac{MSE} values are compiled in Table \ref{tab:MAE_MSE_comparison}. For the isolated complex, the \ac{MAE} and \ac{MSE} values suggest a small blue shift for the energies (both values are 0.01 eV) when using PBEh-3c as the underlying structure compared to the PBE reference. Visually, the overall shape of the spectrum is preserved (see Figure  \ref{fig:iso_method_comparison_PBEh-3c_and_PBE}). This is also seen for the  \ac{MAE} and \ac{MSE} for the intensities, where the obtained values are 0.04 eV and –0.01 eV, indicating a small, non-systematic shift in the intensities.

For the intercalated complex (Figure \ref{fig:method_comparison_PBEh-3c_and_PBE}b), the \ac{MAE} and \ac{MSE} values for the energies are slightly larger (0.07 eV and 0.07 eV, respectively), but also show a blue shift in the spectrum when PBEh-3c is used as the underlying structure. The intensities are more systematically overestimated than for the isolated complexes (with \ac{MAE} and \ac{MSE} values of 0.09 eV and 0.07 eV, respectively). As seen from Figure \ref{fig:method_comparison_PBEh-3c_and_PBE}, this is mainly driven by the peak above 4.5 eV. 

\begin{table}[htb!]
\centering
\caption{Compilation of \ac{MAE} and \ac{MSE} values for excitation energies ($\Delta E$) and intensities ($\Delta I$) for both isolated and intercalated complexes). The first three entries are based on UV-Vis spectra calculated with PBE0/x2c-SVPall, using different methods for the geometry optimization, whereas the two last use different methods to obtain the UV-vis spectra on geometries optimized with PBE/def2-SVP  \label{tab:MAE_MSE_comparison} }
\begin{tabular}{l c c c c c c c c}
\hline
\hline
\textbf{Method}  &  \multicolumn{2}{c}{\textbf{Isolated  ($\Delta E$)}}  &  \multicolumn{2}{c}{\textbf{Isolated  ($\Delta I$)}} & \multicolumn{2}{c}{\textbf{Intercalated ($\Delta E$)}} & \multicolumn{2}{c}{\textbf{Intercalated  ($\Delta I$)}}  \\
 \hline
 \hline
                                      &  MAE & MSE  & MAE  & MSE & MAE  & MSE & MAE  & MSE      \\ 
 \multicolumn{1}{l}{\textbf{PBEh-3c}} & 0.01 & 0.01 & 0.04 & $-$0.01 & 0.07 & 0.07 & 0.09 & 0.07                                \\ 
 \multicolumn{1}{l}{\textbf{GFN1}}    & -    & -    & -    & -   & 0.10  & 0.08   & 0.10  & 0.08  \\
 \multicolumn{1}{l}{\textbf{GFN2}}    & -    & -    & -    & -   & 0.12  & 0.10  & 0.12   & 0.10   \\
\hline
 \multicolumn{1}{l}{\textbf{PBE0 (TDA)}} & 0.07 & 0.07 & 0.03  & 0.01 & 0.03 & 0.03 & 0.07 & 0.06 \\
  \multicolumn{1}{l}{\textbf{LC-PBE }} & 0.63 & 0.62  & 0.09  & $-$0.03 & 1.49 & 1.49 & 0.13 & 0.08 \\  
\hline
\hline
\end{tabular}
\end{table}

Before discussing the tight-binding methods, we briefly discuss the results compared to experiment: The calculated spectra of the isolated complex in water (Figure \ref{fig:method_comparison_PBEh-3c_and_PBE}) resemble the experimental spectrum, which is also comprised of two bands\cite{Peyratout1995}: the first (less intense) band around 3.2 eV is assigned as a charge-transfer band, whereas the second (more intense) band (at 3.7 eV) is assigned as an intra-ligand transition. An analysis of the orbitals (See Figures S8 and S9) involved in the two peaks in the theoretical spectra confirms that the first band is of \ac{MLCT} type, whereas the second band is mainly within the ligand.

From Figure \ref{fig:iso_method_comparison_PBEh-3c_and_PBE}, we see that the spectrum based on the PBE structure estimates the first charge-transfer band for the isolated complex (in water) to be around 2.5--3.3 eV with maximum at 3.0 eV, in reasonable agreement with experiment (3.2 eV). The more intense intra-ligand band has the first maximum at 3.7 eV, which is in excellent agreement with experiment (3.7 eV). The corresponding values for the spectrum based on the PBEh-3c structure are 3.1 eV and 3.9 eV, which is also in good agreement with experiment. We emphasize (as also stated in the Introduction), that whether the methods provide an excellent fit with experiment may change when including dynamics and explicit solvation, and without including these factors,  we have preferred to use a best theoretical estimate for the underlying structure, choosing the PBE structure as the best theoretical estimate.

Moving now to the intercalated complex, the experimental  spectrum has the same shape as the isolated one, but the first charge-transfer excitation has maximum at 3.0 eV\cite{Peyratout1995}, i.e., slightly red shifted  compared to the isolated complex (3.2 eV). The spectrum based on structures from PBE does not obtain this red shift upon intercalation, the obtained spectrum for the intercalated complex is rather close to experiment: the first charge-transfer for the intercalated complex (black spectrum in Figure \ref{fig:inter_method_comparison_PBEh-3c_and_PBE}) has a maximum at 3.1 eV, slightly blue shifted compared to the isolated one (at 3.0 eV; black spectrum in Figure \ref{fig:iso_method_comparison_PBEh-3c_and_PBE}). With the structure obtained with PBEh-3c, we obtain a maximum at 3.1 eV for the \ac{MLCT} band, which is essentially identical to the isolated complex. Only a closer investigation reveals a small red shift of approximately 0.04 eV (using the transitions at 3.13 eV and 3.09 eV for isolated and intercalated complexes, respectively). Although this is in the right direction, it is below the experimental 0.2 eV. We  discuss this further in a section below.

\begin{figure}[htb!]
    \centering
    \begin{subfigure}[b]{0.48\textwidth}
    \centering
    \includegraphics[width=\linewidth]{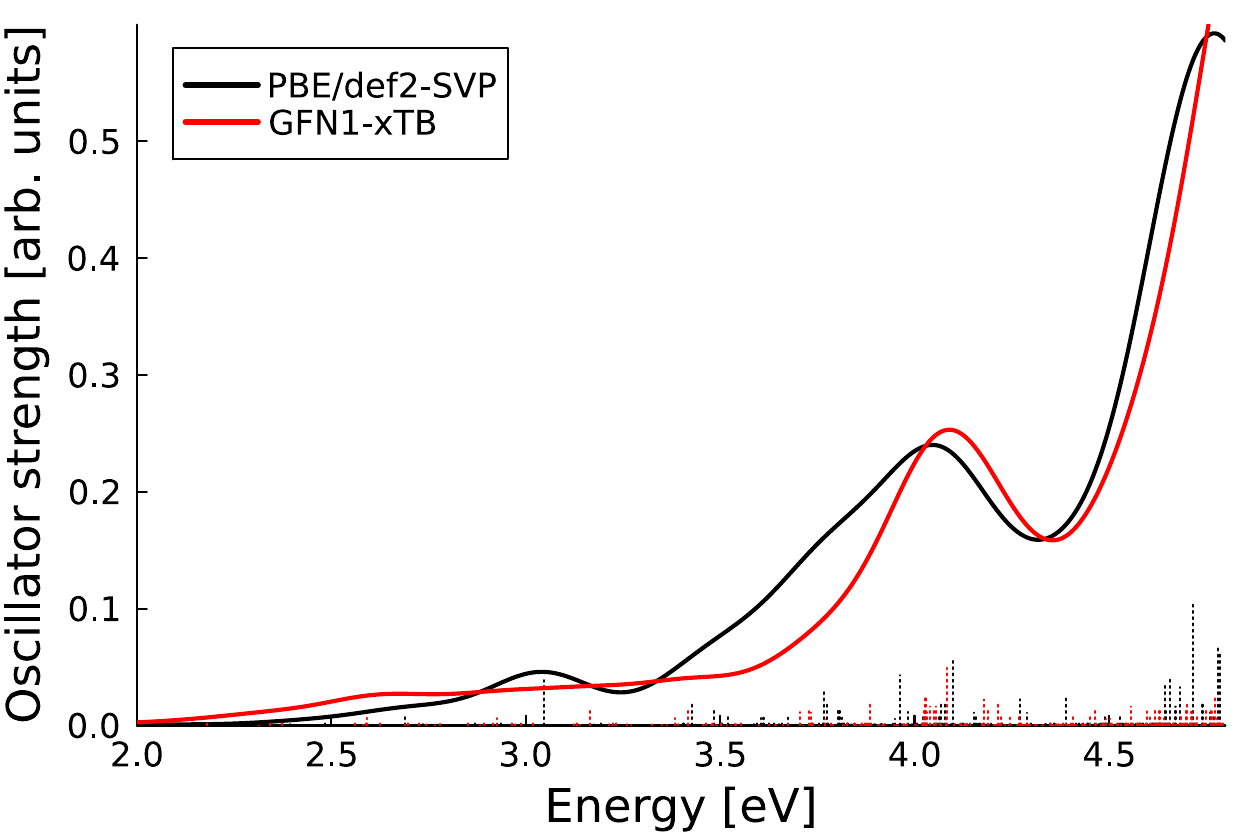}
        \caption{}
        %\label{xtb1}
     \end{subfigure}
     \begin{subfigure}[b]{0.48\textwidth}
    \centering
    \includegraphics[width=\linewidth]{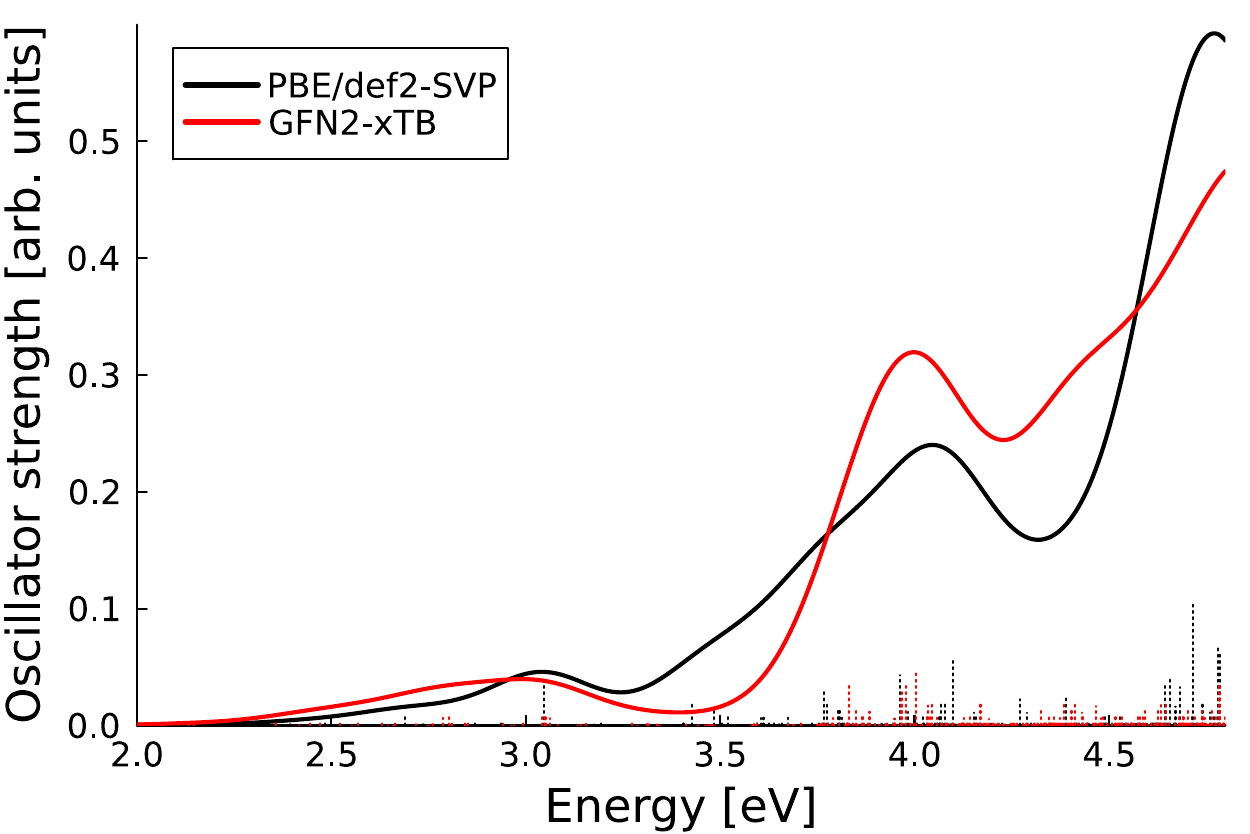}
        \caption{}
        %\label{xtb2}
     \end{subfigure}
    \caption{UV-Vis spectra for the intercalated complex calculated with PBE0/x2c-SVPall and \ac{SOC}, but without \ac{TDA} and the \ac{RI}. The compared structures were optimized with PBE/def2-SVP in black and in red (a) GFN1-xTB and (b) GFN2-xTB. A FWHM of 0.3 eV was used to smear the spectra.}
    \label{fig:UV-Vis_RFS_intercalated}
\end{figure}

The geometries obtained with the two tight-binding methods were also employed in \ac{TD-DFT} calculations. The results are shown in Figure \ref{fig:UV-Vis_RFS_intercalated} and Table \ref{tab:MAE_MSE_comparison}. 
In both cases, the overall spectrum is still maintained with respect to the reference (PBE/def2-SVP), although the differences are larger than for PBEh-3c, both for the \ac{MAE} and \ac{MSE} values for the energies/intensities. Combined with the higher \ac{RMSD}-values for the underlying structures, we recommend using PBEh-3c for fast structure optimization. In fact, the RMSD is lower for PBEh-3c than if we use the PBE0 functional as shown in Figure \ref{pbe0}.

\subsection{Level of Theory for Obtaining UV-Vis Spectra}

We now discuss three factors related to the electronic structure of the isolated and intercalated \ce{[Pt(OH)(terpy)]+} complex. Namely, 1) the effect of \ac{SOC}, 2) the use of \ac{TDA}, 3) the effects of changing to a range-separated version of PBE, specifically LC-PBE. We discuss the factors individually below. 

\begin{figure}[htb!]
\begin{subfigure}{.475\linewidth}
  \includegraphics[width=\linewidth]{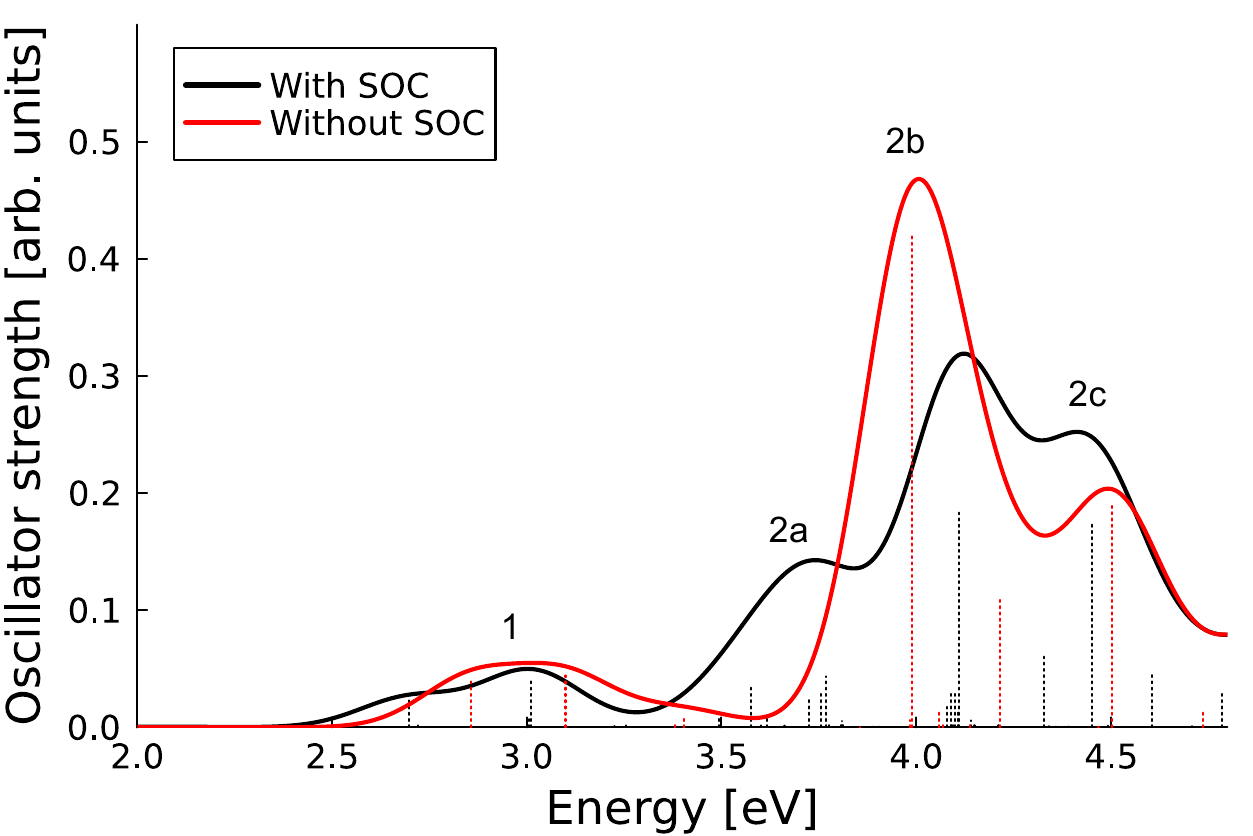}
  \caption{}
  \label{fig:SOC_test_isolated}
\end{subfigure}\hfill % <-- "\hfill"
~ % optional tilde b/t figures for readability.
  % this solution will not work w/ empty lines b/t subfigures
\begin{subfigure}{.475\linewidth}
  \includegraphics[width=\linewidth]{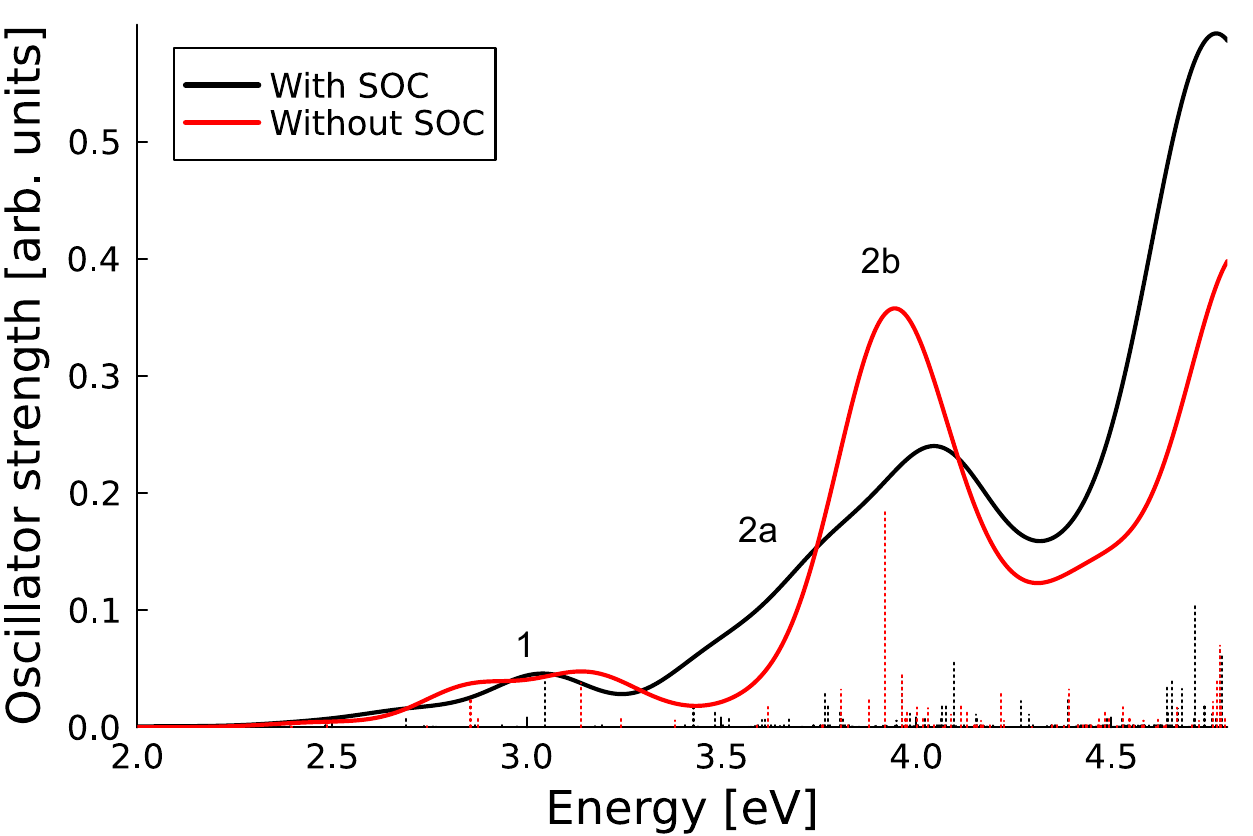}
  \caption{}
  \label{fig:SOC_test_intercalated}
\end{subfigure}
\medskip % create some *vertical* separation between the graphs
\begin{subfigure}{.475\linewidth}
  \includegraphics[width=\linewidth]{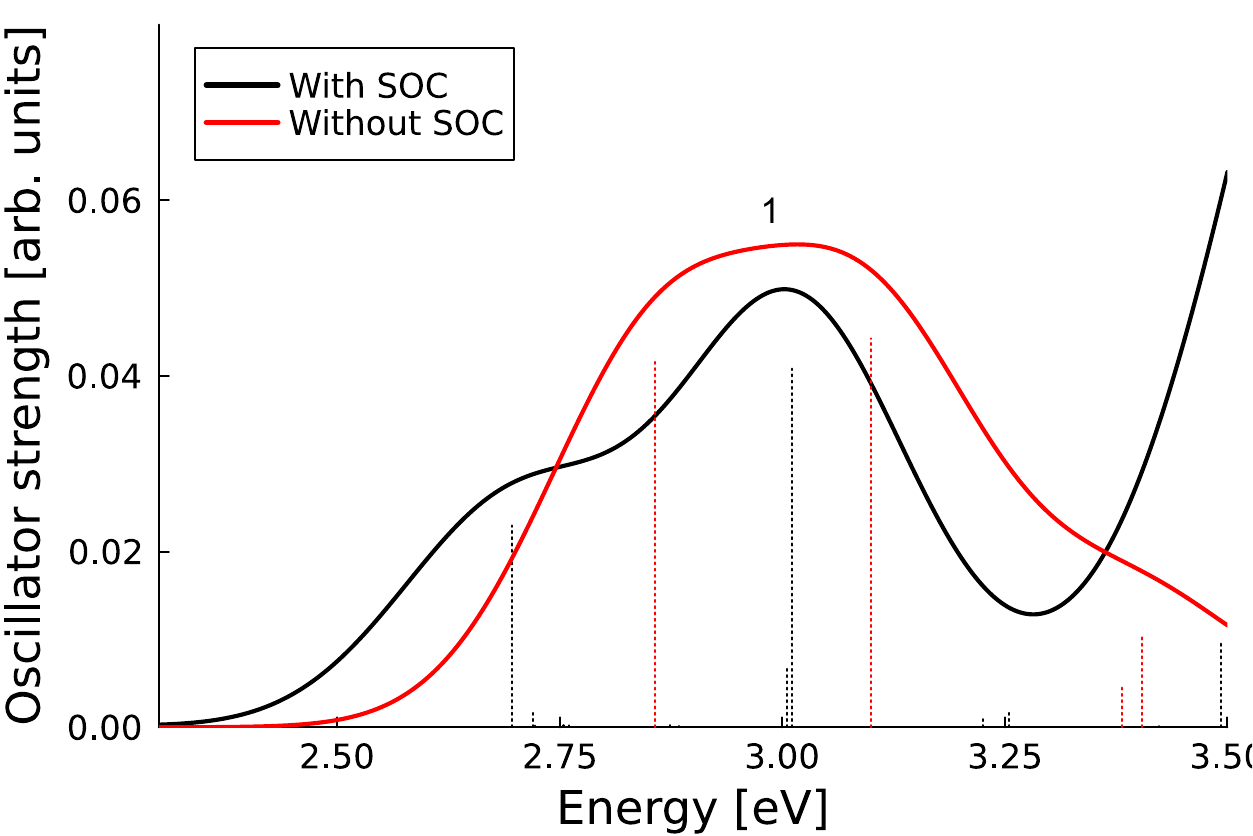}
  \caption{}
  \label{fig:SOC_test_isolated_zoom}
\end{subfigure}\hfill % <-- "\hfill"
% or just a comment and no tilde
\begin{subfigure}{.475\linewidth}
  \includegraphics[width=\linewidth]{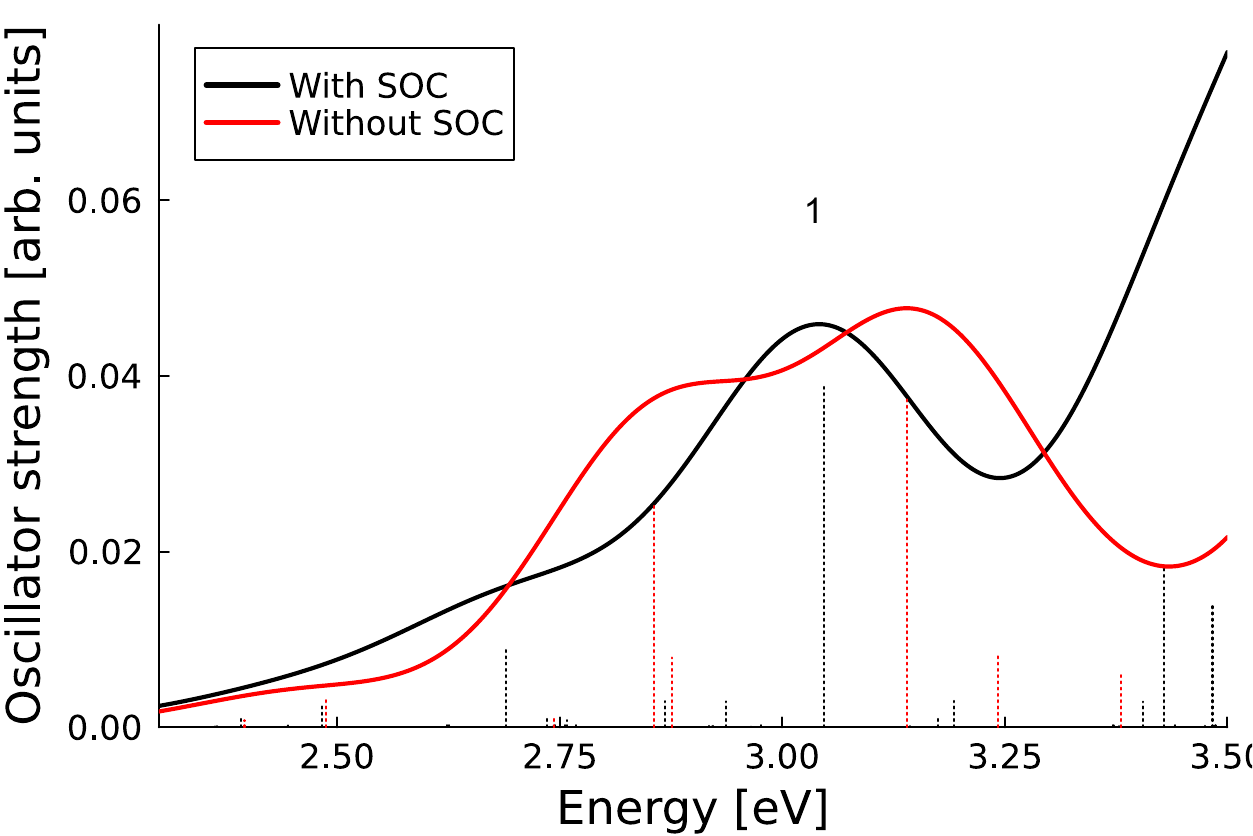}
  \caption{}
  \label{fig:SOC_test_intercalated_zoom}
\end{subfigure}
\caption{UV-Vis spectra calculated with PBE0/x2c-SVPall without \ac{TDA} and the \ac{RI}. Calculated with (black) and without (red) \ac{SOC} for (a) the isolated complex and  (b) the intercalated complex. (c) and (d) are enlargements of band \textbf{1} (2.3–3.5 eV). The structures were optimized with PBE/def2-SVP. A \ac{FWHM} of 0.3 eV was used to smear the spectra.}
\label{fig:SOC}
\end{figure}

\noindent \textbf{Effect of Spin-Orbit Coupling.}
The UV-Vis spectra with and without \ac{SOC} are shown in Figures \ref{fig:SOC_test_isolated} and \ref{fig:SOC_test_intercalated} with an enlargement on the first band (\textbf{1}, 2.3--3.5 eV) in  Figures \ref{fig:SOC_test_isolated_zoom} and \ref{fig:SOC_test_intercalated_zoom}. 
Visually, the two spectra keep their overall form with and without \ac{SOC}. There are, however, large changes in the individual bands: The first band (\textbf{1}) is both with and without \ac{SOC} comprised of two main transitions, but the energies are red shifted (and oscillator strengths are slightly lowered) through the \ac{SOC}. This is seen for both the isolated and intercalated complex. The changes are even larger for band \textbf{2} around 4.1--4.5 eV. The spectrum obtained without \ac{SOC} is mainly comprised of two strong transitions (\textbf{2b} and \textbf{2c}), while the spectrum obtained with \ac{SOC} has several additional transitions that contribute significantly. This is expected as \ac{SOC} lifts the degeneracy of the triplet state and allows singlet-triplet transitions to have non-zero oscillator strengths. The result is that contributions to peak \textbf{2b} are distributed among multiple less intense states in the \ac{SOC}  corrected spectrum. Moreover, employing \ac{SOC}  results in an additional peak (\textbf{2a}). Clearly, the  \ac{SOC} should not be neglected, and we recommend always using SOC when describing these types of platinum complexes. 

\begin{figure}[htb!]
\begin{subfigure}{.475\linewidth}
  \includegraphics[width=\linewidth]{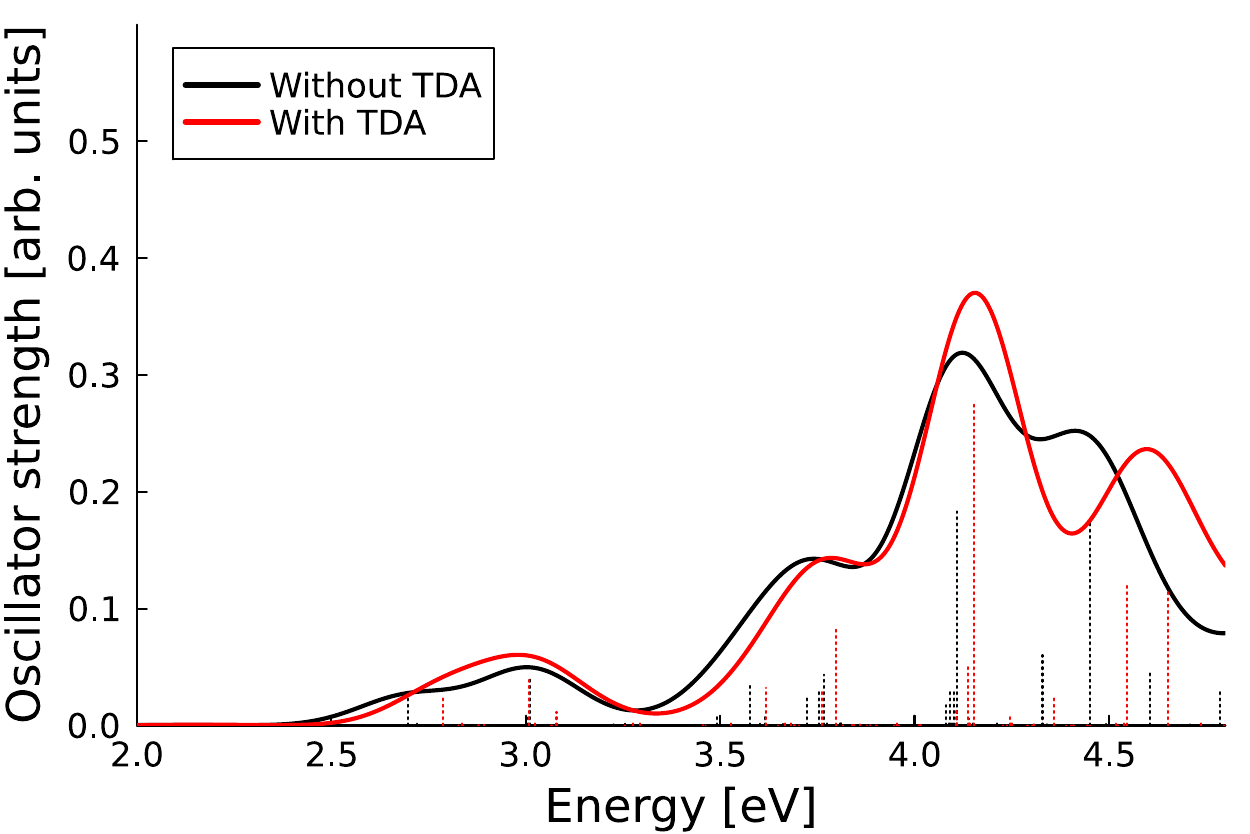}
  \caption{}
  \label{fig:TDA_test_isolated}
\end{subfigure}\hfill % <-- "\hfill"
~ % optional tilde b/t figures for readability.
  % this solution will not work w/ empty lines b/t subfigures
\begin{subfigure}{.475\linewidth}
  \includegraphics[width=\linewidth]{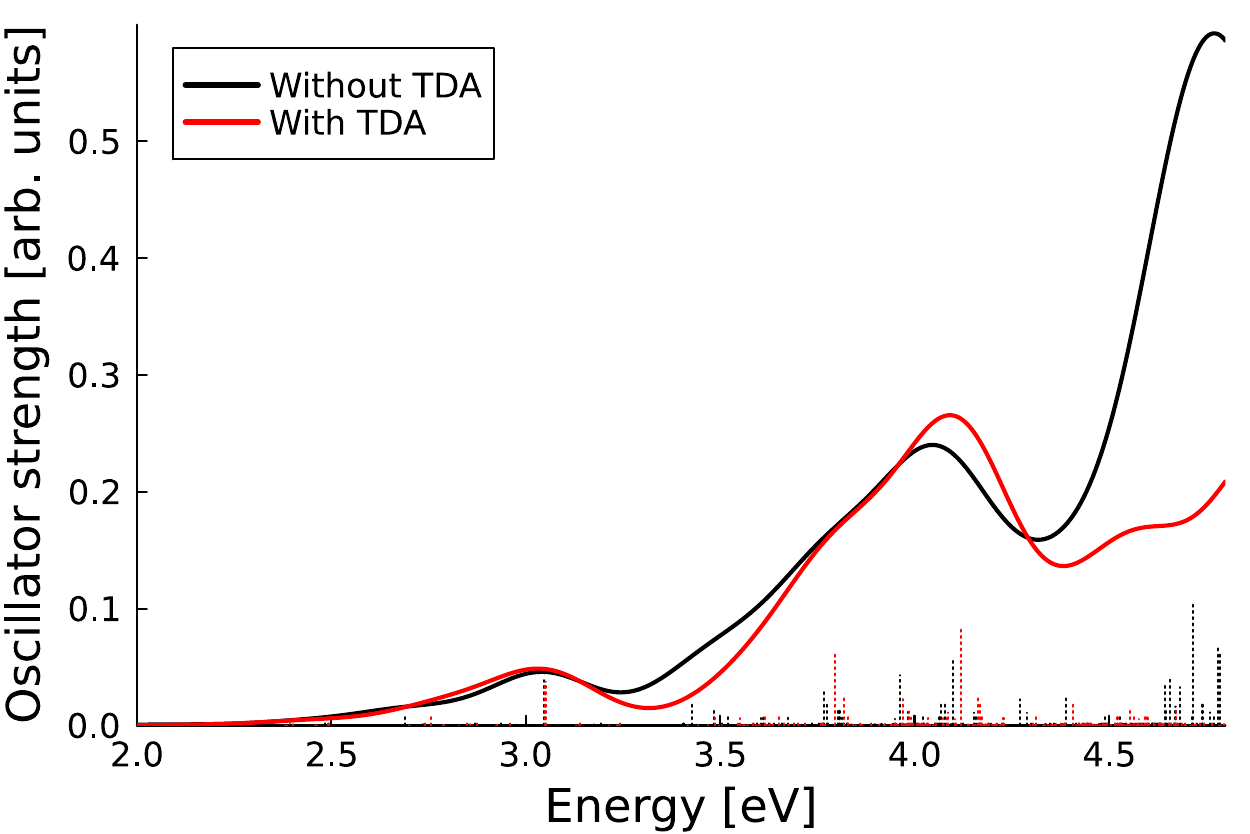}
  \caption{}
  \label{fig:TDA_test_intercalated}
\end{subfigure}

\medskip % create some *vertical* separation between the graphs
\begin{subfigure}{.475\linewidth}
  \includegraphics[width=\linewidth]{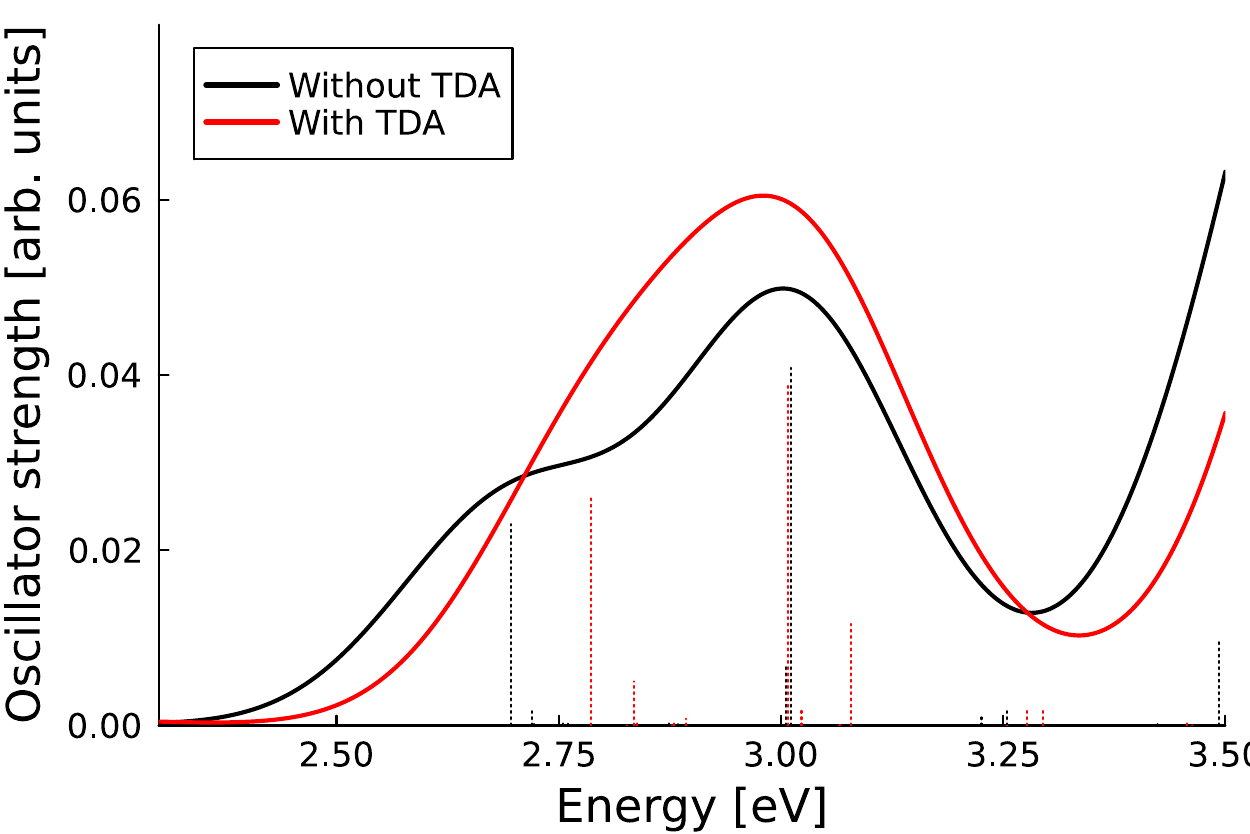}
  \caption{}
  \label{fig:TDA_test_isolated_zoom}
\end{subfigure}\hfill % <-- "\hfill"
% or just a comment and no tilde
\begin{subfigure}{.475\linewidth}
  \includegraphics[width=\linewidth]{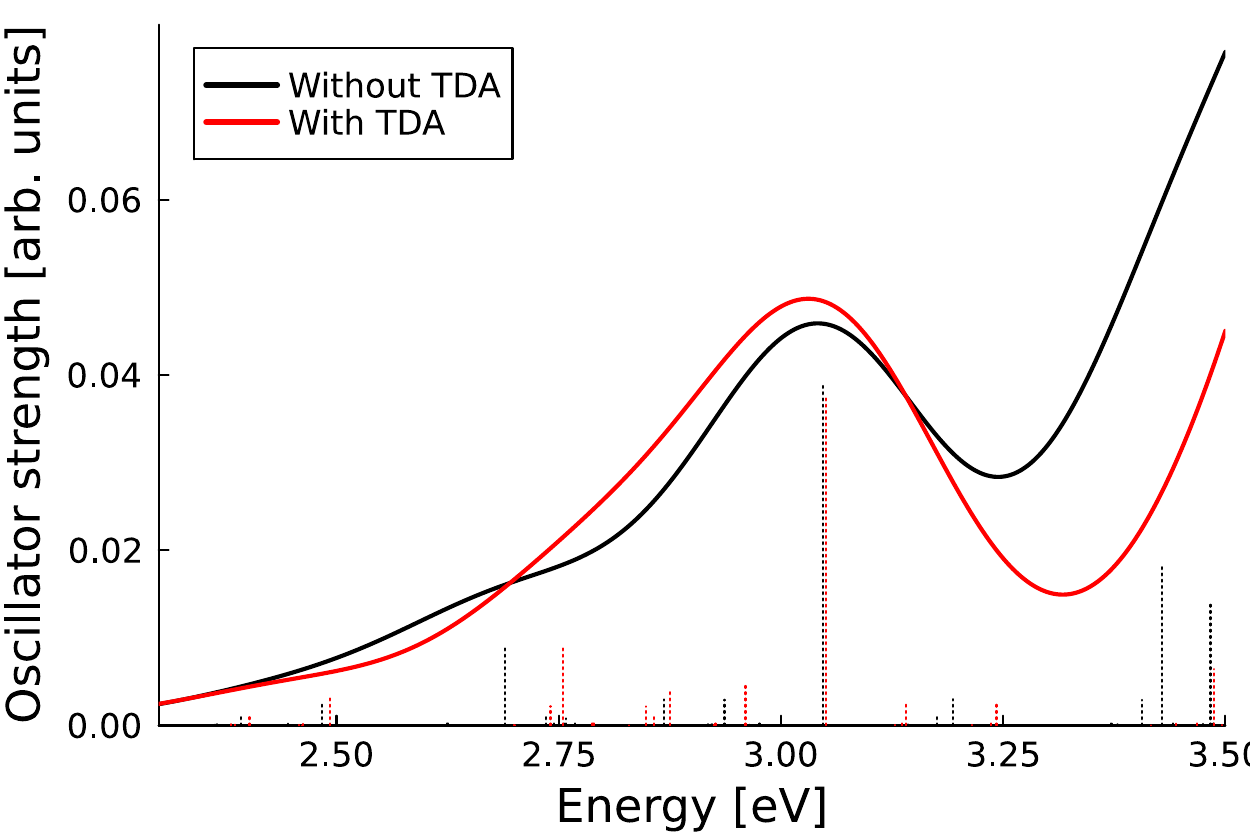}
  \caption{}
  \label{fig:TDA_test_intercalated_zoom}
\end{subfigure}
\caption{UV-Vis spectra calculated with PBE0/x2c-SVPall and \ac{SOC}, but without the \ac{RI}. Spectra calculated with (black) and without (red) TDA for (a) the isolated complex and (b) the intercalated complex. (c) and (d) are an enlargement of band \textbf{1} (2.3–3.5 eV). The structures were optimized with PBE/def2-SVP. A FWHM of 0.3 eV was used to smear the spectra.}
\label{fig:TDA}
\end{figure}

\noindent \textbf{Tamm-Dancoff Approximation.}
Next, we calculated the spectra with and without \ac{TDA}. The spectra for the isolated and intercalated complexes are shown in Figures \ref{fig:TDA_test_isolated} and \ref{fig:TDA_test_intercalated}, respectively, with an enlargement on the first band in Figures \ref{fig:TDA_test_isolated_zoom} and \ref{fig:TDA_test_intercalated_zoom}.

 Visually, the overall spectra are maintained when \ac{TDA} is used. For the isolated system, \ac{TDA}  globally blue shifts the spectrum  (MAE and MSE are both 0.07 eV), which is most pronounced for higher energies (see Figure \ref{fig:TDA_test_isolated}). The intensities are reasonably well reproduced, being comparable to changes seen upon using PBEh-3c as the underlying structure (cf.~Table \ref{tab:MAE_MSE_comparison}). The blue shift is still present for the intercalated system, but both  \ac{MAE} and \ac{MSE} are smaller (both 0.03 eV). The intensities have the \ac{MAE} and \ac{MSE} values of  0.07 and 0.06, i.e., slightly larger than for the isolated complex (the main differences are again at higher energies, see Figure \ref{fig:TDA_test_intercalated}). However, it should again be noted that the changes in both energies and intensities are smaller than the change seen when using the PBEh-3c structure (or one of the tight-binding methods) for the intercalated complex. Thus, TDA is recommended if the structures are obtained with one of the semi-empirical methods.   

To summarize, \ac{TDA} gives reasonable spectra with changes less pronounced than the changes seen using semi-empirical methods for the underlying structures. We additionally investigated using \ac{RI} (see Figure S4), but here, the obtained spectra were close to identical. Combined with the above subsection, we can thus generally recommend using \ac{SOC} on PBEh-3c structures, combined with \ac{TDA} and \ac{RI} to increase the computational efficiency.

\begin{figure}[htb!]
    \centering
    \begin{subfigure}[b]{0.45\textwidth}
        \centering
        \includegraphics[width=\textwidth]{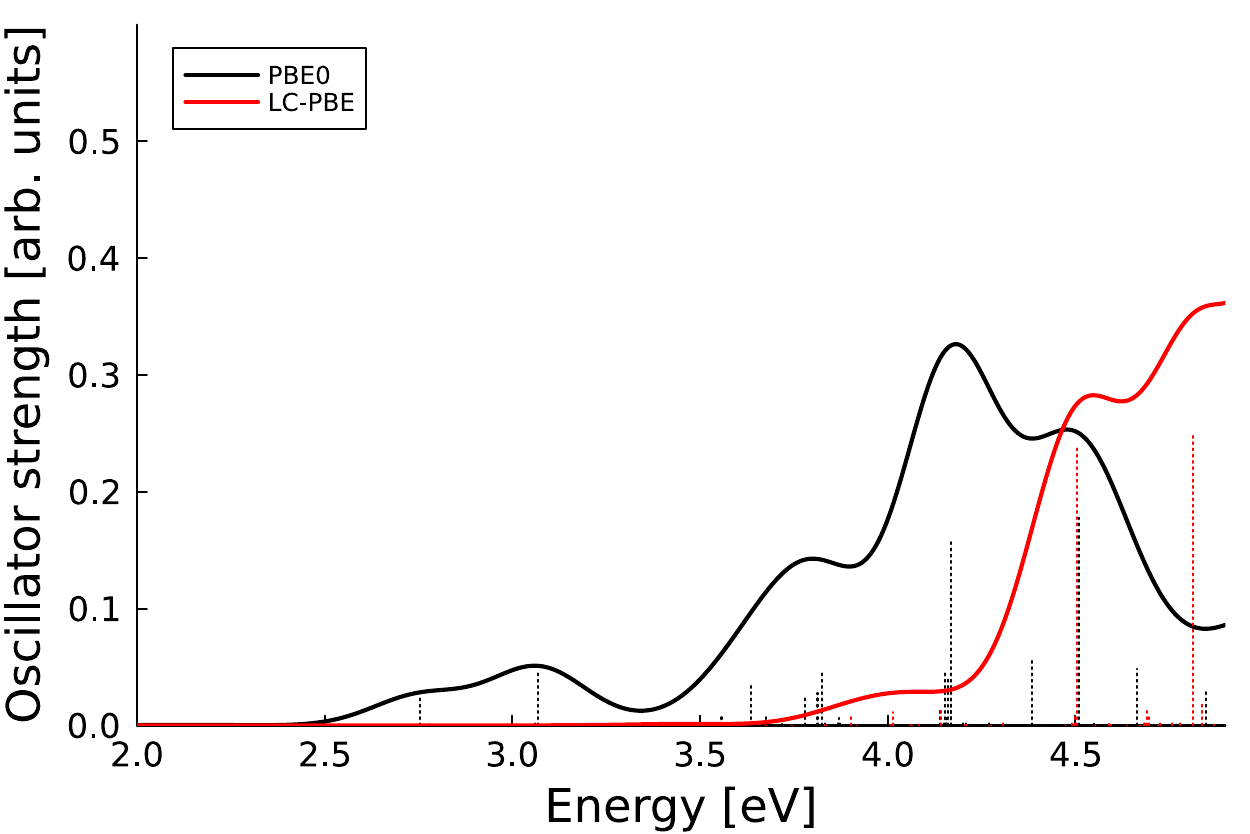}
        \caption{}
        \label{fig:RFS_method_comparison_isolated}
    \end{subfigure}
    \hfill
    \begin{subfigure}[b]{0.45\textwidth}
        \centering
        \includegraphics[width=\textwidth]{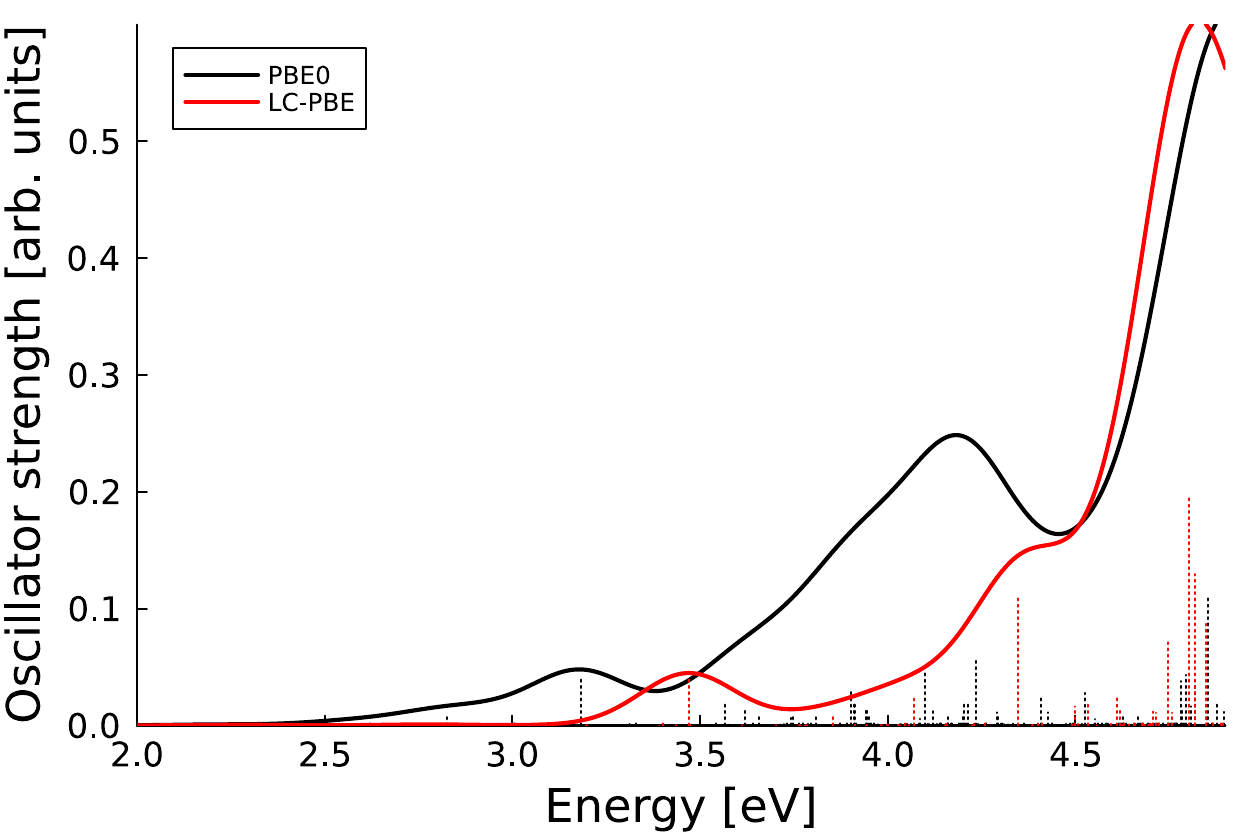}
        \caption{}
        \label{fig:RFS_method_comparison_intercalated}
    \end{subfigure}
    \caption{UV-Vis spectra calculated with PBE0/x2c-SVPall (black) and LC-PBE/x2c-SVPall (red) for (a) the isolated and (b) the intercalated complex without \ac{TDA} and the \ac{RI}. A \ac{FWHM} of 0.3 eV was used to smear the spectra. All geometries were optimized with PBE/def2-SVP.}
    \label{fig:RSF_comparison_between_methods_for_each_config}
\end{figure}

\begin{figure}[htb!]
    \centering
    \begin{subfigure}[b]{0.45\textwidth}
        \centering
        \includegraphics[width=\textwidth]{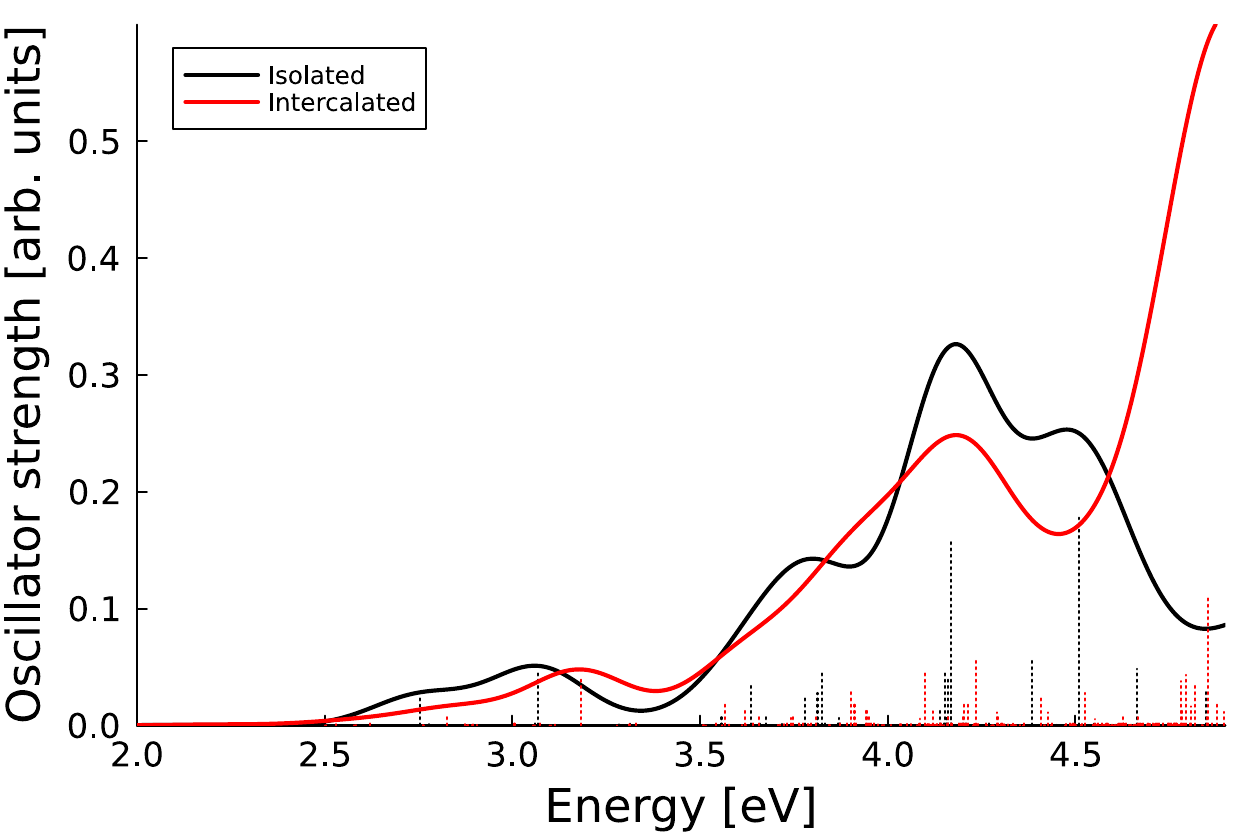}
        \caption{}
        \label{fig:RFS_A1}
    \end{subfigure}
    \hfill
    \begin{subfigure}[b]{0.45\textwidth}
        \centering
        \includegraphics[width=\textwidth]{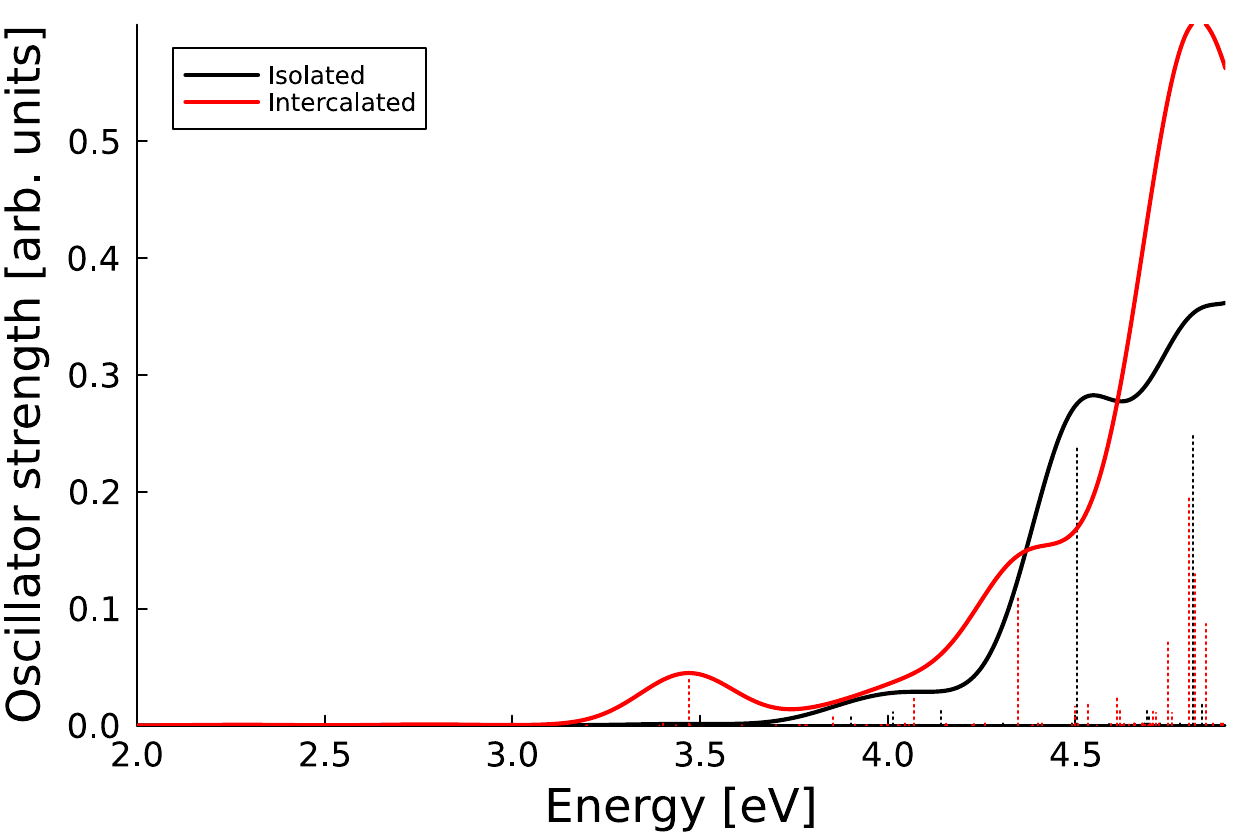}
        \caption{}
        \label{fig:RFS_A2}
    \end{subfigure}
    \caption{UV-Vis spectra calculated with (a) PBE0/x2c-SVPall and (b) LC-PBE/x2c-SVPall for the isolated (black) and intercalated (red) complex without \ac{TDA} and the \ac{RI}. A \ac{FWHM} of 0.3 eV was used to smear the spectra. All geometries were optimized with PBE/def2-SVP.}
\label{fig:RSF_comparison_between_intercalated_and_isolated_for_both_methods}
\end{figure}

\noindent \textbf{Range-Separated Functionals} The charge-transfer nature of the first transition suggests that using a range-separated functional may be beneficial, although the spectrum obtained with the PBE0 functional already matches the experimental spectra quite well. Our goal was to see if LC-PBE can more convincingly reproduce the red shift of the charge-transfer band  upon intercalation, described in ref.~\citenum{Peyratout1995} (from  3.2 eV to 3.0 eV). For the PBE0 functional, this red shift was only obtained from underlying PBEh-3c structures, albeit it was underestimated compared to the experimental value (See Figure \ref{fig:method_comparison_PBEh-3c_and_PBE}).
In this section, we therefore evaluate the resulting UV-Vis spectra obtained through the use of LC-PBE/x2c-SVPall for both the isolated and intercalated complexes, using  PBE0/x2c-SVPall as reference. These two methods are compared for each system in  Figures \ref{fig:RFS_method_comparison_isolated} and \ref{fig:RFS_method_comparison_intercalated}. 
For both the isolated and intercalated complex, there is a large discrepancy between the two functionals, resulting in a significant blue shift. The discrepancy is evident from their \ac{MAE} and \ac{MSE} values of 0.63 eV and 0.62 eV for the isolated complex and \ac{MAE} and \ac{MSE} values of 1.49 eV (in both cases) for the intercalated complex. The intensities are less different, but as seen from Table \ref{tab:MAE_MSE_comparison}, the effect of using a long-range functional is far larger than any effect of the underlying structures. For the LC-PBE functional (Figure \ref{fig:RFS_A2}), the \ac{MLCT} band for the isolated complex is predicted to have its maximum at 4.0 eV, whereas the first maximum of the intra-ligand band is at 4.3 eV. The \ac{MLCT} band for the intercalated complex appears at 3.3 eV while the first maximum of the intra-ligand band is at 4.2 eV. Thus, although the LC-PBE functional is significantly blue shifted, the red shift upon intercalation is in fact reproduced. However, this red shift is 0.7 eV and thus significantly overestimated compared to the experimental shift of 0.2 eV. 

\begin{figure}[H]
%    \centering
%    \begin{subfigure}[b]{0.45\textwidth}
        \centering
        \includegraphics[width=0.45\textwidth]{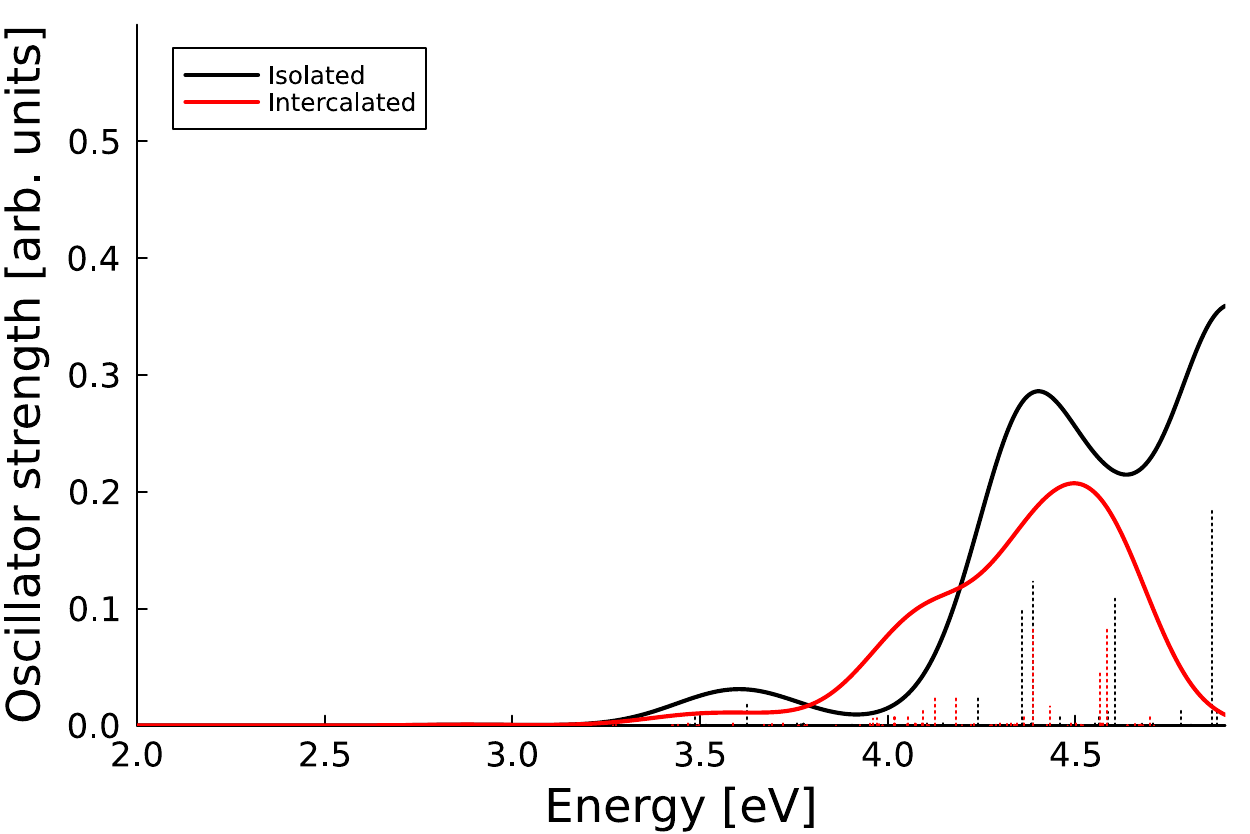}
%        \caption{PBE0/x2c-SVPall}
%        \label{fig:RFS_A1}
%    \end{subfigure}
%    \hfill
%    \begin{subfigure}[b]{0.45\textwidth}
%        \centering
%        \includegraphics[width=\textwidth]{figures_resultsanddiscussion/RSF/LC_PBE_isoVSinter.pdf}
%        \caption{LC-PBE/x2c-SVPall}
%        \label{fig:RFS_A2}
%    \end{subfigure}
    \caption{UV-Vis spectra calculated with LC-PBE$_{\omega=0.35}$/x2c-SVPall with \ac{SOC}, \ac{TDA}, and the \ac{RI}. A \ac{FWHM} of 0.3 eV was used to smear the spectra. All geometries were optimized with PBE/def2-SVP.}
\label{fig:RSF035_intercalated_and_isolated}
\end{figure}

The different behaviors of the two functionals (in particular, the blue shift observed for LC-PBE) prompted us to investigate the range-separation parameter $\omega$ of the LC-PBE functional in more detail. Previous TD-DFT investigations with LC-PBE have shown that $\omega$ can affect the opto-electronic properties of conjugated systems.\cite{Wong2010} The $\omega$-parameter investigation (see  Figures S10--S12) shows that the blue shift of the first \ac{MLCT} is less severe for lower $\omega$-values. For the remaining part, we chose an $\omega$-value of 0.35, which we denote here as LC$_{\omega=0.35}$-PBE (lower values are possible, but these do not preserve the form of the spectrum as well (see Figures S10--S11a). We additionally calculated the spectrum for the intercalated complex with LC$_{\omega=0.35}$-PBE, and compared this to the isolated form (Figure \ref{fig:RSF035_intercalated_and_isolated}).  For the isolated complex with LC$_{\omega=0.35}$-PBE, the first \ac{MLCT} transition occurs at 3.6 eV while the intra-ligand transition occurs at 4.4 eV. While this is an improvement compared to the original LC-PBE functional, it is still somewhat blue shifted. Yet, the \ac{MLCT} of the intercalated complex has a maximum around 3.5 eV, which, compared to the 3.6 eV of the isolated complex, fits well with the experimental red shift of 0.2 eV. We conclude that although the red shift is now reproduced, the blue shift of the spectra is still too large to recommend the usage of the LC$_{\omega=0.35}$-PBE functional. Rather, we recommend using PBE0 based on PBEh-3c structures. 

\section{Conclusions}

We have investigated how best to calculate UV-Vis spectra of a Pt(II) pincer complex that changes its luminescence upon intercalation between DNA base pairs. We investigate both an isolated form and one intercalated into a small DNA model. This can be seen as a model for probes that use exactly these abilities to target specific molecular abnormalities in DNA and thereby serve as cancer-associated targets.

Our calculations show that the largest source of error is likely to be introduced from the exchange-correlation functional; both the \ac{TDA} and \ac{RI} can be employed to accelerate the \ac{TD-DFT} calculations. The best results are obtained with PBE0, although the small, experimentally observed red shift of the first charge-transfer band is only reproduced when using underlying PBEh-3c structures. The red shift is smaller than the experimental one, and a quantitative reproduction of this shift will likely require explicit solvation and dynamics. Significant computational time is often spent optimizing the underlying structures, and our results show that the composite method PBEh-3c is a good alternative to \ac{DFT} methods. In fact, the error introduced from \ac{TDA} is comparable to the error introduced from using PBEh-3c.  The tight-binding methods GFN1-xTB and GFN2-xTB are significantly faster than PBEh-3c and offer reasonable accuracy, albeit with some error (both in the underlying structure and the resulting UV-Vis spectra). Unless a very large number of structure optimizations are required, PBEh-3c is preferred.   

%%%%%%%%%%%%%%%%%%%%%%%%%%%%%%%%%%%%%%%%%%%%%%%%%%%%%%%%%%%%%%%%%%%%%
%% The "Acknowledgement" section can be given in all manuscripts
%% classes.  This should be given within the "acknowledgement."
\begin{acknowledgement}

The authors thank the Villum Foundation, Young Investigator Program (grant no.~29412)  and Independent Research Fund Denmark (grant no.~2064-00002B) for support. The computations were performed on computer resources provided by the Danish e-infrastructure Cooperation at GenomeDK (Århus University). E.K.W. thanks the European Commission (LyticPol project;
project 101106997), the Ministry of Culture and Science of the Federal State of North Rhine-Westphalia (NRW Return Grant) and the Wübben Stiftung Wissenschaft (0168-2025-TT) for support.

\end{acknowledgement}

\begin{suppinfo}

The supporting information contains all structures given in xyz-format and additional spectra calculated with RI, different $\omega$-values, intercalated complexes obtained with different total charge, and calculations with larger basis sets.   Additional figures with molecular orbital plots and comparisons of structures for the isolated \ce{[Pt(OH)(terpy)]+} complex are also given. 

\end{suppinfo}

\bibliography{bib}

\newpage
\appendix
\renewcommand{\thefigure}{S\arabic{figure}}
\renewcommand{\thetable}{S\arabic{table}}
\setcounter{figure}{0}
\setcounter{table}{0}

\section{Supporting Information}
\section*{Intercalated Complex}

\begin{figure} [H]
    \centering
    \includegraphics[width=0.5\linewidth]{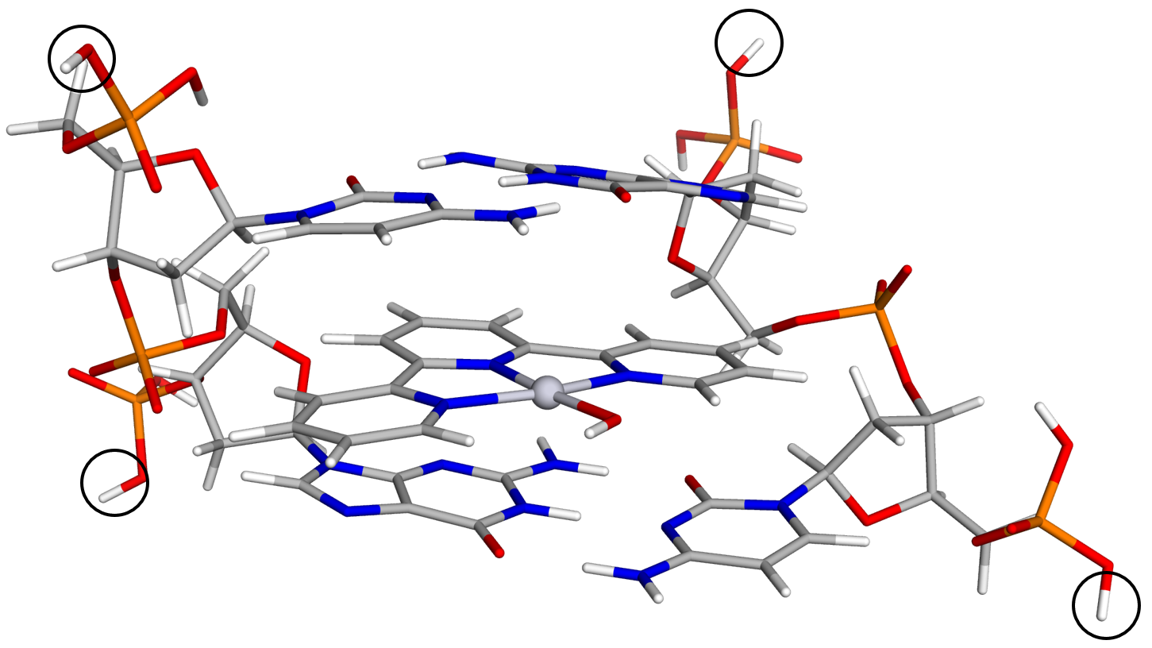}
    \caption{Four added hydrogen atoms on the outermost phosphate groups to reduce the charge from -5 to -1. }
    \label{added hydrogens}
\end{figure}

\section*{Different Charges for Intercalated Complex}

\begin{figure}[H]
    \centering
    \includegraphics[width=0.7\linewidth]{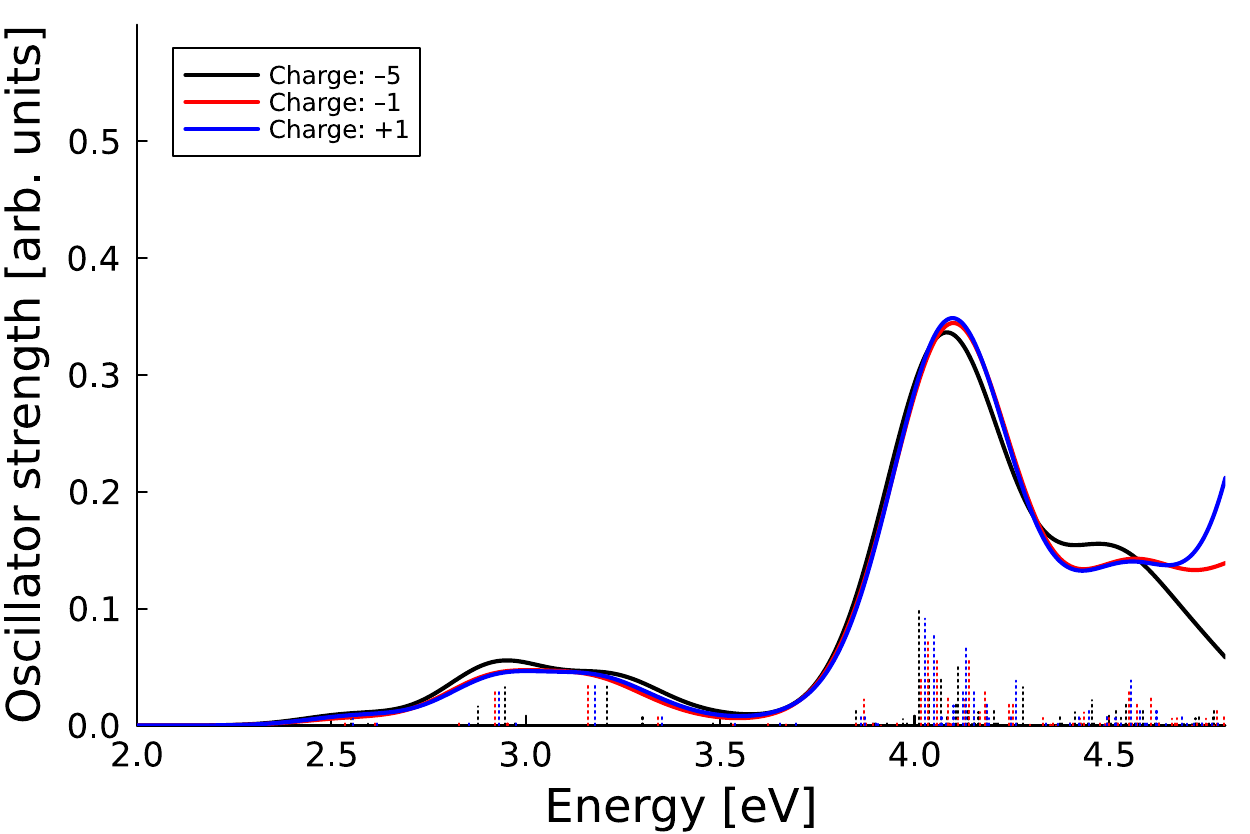}
    \caption{UV-Vis spectra for the intercalated structure with three different charges, –5, –1, and +1. The spectra were calculated with PBE0/x2c-SVPall and without  perturbative \ac{SOC}. The compared structures were optimized with PBEh-3c. A FWHM of 0.3 eV was used to smear the spectra.}
    \label{figure:different_charges_intercalated}
\end{figure}

Two structures with different charge states were optimized with PBEh-3c; one with hydrogen added to all six phosphate groups (charge: +1) and one with no extra hydrogen (charge: -5). UV-Vis spectra were calculated with PBE0/x2c-SVPall using the PBEh-3c structures for the three optimized structures with charges of –5, –1, and +1. The comparison (Figure \ref{figure:different_charges_intercalated}) showed that the spectra did not differ significantly, thus only the structure with a charge of -1 was considered for the remaining spectra calculations. 

\newpage
\section*{xyz-Files}
\subsection*{Isolated Complex}
\subsubsection*{Input Geometry}
\begin{tcolorbox}[breakable]
\scriptsize
\begin{lstlisting}
32
OH
Pt   7.63500  14.90600  13.19100
N    8.64500  13.15800  13.40800
N    7.56100  14.17000  11.33000
N    6.47400  16.36200  12.33000
C    9.29800  12.72500  14.51400
C   10.01000  11.56300  14.49600
C   10.11000  10.81300  13.30900
C    9.51500  11.31700  12.17800
C    8.76900  12.49400  12.21000
C    8.12300  13.10200  11.04200
C    8.09300  12.56600   9.71500
C    7.36800  13.33800   8.82900
C    6.76800  14.55600   9.16700
C    6.94800  15.04100  10.43600
C    6.29900  16.26200  11.01200
C    5.45200  17.07400  10.27000
C    4.96100  18.25100  10.87900
C    5.23100  18.42300  12.26000
C    6.00500  17.51000  12.90700
H    9.25400  13.30900  15.42200
H   10.49900  11.21700  15.39500
H   10.63800   9.87100  13.28800
H    9.62800  10.78900  11.24300
H    8.58900  11.64800   9.43400
H    7.25800  12.98100   7.81600
H    6.17700  15.10100   8.44500
H    5.17700  16.81400   9.25800
H    4.40500  18.98700  10.31700
H    4.82600  19.27000  12.79300
H    6.26400  17.70300  13.93700
O    7.32844  15.72532  15.25492
H    7.28604  16.29585  16.48636
\end{lstlisting}
\end{tcolorbox}

\subsubsection*{Optimized Geometries}
The coordinates here were obtained through geometry optimizations using the stated method written in bold. 
\begin{tcolorbox}[breakable]
\textbf{HF-3c}
\scriptsize
\begin{lstlisting}
32
Coordinates from ORCA-job opt E -926.441277034866
  Pt          7.54786291174869     14.95622086861564     13.19958375190492
  N           8.65538295267920     13.24962648784280     13.43947676703456
  N           7.48640851523008     14.27509558648156     11.35535600428734
  N           6.42135572029535     16.42626112526132     12.35581329397663
  C           9.21124977113366     12.79360914986981     14.54846664432913
  C           9.94501167599829     11.60310095311320     14.57422496059873
  C          10.08815770196557     10.88694685238003     13.39891776092901
  C           9.49515609480136     11.37394696951682     12.22919949648671
  C           8.78627437580927     12.55519447289224     12.28368631170764
  C           8.11735287942228     13.14676001459883     11.09245645388537
  C           8.10006009704757     12.63846333837640      9.79894176085450
  C           7.40529242277482     13.36063221763193      8.82796326861469
  C           6.75234355661117     14.55177724265332      9.15026446854943
  C           6.81839883627591     14.98939826534985     10.46696044282964
  C           6.20424341906744     16.22143729416463     11.03618906190825
  C           5.46030746020920     17.10824671600453     10.28807257391305
  C           4.92593671076971     18.23799380122356     10.91951952690694
  C           5.15698956327933     18.43261968841997     12.26992149378321
  C           5.92238298862191     17.48666893801474     12.96178522444504
  H           9.07662495289334     13.38160347465729     15.45100418663228
  H          10.38536081270183     11.26408973807627     15.50480746248779
  H          10.65148013420048      9.95719507439019     13.37833135861301
  H           9.58764206615806     10.83524568589933     11.29435114168239
  H           8.60749221072108     11.71410807719378      9.54834785992627
  H           7.37235039884209     12.99085624607687      7.80601813906812
  H           6.21365611159068     15.11355233376419      8.39598581347603
  H           5.29379259205522     16.93109990290986      9.23265000844256
  H           4.33617229895901     18.95104823217075     10.34887567560529
  H           4.76094505911792     19.29352816384624     12.79572731858102
  H           6.13855294082942     17.58329136465463     14.02160345353682
  O           7.53529903910502     15.72501523071825     14.98275175794701
  H           8.13494372908459     15.30753649323031     15.61502655705585
\end{lstlisting}
\end{tcolorbox}

\begin{tcolorbox}[breakable]
\textbf{PBEh-3c}
\scriptsize
\begin{lstlisting}
32
Coordinates from ORCA-job opt E -935.795516416477
  Pt          7.54427249360236     14.93855108246919     13.19072707128642
  N           8.63982135141939     13.24277388600618     13.42914163000773
  N           7.48262692544335     14.27276793924214     11.36976890329245
  N           6.42272949480741     16.40612276645856     12.35726004099109
  C           9.17791821465054     12.77045579863776     14.54950298121284
  C           9.90262339705784     11.59167653153284     14.57566915754516
  C          10.06858571851201     10.88280091143444     13.40062218900999
  C           9.50028583212520     11.36934347768632     12.23279670656270
  C           8.78707114672329     12.55114733469909     12.26884337980071
  C           8.13217519741512     13.14168259885863     11.09107131362214
  C           8.11924573359257     12.65761016209052      9.79331001738515
  C           7.42463688935482     13.37573872526937      8.82983938623879
  C           6.76079533696694     14.55045588929096      9.15583690572765
  C           6.80796476241861     14.98964128613732     10.46807488561906
  C           6.19661584800292     16.20291990820685     11.03432911651677
  C           5.44445266813788     17.09892294268613     10.30108281959796
  C           4.91924420235803     18.21772646393867     10.93200845697088
  C           5.16163715088793     18.41032905706959     12.27956808906990
  C           5.92349959223746     17.47776073044260     12.96297834427116
  H           9.02121666485679     13.34833330822924     15.44806587820604
  H          10.32319900137702     11.24763975159397     15.50913478122067
  H          10.62902942433995      9.95847039158005     13.38630617894890
  H           9.61056689719539     10.82729085883663     11.30481585420936
  H           8.63292531517093     11.74504840442854      9.52767337360085
  H           7.40108407711961     13.01574095241000      7.81099250396169
  H           6.22557196779809     15.10379497857662      8.39797896292368
  H           5.26542951997564     16.93116355369507      9.24861537407321
  H           4.32836955001994     18.92794081958164     10.37053927413562
  H           4.77174334488751     19.26832876589648     12.80734379471440
  H           6.14647914900045     17.58104936753379     14.01523703426804
  O           7.51013359382174     15.70363355542661     15.00971009141607
  H           8.31252953872301     15.53530780005363     15.51343550359239
\end{lstlisting}
\end{tcolorbox}

\begin{tcolorbox}[breakable]
\textbf{PBE/def2-SVP}
\scriptsize
\begin{lstlisting}
32
Coordinates from ORCA-job opt E -936.146184574685
  Pt          7.54981899198702     14.94929671333991     13.18583007523018
  N           8.65432049502379     13.24763330276535     13.43228374746921
  N           7.48872862486957     14.27502290216326     11.36655038513179
  N           6.42838628447309     16.42178313552706     12.35095166001216
  C           9.21939053276983     12.78676196136253     14.56874727857989
  C           9.95918455441059     11.60120577026513     14.60225029614736
  C          10.12058447896872     10.86540353561854     13.41974592631039
  C           9.53358099214754     11.33641879576077     12.23932033852541
  C           8.79940576276548     12.53012806674647     12.25769134271370
  C           8.13630429625801     13.11166996369119     11.07975308624213
  C           8.10129141175751     12.61753437273255      9.76723234930414
  C           7.39851877973207     13.34139625777734      8.79019846756808
  C           6.74203466555394     14.53963632937777      9.11858040257240
  C           6.79885938594129     15.00220889950140     10.44088395557459
  C           6.19484742504865     16.22188416823363     11.00498957907589
  C           5.43312707238713     17.14798227197928     10.27929537733212
  C           4.91398699859810     18.27425106037882     10.93128938101160
  C           5.16585082912181     18.45598840931159     12.29953254574918
  C           5.93114805204829     17.50431241336391     12.98102291947904
  H           9.06685931253141     13.39305329333296     15.47261266791360
  H          10.39966257277285     11.26953699574364     15.55257613972415
  H          10.69809724329371      9.92984836383059     13.41435017292741
  H           9.64099587174954     10.78017373170704     11.29838159332128
  H           8.61435777836255     11.68119423924036      9.50991227525244
  H           7.36213432046909     12.96706055986075      7.75757448642863
  H           6.19333046063977     15.10673101540132      8.35472767225581
  H           5.24968235812543     16.98257163331986      9.20891073996341
  H           4.31506024000867     19.00644081868458     10.37073281293907
  H           4.77563802549996     19.32643682533061     12.84481678542303
  H           6.17346721611031     17.57472174880410     14.05265354362128
  O           7.50514458433917     15.79348573895011     14.97994964666550
  H           8.03068038223473     15.30039670589692     15.63893234953461
\end{lstlisting}
\end{tcolorbox}

\begin{tcolorbox}[breakable]
\textbf{PBE0/def2-SVP}
\scriptsize
\begin{lstlisting}
32
Coordinates from ORCA-job opt E -936.121904542629
  Pt          7.54737492043444     14.94725564679969     13.17890150179488
  N           8.64886241163809     13.25163107396925     13.42481191528525
  N           7.48832776315234     14.27378303245731     11.36186538240200
  N           6.43233264560678     16.41080962615730     12.34704290788071
  C           9.20689202646614     12.79543851406753     14.54985393875580
  C           9.94189996202907     11.61625678911267     14.58363112075248
  C          10.10041039690234     10.88905429539970     13.40908018126442
  C           9.51695180386579     11.36003492710051     12.23602476621546
  C           8.79143353103382     12.54497076161433     12.26324743610278
  C           8.12910557198379     13.12641042088891     11.08387936732483
  C           8.09951795231870     12.63042846967121      9.78242548054673
  C           7.40155016488479     13.35035423414718      8.81301105437200
  C           6.74802538890032     14.53995604288532      9.13675205407221
  C           6.80705289272296     14.99375248826353     10.45161226271127
  C           6.20239631459034     16.21302889483363     11.01686285323576
  C           5.44794072685903     17.12682061852178     10.29107061778871
  C           4.93081676882944     18.24678610165101     10.93918576047512
  C           5.17913929166726     18.42840082515687     12.29629661115287
  C           5.94043781230222     17.48145261756195     12.97248873077782
  H           9.05648466410827     13.39533551839944     15.44860495177671
  H          10.37934719620330     11.28450629774300     15.52622798196342
  H          10.67252016867566      9.95923553972146     13.40113727240901
  H           9.62278664889088     10.80909233260867     11.30115172642852
  H           8.60909278539661     11.70135390298961      9.52558327720097
  H           7.36643502789914     12.97858840303740      7.78748852978010
  H           6.20400694863709     15.10175751518322      8.37700588646882
  H           5.26604750217605     16.96269308756493      9.22837303309682
  H           4.33628111541387     18.97317823360557     10.38146248776402
  H           4.79078591818052     19.29278768064858     12.83669877915307
  H           6.18094096859413     17.55155017591690     14.03635331865020
  O           7.49787481438057     15.77955839254241     14.96957956252667
  H           8.05140789525575     15.32590753977806     15.61456924986963
\end{lstlisting}
\end{tcolorbox}

\begin{tcolorbox}[breakable]
\textbf{B3LYP/def2-SVP}
\scriptsize
\begin{lstlisting}
32
Coordinates from ORCA-job opt E -936.592145147154
  Pt          7.54890425017074     14.95420832846899     13.19923223177974
  N           8.66097987632935     13.24022848200987     13.43597118736735
  N           7.48893540926138     14.27534536794900     11.36471228890948
  N           6.42289214226026     16.42851261628012     12.34940608123383
  C           9.22833412762419     12.77985545369830     14.55991275296746
  C           9.96407045708222     11.59678439902714     14.58525373705704
  C          10.11402045478056     10.87049616475070     13.40451670051057
  C           9.52268064288064     11.34652722935831     12.23317043633254
  C           8.79635077405024     12.53498259432865     12.26559892276577
  C           8.12925794598006     13.12164691486534     11.08677891371216
  C           8.09586820424681     12.62439668451756      9.78210792154313
  C           7.39723405881837     13.34670546693583      8.80960380614952
  C           6.74549582408065     14.54143793515110      9.13398024003020
  C           6.80564778116752     14.99888605376200     10.45130512782313
  C           6.19867374506705     16.22310848821065     11.01315948541969
  C           5.44502808250569     17.13687536438105     10.27968883175055
  C           4.92287389399390     18.26259121401924     10.92131314672913
  C           5.16475944836236     18.45097123621187     12.28249816767156
  C           5.92474476909311     17.50530481372357     12.96909092266086
  H           9.08507710187768     13.37916435154890     15.45972117723835
  H          10.40790988600403     11.26168995758439     15.52371821021720
  H          10.68487006481428      9.93998662014884     13.39061040539226
  H           9.62339529389600     10.79746875377687     11.29704469222634
  H           8.60251327844658     11.69401907186386      9.52575891901694
  H           7.36020828247377     12.97394504187507      7.78453471221623
  H           6.20240887322387     15.10314661394475      8.37406064402142
  H           5.26782930781374     16.96876558532748      9.21719495891965
  H           4.33091583963290     18.98613269479344     10.35722322217979
  H           4.77304123927974     19.31773268128809     12.81671683716589
  H           6.15685295537821     17.58278661720670     14.03357426566437
  O           7.50218047600656     15.79196436994915     15.00433734248048
  H           8.02052551339708     15.30650283304238     15.66048371084668
\end{lstlisting}
\end{tcolorbox}

%%%%%%%%%%%%%%%%%%%%%%%%%%%%%%%%%%%% x2c-SVPall  %%%%%%%%%%%%%%%%%%%%%%%%%%%%%%%%%%%%%%

\begin{tcolorbox}[breakable]
\textbf{PBE/x2c-SVPall}
\scriptsize
\begin{lstlisting}
32
Coordinates from ORCA-job opt E -19234.919057959327
  Pt          7.54916116344037     14.94259714463947     13.16585650806130
  N           8.64606677617398     13.25725818928188     13.42538495965133
  N           7.48834947962593     14.27422121553310     11.36193555914878
  N           6.43555793737160     16.40990662200298     12.34977723827015
  C           9.20668388837540     12.80138663438791     14.56599636202195
  C           9.94751187040045     11.61682122555324     14.60625069467997
  C          10.11535166340993     10.87596958941501     13.42789732189097
  C           9.53214883578969     11.34216752718794     12.24370585843318
  C           8.79698603360298     12.53434808950055     12.25513561095611
  C           8.13647842819287     13.11207451152372     11.07594643033535
  C           8.10246251914012     12.61756888380252      9.76396599334057
  C           7.39952220078053     13.34016611944630      8.78619642199798
  C           6.74226682739625     14.53790357508409      9.11467269907522
  C           6.79861104393270     15.00053394603930     10.43649701232720
  C           6.19659876198598     16.21756919742056     11.00399094371918
  C           5.43358082956521     17.14692205017450     10.28550664376294
  C           4.91757034338267     18.27062859859346     10.94445760080201
  C           5.17536811153188     18.44452675450056     12.31256068267313
  C           5.94221939368846     17.48950966981330     12.98743176426862
  H           9.05060838887269     13.40922237195814     15.46808623409326
  H          10.38410116736797     11.28991984504821     15.55988818306798
  H          10.69407889478195      9.94128732403267     13.42870452697135
  H           9.64323668211270     10.78319713910421     11.30486007081584
  H           8.61656276105034     11.68149983370071      9.50809673473320
  H           7.36365343407903     12.96557710013667      7.75376002742987
  H           6.19305894963819     15.10498522685206      8.35128438761864
  H           5.24656558304273     16.98600134279958      9.21512563498783
  H           4.31719138896489     19.00610278874859     10.39001352231049
  H           4.78833693949083     19.31243954134829     12.86390715256510
  H           6.18851710430300     17.55527134612103     14.05825545147550
  O           7.50635671877345     15.78418939270044     14.95252466930103
  H           8.03971587973494     15.29039720354837     15.60460709921337
\end{lstlisting}
\end{tcolorbox}

\begin{tcolorbox}[breakable]
\textbf{PBE0/x2c-SVPall}
\scriptsize
\begin{lstlisting}
32
Coordinates from ORCA-job opt E -19234.705419542996
  Pt          7.54646676311464     14.94060767636588     13.15972841003807
  N           8.64029138443607     13.26068250894281     13.41811924869580
  N           7.48744951080455     14.27276692623466     11.35773307934197
  N           6.43910043462403     16.39901269672598     12.34563683287183
  C           9.19395123624747     12.80960537003030     14.54723368414780
  C           9.93006532523340     11.63152153990255     14.58775481733192
  C          10.09490481866792     10.89926122428741     13.41726289769790
  C           9.51513716662744     11.36529137524498     12.24046425816357
  C           8.78851853678771     12.54874064692770     12.26075568450237
  C           8.12898469645161     13.12650842741798     11.08032478658034
  C           8.10085691923939     12.63048972624210      9.77935411744380
  C           7.40299909269214     13.34929161868972      8.80906830918886
  C           6.74862460937473     14.53840157012373      9.13289642617050
  C           6.80668772343437     14.99205349493435     10.44728007629080
  C           6.20395361626938     16.20857724359551     11.01563483608061
  C           5.44824514083791     17.12564010005119     10.29688996718480
  C           4.93411755927371     18.24298827970798     10.95191489222408
  C           5.18819041332600     18.41686171998229     12.30895637449121
  C           5.95102272386477     17.46661900014869     12.97856579453616
  H           9.04020334459409     13.41121115161081     15.44412653882500
  H          10.36382661972170     11.30452690424103     15.53357667839599
  H          10.66835476044951      9.97036498300571     13.41550601122615
  H           9.62456546693130     10.81156747659871     11.30770698332660
  H           8.61165606581394     11.70177956381401      9.52398920727757
  H           7.36876715868724     12.97745845321149      7.78365339267558
  H           6.20437263418729     15.10040252262012      8.37356768975933
  H           5.26305664324657     16.96590349943512      9.23415254479634
  H           4.33814078594371     18.97264347794320     10.40023790575289
  H           4.80286149071787     19.27872636948046     12.85529454340984
  H           6.19559137947861     17.53217797421377     14.04164345709010
  O           7.50189743118524     15.77220247322017     14.94189289011723
  H           8.06161854773520     15.31828400504851     15.58135766436416
\end{lstlisting}
\end{tcolorbox}

\begin{tcolorbox}[breakable]
\textbf{B3LYP/x2c-SVPall}
\scriptsize
\begin{lstlisting}
32
Coordinates from ORCA-job opt E -19235.392809863552
  Pt          7.54783585348012     14.94740846216085     13.17961895461106
  N           8.65164029155255     13.24901232998507     13.42952553771332
  N           7.48728289626076     14.27413462144635     11.36080174761885
  N           6.42888854698471     16.41660572430828     12.34861739859656
  C           9.21209915989288     12.79222653415083     14.55843320876397
  C           9.94869637477363     11.61011824264256     14.59058897478929
  C          10.10722981234383     10.88013190359364     13.41321788972915
  C           9.52152031944241     11.35229709792828     12.23756269706469
  C           8.79408267567985     12.53927786857418     12.26288209024527
  C           8.13026723873630     13.12260742261217     11.08295153616023
  C           8.09991296421183     12.62651938713533      9.77830392632350
  C           7.40085578264756     13.34746111101699      8.80509107825507
  C           6.74636203383681     14.54056810120350      9.13011878109251
  C           6.80428381576082     14.99703158645310     10.44725368871310
  C           6.19870651167576     16.21802216281222     11.01251357524421
  C           5.44301309515179     17.13449079975623     10.28615788113191
  C           4.92352078337814     18.25756698962911     10.93450281990710
  C           5.17178769951066     18.43887951266621     12.29551033016137
  C           5.93429040126049     17.49069155047232     12.97549499342130
  H           9.06279708733906     13.39216506079428     15.45681068226489
  H          10.38693469853870     11.27847413518184     15.53277511146007
  H          10.67936899016332      9.95047339775964     13.40551706681619
  H           9.62714864677927     10.80108275028210     11.30328335396780
  H           8.60937809964503     11.69752197163246      9.52284843647476
  H           7.36585704354407     12.97550286182563      7.77977510428097
  H           6.20289943600833     15.10228942864164      8.37055862568137
  H           5.26213594725326     16.97028973208290      9.22371179347105
  H           4.32940444561059     18.98388909609792     10.37650199285707
  H           4.78283276857954     19.30314175959051     12.83555066663901
  H           6.17094050137613     17.56503025480796     14.03899567504333
  O           7.50847875252632     15.78151330362292     14.97786959196143
  H           8.05402732605507     15.30574483913211     15.61893478953887
\end{lstlisting}
\end{tcolorbox}
%%%%%%%%%%%%%%%%%%%%%%%%%%%%%%%%% Isolated explicit solvation %%%%%%%%%%%%%%%%%%%%%%%%%%%%%%%%%
\begin{tcolorbox}[breakable]
\textbf{6Å Cutout of 20 Å Water Sphere}
\scriptsize
\begin{lstlisting}
284
water_6A_cutout
 Pt        0.03100        0.47000        1.23100
  N        1.13500       -1.23100        1.47700
  N       -0.03000       -0.20400       -0.58800
  N       -1.09100        1.94300        0.39600
  C        1.70000       -1.69200        2.61400
  C        2.44000       -2.87800        2.64700
  C        2.60200       -3.61400        1.46500
  C        2.01500       -3.14300        0.28400
  C        1.28000       -1.94900        0.30300
  C        0.61700       -1.36700       -0.87500
  C        0.58200       -1.86100       -2.18800
  C       -0.12000       -1.13800       -3.16500
  C       -0.77700        0.06100       -2.83600
  C       -0.72000        0.52300       -1.51400
  C       -1.32400        1.74300       -0.95000
  C       -2.08600        2.66900       -1.67600
  C       -2.60500        3.79500       -1.02400
  C       -2.35300        3.97700        0.34500
  C       -1.58800        3.02500        1.02600
  O       -0.01400        1.31500        3.02500
  H        1.54800       -1.08600        3.51800
  H        2.88100       -3.20900        3.59800
  H        3.17900       -4.54900        1.45900
  H        2.12200       -3.69900       -0.65700
  H        1.09500       -2.79800       -2.44500
  H       -0.15700       -1.51200       -4.19700
  H       -1.32600        0.62800       -3.60000
  H       -2.26900        2.50400       -2.74600
  H       -3.20400        4.52700       -1.58400
  H       -2.74300        4.84700        0.89000
  H       -1.34500        3.09600        2.09800
  H        0.51200        0.82100        3.68400
  H        4.81897       -3.06554       -2.63447
  H        6.44011       -2.90257       -2.68841
  O        5.58471       -2.46643       -2.36943
  H        1.33852        4.65986       -3.94503
  H        0.64424        3.97976       -5.27780
  O        0.77647        4.84423       -4.77268
  H       -7.45205       -2.45985       -3.42111
  H       -6.39404       -1.25190       -3.81563
  O       -6.53498       -2.25396       -3.80177
  H        5.20930       -3.16598       -6.68846
  H        4.07090       -1.99985       -6.63389
  O        4.24761       -2.96155       -6.90485
  H        3.01605        4.32814        3.56680
  H        2.51643        4.92532        2.11606
  O        3.30688        4.54861        2.61772
  H       -1.23327        8.81611        4.50104
  H       -1.74111        7.46840        3.72777
  O       -1.31468        8.37376        3.59657
  H        1.07868        9.06646        1.70499
  H        2.10579        7.86665        2.14033
  O        1.85092        8.82195        2.31392
  H        1.91409       -7.81185        2.00342
  H        1.80215       -9.41995        2.27633
  O        2.45410       -8.65362        2.18799
  H       -0.82909        1.97378       -6.58195
  H       -0.79128        3.56424       -6.99688
  O       -0.26929        2.82057       -6.55736
  H        7.65451       -4.31560        4.63074
  H        8.60771       -3.81311        5.86246
  O        7.91203       -4.49803        5.58289
  H        3.70320       -6.66109       -6.08847
  H        3.14367       -5.42962       -7.00636
  O        2.86735       -6.19319       -6.41223
  H        3.35133       -5.34760       -2.53943
  H        3.75647       -6.87578       -2.13893
  O        3.05028       -6.16885       -2.03580
  H       -1.78368        5.94299        5.39990
  H       -1.31523        5.26326        6.82450
  O       -1.28322        6.09935        6.26308
  H        1.05301       -5.52497        7.30235
  H        0.39373       -4.84346        5.96834
  O        0.76911       -4.65098        6.88406
  H        3.91089        3.06127       -4.92385
  H        5.15046        2.01482       -5.13817
  O        4.43428        2.31911       -4.48843
  H        6.49595       -3.21500       -0.63675
  H        7.69885       -3.31305        0.48158
  O        7.07185       -3.84766       -0.10605
  H        5.14363        3.02551        0.12153
  H        6.60809        3.63751        0.55758
  O        6.06123        3.29693       -0.21874
  H        0.36993       -0.85745       -8.54730
  H        1.82496       -0.59064       -7.84405
  O        0.86390       -0.83605       -7.66364
  H       -4.96237        8.85806        4.34605
  H       -4.23690        8.64334        5.80954
  O       -4.61935        8.16515        5.00899
  H       -7.42501       -1.09949        0.58835
  H       -5.96858       -1.64680        0.06865
  O       -6.83536       -1.91411        0.50010
  H        7.29235       -6.81263       -1.65386
  H        6.09857       -7.77625       -2.22463
  O        6.69708       -6.99682       -2.44464
  H        7.10430       -0.68848        5.04033
  H        6.22536       -1.80608        5.86851
  O        6.84488       -1.02235        5.96232
  H       -6.02823        0.38319        5.49555
  H       -6.35014        1.39443        4.25089
  O       -6.10027        1.35763        5.22821
  H       -0.92534       -1.02523        9.08693
  H       -1.00686        0.47812        8.44206
  O       -0.86953       -0.49700        8.22480
  H       -1.12286        5.66682       -7.43187
  H       -0.17563        5.10721       -8.65535
  O       -1.05153        4.96809       -8.16241
  H       -8.28160        1.02191       -0.68394
  H       -8.37359        2.31067       -1.70032
  O       -7.74633        1.76492       -1.11567
  H        0.36945       -9.08900       -0.51246
  H       -1.12192       -9.65037       -0.89896
  O       -0.46945       -9.50950       -0.13508
  H       -6.28744        1.17356       -4.13818
  H       -6.78634        0.90815       -2.59272
  O       -6.56926        0.45925       -3.47141
  H        2.36988        0.11747       10.02142
  H        2.68214       -0.85961        8.73640
  O        3.05994       -0.12547        9.32119
  H        2.30661        3.78663       -2.01787
  H        3.14350        5.18262       -2.18033
  O        2.37862        4.63466       -2.55296
  H        3.42483       -3.41996       -8.46589
  H        3.19167       -2.99670      -10.03209
  O        3.16789       -3.76610       -9.38048
  H       -7.06428        4.00161        1.27362
  H       -7.48533        2.67578        0.41893
  O       -7.53369        3.11079        1.33066
  H       -6.62240        5.66814        3.78807
  H       -7.93582        5.62968        2.81725
  O       -6.93350        5.74859        2.83459
  H        1.56388        6.35778       -4.93434
  H        1.16819        7.92095       -5.23616
  O        1.93295        7.29270       -5.06187
  H       -8.58090        7.03702       -0.47369
  H       -7.07409        6.73695        0.09368
  O       -7.86351        7.35809        0.15963
  H        3.47218       -6.42503        7.98892
  H        4.76962       -6.72159        8.94099
  O        4.37057       -6.07361        8.28299
  H       -6.95196        2.19820       -5.91456
  H       -5.57401        3.07790       -5.83718
  O       -6.08372        2.31071       -5.41620
  H       -2.27166       -0.04267       -7.38147
  H       -0.68544       -0.12488       -6.96143
  O       -1.53872        0.40386       -6.85350
  H       -6.19188       10.65575       -0.32185
  H       -6.01452        9.17182        0.33326
  O       -5.81867       10.14648        0.46895
  H        8.05162       -7.28271        2.11728
  H        6.55873       -7.75617        1.60519
  O        7.51043       -7.56533        1.31251
  H        2.22922       -7.37103       -1.17134
  H        2.53926       -8.59380       -0.12391
  O        1.86022       -8.22686       -0.77004
  H       -1.93001        3.13739        8.66880
  H       -2.37710        3.29828        7.11059
  O       -1.74374        3.64407        7.80932
  H        8.27971       -6.65973        0.12902
  H        8.03621       -5.15097       -0.46628
  O        8.61982       -5.97597       -0.53980
  H        5.40080       -9.53543        1.68356
  H        4.11217       -8.57554        2.02087
  O        5.12701       -8.60724        1.98165
  H       -0.04997       -6.39531       -2.27666
  H       -1.43988       -7.10855       -1.76434
  O       -0.75652       -7.06812       -2.50663
  H        6.24842        0.53860       -3.64087
  H        5.83590       -1.03113       -3.42155
  O        5.99057       -0.31258       -4.11762
  H       -0.34473       -4.44904        3.45760
  H       -1.79638       -4.26148        4.19111
  O       -0.84691       -4.59886        4.31210
  H       -4.61105        1.91137        5.96885
  H       -3.95337        1.35970        7.36964
  O       -3.93765        2.11613        6.69942
  H        3.58177        3.14863        1.68746
  H        3.22063        1.61261        1.23425
  O        3.77858        2.42097        1.01068
  H        2.61013       -3.27813        8.49451
  H        3.56354       -4.55319        8.90469
  O        3.20504       -3.66181        9.21497
  H       -1.86518        6.72545       -5.34381
  H       -0.41686        5.95332       -5.34662
  O       -1.04417        6.52446       -5.89680
  H       -1.23823       10.55735        1.49138
  H       -1.05298       11.93848        2.35752
  O       -1.61703       11.47884        1.65011
  H        6.16669       -7.78531        4.94963
  H        5.20245       -8.46787        3.80861
  O        5.25636       -8.19259        4.77789
  H        7.00194       -4.38153        7.05505
  H        5.76318       -4.86269        8.02242
  O        6.49046       -4.17776        7.90742
  H        0.35954       -3.62837       -9.02805
  H        0.49375       -5.16220       -9.58833
  O       -0.07560       -4.32613       -9.60858
  H       -7.08065        6.09716        5.94922
  H       -5.67886        6.77285        5.40990
  O       -6.17554        5.90247        5.53361
  H       -0.40278       -7.10816        2.18041
  H        0.39626       -7.28355        0.76184
  O        0.45667       -6.89764        1.68341
  H       -2.65560        1.29268       -9.28439
  H       -3.84183        0.43239      -10.01398
  O       -2.95152        0.36296       -9.54136
  H        1.11050        2.39546       -9.47227
  H        1.11663        2.42826       -7.83019
  O        1.70227        2.33259       -8.64672
  H        0.86067        5.23137       -0.14773
  H        1.67827        6.62280        0.12271
  O        1.42405        5.72576        0.51926
  H       -7.20767        4.98890       -1.99418
  H       -6.73537        5.79332       -3.34032
  O       -7.54531        5.47192       -2.81288
  H        3.26311       -4.33545       -4.34695
  H        4.87499       -4.57032       -4.49272
  O        4.13308       -4.30739       -3.85086
  H       -1.26826       -0.24562       -9.91265
  H        0.14322        0.17182      -10.64771
  O       -0.31044       -0.53038      -10.07539
  H        2.02441       -9.01754       -2.41140
  H        1.45210      -10.44494       -2.96046
  O        1.94727       -9.60932       -3.22468
  H       -5.48265        5.20917        0.18462
  H       -5.45709        3.83461       -0.72892
  O       -5.95617        4.68742       -0.54842
  H        1.10089       -2.96031        7.38317
  H        0.68866       -1.43660        7.84049
  O        1.45685       -2.08595        7.74935
  H        1.52388        2.70937        7.42519
  H       -0.03632        3.22837        7.30969
  O        0.76655        2.92807        6.78598
  H       -2.61813       -6.77351        3.36432
  H       -1.16149       -6.52890        4.05968
  O       -1.70047       -7.16913        3.49549
  H        2.25527       -8.49648        4.18439
  H        3.38549       -8.23685        5.32507
  O        2.39418       -8.34298        5.16776
  H       -0.80272       -8.19925       -3.79277
  H       -0.10625       -9.02045       -5.02092
  O       -0.87056       -9.03543       -4.36222
  H       -3.63714        3.51653       -8.05295
  H       -5.24580        3.28329       -7.92787
  O       -4.47275        3.84334       -7.59617
  H        3.02374       -0.19899       -5.52676
  H        4.64563       -0.13117       -5.32077
  O        3.90713       -0.33290       -5.98413
  H       -1.17788       -2.24066        5.51847
  H       -1.64843       -1.28217        6.76504
  O       -1.97012       -1.90787        6.04364
  H       -4.06814       -4.13609        6.17488
  H       -3.19524       -2.85884        6.69645
  O       -3.76927       -3.63708        7.00344
  H       -4.36797        7.18000        1.30785
  H       -5.59936        6.24412        1.89628
  O       -4.81389        6.26739        1.25234
  H        8.66564        0.21132        3.60080
  H        7.25132        0.30727        2.78740
  O        7.67689        0.41919        3.68935
  H       -0.76604        8.72806        2.11167
  H       -1.40018        8.67994        0.59884
  O       -0.61693        9.00102        1.14332
  H        1.92275        3.57667        5.62968
  H        3.01881        4.75958        5.92238
  O        2.67943        4.10657        5.22449
  H        3.75596        0.16093       -7.76777
  H        2.74741        1.03958       -8.73163
  O        3.25851        0.16555       -8.64811
  H        4.05518       -9.91235       -1.54505
  H        4.62433       -9.41290       -2.99809
  O        4.63569       -9.21709       -2.00605
  H       -3.66518        6.91519        4.42236
  H       -3.48692        5.86271        3.18158
  O       -3.03398        6.24047        3.99498
  H       -4.42644        6.67145       -4.11396
  H       -5.68083        6.11937       -5.01968
  O       -5.27583        6.14245       -4.08600
  H        3.23012        1.35218        8.45775
  H        3.94753        2.57831        7.63452
  O        3.10048        2.28832        8.08845
  H       -3.84089        9.54791        1.21977
  H       -2.91081        8.81314        2.34995
  O       -3.46902        8.66247        1.52277

\end{lstlisting}
\end{tcolorbox}
%%%%%%%%%%%%%%%%%%%%%%%%%%%%%%%%%%%% Intercalated %%%%%%%%%%%%%%%%%%%%%%%%%%%%%%%%%%%%%%

\subsection*{Intercalated Complex}
Unless otherwise stated, the charge of the system is –1. I.e., four hydrogen atoms have been added to the backbone.

\subsubsection*{Input Geometries}

\begin{tcolorbox}[breakable]
\textbf{Charge –1}
\scriptsize
\begin{lstlisting}
176
deoxy CpG Coordinates from ORCA-job opt E -936.146184574685; charge=-6
  O        4.14922        9.96270       -7.51580
  P        0.75655       -6.82902       -4.36325
  O        0.42301       -7.65045       -3.17811
  O       -0.22472       -6.87054       -5.47074
  O        1.00595       -5.31273       -3.91892
  P        5.71415        9.62671       -7.37762
  O        6.36504       10.72198       -6.62427
  O        6.29238        9.32941       -8.70732
  O        5.69093        8.29698       -6.48881
  C        4.72852        8.18597       -5.42325
  C        5.27013        7.29196       -4.32512
  O        4.95667        5.90617       -4.64850
  C        6.78782        7.30560       -4.14490
  O        7.15382        6.86651       -2.84042
  C        7.25697        6.25340       -5.14937
  C        6.12598        5.23207       -5.08885
  N        5.82472        4.59386       -6.40066
  C        5.43451        3.25730       -6.39787
  O        5.35478        2.66114       -5.31604
  N        5.15536        2.65694       -7.58314
  C        5.25307        3.33759       -8.73173
  N        4.96991        2.70728       -9.86178
  C        5.65244        4.71366       -8.75710
  C        5.92725        5.29381       -7.56157
  H        4.69627        1.75876       -9.84280
  H        5.03133        3.17725      -10.70758
  P        3.12290       -8.38494      -10.22559
  O        2.52781       -9.66567       -9.78265
  O        2.41953       -7.69816      -11.33232
  O        3.26544       -7.39360       -8.97837
  C        4.33324       -7.60652       -8.03580
  C        3.92568       -7.10019       -6.66627
  O        4.23510       -5.67895       -6.57832
  C        2.43417       -7.18439       -6.34313
  O        2.21207       -7.17259       -4.93639
  C        1.88881       -5.87875       -6.92115
  C        3.04230       -4.91413       -6.66675
  N        3.22300       -3.89952       -7.74245
  C        3.03507       -4.02763       -9.10122
  N        3.28355       -2.93396       -9.77965
  C        3.66275       -2.01574       -8.80169
  C        4.05042       -0.65663       -8.92532
  O        4.13937        0.02908       -9.93983
  N        4.35374       -0.09593       -7.67660
  C        4.29139       -0.76138       -6.46769
  N        4.62047       -0.04761       -5.38870
  N        3.92744       -2.03649       -6.35322
  C        3.62892       -2.59566       -7.55546
  H        4.63161        0.84463       -7.67064
  H        4.88951        0.90151       -5.48543
  H        4.59839       -0.44985       -4.52830
  H        4.52518        9.17479       -5.01216
  H        3.80442        7.75712       -5.81083
  H        4.77189        7.52807       -3.38484
  H        7.15254        8.33082       -4.20818
  H        7.36634        6.71188       -6.13220
  H        8.21667        5.84745       -4.82956
  H        6.33994        4.39930       -4.41891
  H        5.72249        5.23834       -9.69849
  H        6.23186        6.32978       -7.54241
  H        4.55538       -8.67174       -7.97216
  H        5.22248       -7.06955       -8.36598
  H        4.51150       -7.60933       -5.90096
  H        2.04234       -8.14388       -6.68072
  H        1.67011       -6.01313       -7.98049
  H        0.97575       -5.60106       -6.39456
  H        2.91593       -4.33039       -5.75495
  H        2.71045       -4.97097       -9.51490
  H        3.71244        9.26024       -8.00301
  H        1.22186       -4.78404       -4.69059
  P        8.64412        6.31941       -2.58866
  O        9.10507        6.73185       -1.24401
  O        9.51364        6.72099       -3.71729
  O        8.41362        4.73685       -2.62077
  C        7.23060        4.19107       -2.00716
  C        7.50193        2.78187       -1.51874
  O        7.26939        1.85045       -2.61504
  C        8.93680        2.49637       -1.07627
  O        8.99046        1.37478       -0.20024
  C        9.62372        2.10962       -2.38576
  C        8.50763        1.37847       -3.12444
  N        8.51675        1.60287       -4.59710
  C        8.85687        2.73567       -5.30350
  N        8.75630        2.60383       -6.60375
  C        8.31889        1.29085       -6.77139
  C        8.03023        0.56804       -7.95759
  O        8.10344        0.94298       -9.12435
  N        7.61457       -0.73991       -7.67094
  C        7.49311       -1.27814       -6.40459
  N        7.08120       -2.54649       -6.34447
  N        7.76434       -0.59883       -5.29259
  C        8.16981        0.67204       -5.55266
  H        7.39444       -1.31126       -8.43713
  H        6.88252       -3.04321       -7.17898
  H        6.97690       -2.97249       -5.50174
  P        6.78773       -6.80692      -15.35314
  O        6.11147       -8.03510      -15.82766
  O        6.35680       -5.54623      -15.99792
  O        6.63363       -6.68227      -13.76588
  C        7.45500       -7.50676      -12.91766
  C        6.73935       -7.78137      -11.60966
  O        7.01429       -6.68992      -10.68419
  C        5.21392       -7.83902      -11.68600
  O        4.67220       -8.56730      -10.58850
  C        4.81193       -6.37287      -11.52797
  C        5.87126       -5.85716      -10.55951
  N        6.29187       -4.45293      -10.82438
  C        6.60672       -3.65214       -9.72969
  O        6.52501       -4.13583       -8.59328
  N        6.99239       -2.36913       -9.94960
  C        7.06875       -1.88430      -11.19504
  N        7.45021       -0.62627      -11.35734
  C        6.74963       -2.69128      -12.33554
  C        6.36596       -3.97009      -12.09302
  H        7.66837       -0.07109      -10.57047
  H        7.51521       -0.25056      -12.24873
  P       10.38869        0.60891        0.01445
  O       10.52018        0.16951        1.42165
  O       11.50127        1.43386       -0.50783
  O       10.16791       -0.66184       -0.93172
  O        8.37722       -6.97285      -15.47294
  H        6.93945        4.81342       -1.16099
  H        6.42046        4.16760       -2.73602
  H        6.80009        2.53100       -0.72338
  H        9.30812        3.33266       -0.48394
  H        9.95838        3.01007       -2.90085
  H       10.48263        1.47348       -2.17202
  H        8.55926        0.29574       -3.00979
  H        9.17162        3.62536       -4.77835
  H        7.66072       -8.45271      -13.41861
  H        8.39432       -6.99387      -12.71096
  H        7.13401       -8.69429      -11.16367
  H        4.91487       -8.40143      -12.57052
  H        4.84508       -5.87976      -12.49949
  H        3.80060       -6.31421      -11.12564
  H        5.53471       -5.85401       -9.52278
  H        6.82038       -2.27399      -13.32915
  H        6.11724       -4.60736      -12.92874
  H       10.94598       -1.22332       -0.90094
  H        8.80723       -6.17234      -15.16330
 Pt        5.99208        0.43954       -9.86994
  N        6.60829        2.20145       -9.03783
  N        6.09993       -0.11657       -8.01332
  N        5.41089       -1.49289      -10.09491
  C        6.83946        3.37486       -9.66453
  C        7.26490        4.51385       -8.97475
  C        7.45808        4.43777       -7.58807
  C        7.21616        3.22475       -6.93249
  C        6.78872        2.11129       -7.66857
  C        6.49934        0.79319       -7.08182
  C        6.58405        0.40045       -5.73759
  C        6.25414       -0.92150       -5.39686
  C        5.84696       -1.83640       -6.38259
  C        5.77410       -1.40757       -7.71545
  C        5.38015       -2.18723       -8.90193
  C        4.99633       -3.53513       -8.88358
  C        4.64630       -4.16796      -10.08363
  C        4.68636       -3.44156      -11.28329
  C        5.07542       -2.09870      -11.25097
  H        6.67416        3.39104      -10.75095
  H        7.44005        5.44486       -9.53124
  H        7.79344        5.31635       -7.01840
  H        7.35657        3.13581       -5.84687
  H        6.90245        1.11501       -4.96677
  H        6.31568       -1.24408       -4.34812
  H        5.59003       -2.87061       -6.11684
  H        4.97265       -4.08172       -7.93098
  H        4.34365       -5.22499      -10.07954
  H        4.41887       -3.90347      -12.24370
  H        5.13316       -1.46174      -12.14718
  O        5.82707        0.82077      -11.80924
  H        6.09777        1.73120      -12.03619
  H       10.53703       -0.69327        2.04682
  H        7.28905       11.20540       -6.47095
  H        0.10053       -8.30216       -2.93344
  H        5.47605       -8.53824      -16.14149
\end{lstlisting}
\end{tcolorbox}

\begin{tcolorbox}[breakable]
\textbf{Charge –5}
\scriptsize
\begin{lstlisting}
172
Avogadro structure
O   3.01920   8.19755  -6.36468
P   0.39864  -7.30206  -4.34091
O   0.86781  -7.95329  -2.96647
O  -0.37062  -8.13262  -5.28790
O  -0.35411  -6.04866  -3.70499
P   4.16260   9.25865  -6.67691
O   3.33769  10.59241  -6.37947
O   4.82064   9.19061  -7.99557
O   5.25937   9.10241  -5.51725
C   4.91663   8.79059  -4.16858
C   5.21000   7.33607  -3.86527
O   4.47062   6.49277  -4.75600
C   6.69331   6.92824  -4.00526
O   7.02070   6.17866  -2.85449
C   6.69623   6.12282  -5.29580
C   5.33172   5.46897  -5.24851
N   4.80668   4.95211  -6.49076
C   4.10223   3.73605  -6.47850
O   3.93644   3.15068  -5.41231
N   3.64701   3.24591  -7.64623
C   3.76825   3.92717  -8.77959
N   3.28482   3.40095  -9.89220
C   4.40448   5.20620  -8.80758
C   4.91219   5.65845  -7.64948
H   2.88879   2.46187  -9.89870
H   3.39778   3.87648 -10.77175
P   3.35997  -8.17225  -9.73835
O   3.02568  -9.31575  -8.83637
O   2.48150  -7.72602 -10.86030
O   3.61512  -6.81943  -8.82848
C   4.28826  -6.88848  -7.59119
C   3.48765  -6.20612  -6.50343
O   3.36103  -4.81487  -6.77517
C   2.05692  -6.73726  -6.35003
O   1.76702  -6.72137  -4.94601
C   1.23134  -5.71298  -7.10922
C   2.00512  -4.43396  -6.83394
N   1.84923  -3.39340  -7.81431
C   1.60847  -3.49368  -9.16622
N   1.66466  -2.35347  -9.77766
C   1.97988  -1.44776  -8.79850
C   2.27724  -0.06010  -8.87303
O   2.28683   0.65147  -9.87597
N   2.60861   0.46030  -7.62749
C   2.64217  -0.24501  -6.45611
N   2.91673   0.43812  -5.34426
N   2.40949  -1.53805  -6.39144
C   2.09812  -2.07539  -7.57302
H   2.89233   1.45341  -7.60749
H   3.24637   1.40173  -5.36461
H   3.05743  -0.08795  -4.49867
H   5.52589   9.41614  -3.51532
H   3.87096   9.02227  -3.94882
H   4.88610   7.15496  -2.83521
H   7.34480   7.80303  -4.08095
H   6.79978   6.81166  -6.13574
H   7.50634   5.40422  -5.35607
H   5.34263   4.63293  -4.55020
H   4.50215   5.77687  -9.71835
H   5.42683   6.60677  -7.61050
H   4.47085  -7.91984  -7.27555
H   5.26368  -6.39727  -7.67401
H   4.03684  -6.34344  -5.56548
H   1.94387  -7.74859  -6.74434
H   1.24522  -5.98070  -8.16481
H   0.19016  -5.64933  -6.79718
H   1.69625  -3.99876  -5.87858
H   1.39346  -4.43784  -9.64191
H   3.32412   7.41424  -5.86274
H  -0.95300  -5.59566  -4.31353
P   8.46976   5.40871  -2.66779
O   8.73182   5.41197  -1.19604
O   9.45819   5.90051  -3.67370
O   8.06830   3.88459  -3.15155
C   7.26764   3.06335  -2.32802
C   7.99265   1.79823  -1.91317
O   8.26416   0.97295  -3.03712
C   9.34641   2.02349  -1.24000
O   9.51234   0.95760  -0.29013
C  10.32554   1.90073  -2.39510
C   9.64363   0.87758  -3.29569
N   9.87963   1.05757  -4.70914
C  10.23225   2.18852  -5.41051
N  10.23403   2.01104  -6.69372
C   9.84344   0.70780  -6.86496
C   9.55659  -0.02585  -8.04643
O   9.66432   0.36115  -9.20946
N   9.08226  -1.29881  -7.75822
C   8.90011  -1.80944  -6.50218
N   8.40882  -3.04495  -6.42196
N   9.16927  -1.13378  -5.40530
C   9.61235   0.10261  -5.64550
H   8.82509  -1.88750  -8.56812
H   8.17813  -3.60463  -7.24188
H   8.25677  -3.43818  -5.50947
P   6.49286  -7.72194 -15.42876
O   5.13702  -8.56131 -15.37408
O   6.44440  -6.37095 -16.01418
O   7.04575  -7.62451 -13.93215
C   7.24349  -8.74506 -13.07190
C   6.82748  -8.38205 -11.66391
O   7.61800  -7.29766 -11.18930
C   5.35627  -7.95015 -11.51756
O   4.88210  -8.48687 -10.29674
C   5.47473  -6.43559 -11.49604
C   6.78416  -6.25145 -10.74663
N   7.44403  -4.97672 -10.96809
C   7.83989  -4.19215  -9.87834
O   7.70879  -4.63134  -8.73736
N   8.34335  -2.96631 -10.11844
C   8.51629  -2.52113 -11.35649
N   8.98654  -1.29568 -11.53095
C   8.21416  -3.34097 -12.48685
C   7.68579  -4.55151 -12.23530
H   9.21638  -0.70172 -10.73595
H   9.13783  -0.93361 -12.45706
P  10.67216   0.98536   0.80698
O  10.03850   0.20027   2.03989
O  11.15839   2.31671   1.20959
O  11.73675   0.00223   0.12724
O   7.35959  -8.83304 -16.17567
H   6.94233   3.57590  -1.41831
H   6.36490   2.78174  -2.87961
H   7.32221   1.25847  -1.23607
H   9.40542   2.98874  -0.73601
H  10.41403   2.87594  -2.86775
H  11.32597   1.58871  -2.10122
H   9.98583  -0.13307  -3.05842
H  10.48421   3.11602  -4.92184
H   6.66515  -9.61611 -13.39037
H   8.30016  -9.01981 -13.09331
H   7.01013  -9.26918 -11.04715
H   4.74264  -8.30303 -12.35097
H   5.53291  -6.06194 -12.51851
H   4.63614  -5.94151 -11.01566
H   6.61915  -6.32273  -9.67219
H   8.39596  -3.00873 -13.49765
H   7.43362  -5.23042 -13.03718
H   9.80706  -0.72628   1.88825
H   8.24223  -8.53838 -16.43815
Pt  6.14339   1.08657 -10.31518
N   6.91841   2.90967  -9.84455
N   6.31101   0.86950  -8.39766
N   5.49365  -0.82176 -10.15846
C   7.23458   3.90479 -10.66810
C   7.86190   5.05636 -10.22619
C   8.18677   5.16745  -8.88730
C   7.85113   4.13419  -8.02743
C   7.19623   3.02437  -8.52016
C   6.78003   1.89199  -7.68220
C   6.82073   1.79039  -6.30207
C   6.38452   0.60988  -5.71857
C   5.93320  -0.44651  -6.49490
C   5.90623  -0.28692  -7.86850
C   5.47943  -1.26464  -8.87683
C   5.10040  -2.55745  -8.57795
C   4.73320  -3.40701  -9.60828
C   4.76009  -2.94169 -10.90955
C   5.14723  -1.63646 -11.14879
H   6.98652   3.77640 -11.71090
H   8.09664   5.83753 -10.93415
H   8.69582   6.04387  -8.51088
H   8.10800   4.18579  -6.98129
H   7.17885   2.60185  -5.68405
H   6.41352   0.50710  -4.64497
H   5.61180  -1.36661  -6.02876
H   5.09711  -2.90737  -7.55650
H   4.42518  -4.41795  -9.38199
H   4.47865  -3.56963 -11.74218
H   5.18160  -1.22355 -12.14663
O   5.87029   1.17079 -12.26733
H   6.24418   1.94544 -12.69730
 $END

\end{lstlisting}
\end{tcolorbox}

\begin{tcolorbox}[breakable]
\textbf{Charge +1}
\scriptsize
\lstinputlisting{xyz_files/intercalated/6_extrahydrogens_avogadro.xyz}
\end{tcolorbox}
\newpage
%%%%%%%%%%%%%%%%%%%%%%%%%%%%%%%%%%%% Optimized %%%%%%%%%%%%%%%%%%%%%%%%%%%%%%%%%%%%%%
\subsubsection*{Optimized Geometries}
The coordinates here were obtained through geometry optimizations using the stated method written in bold. 

\begin{tcolorbox}[breakable]
\textbf{PBEh-3c (Charge –1)}
\scriptsize
\lstinputlisting{xyz_files/intercalated/geo_opts/opt_negfreq-7.xyz}
\end{tcolorbox}

\begin{tcolorbox}[breakable]
\textbf{PBEh-3c (Charge –5)}
\scriptsize
\lstinputlisting{xyz_files/intercalated/geo_opts/PBEh-3c_0_extrahydrogens.xyz}
\end{tcolorbox}

\begin{tcolorbox}[breakable]
\textbf{PBEh-3c (Charge +1)}
\scriptsize
\lstinputlisting{xyz_files/intercalated/geo_opts/PBEh-3c_6_extrahydrogens.xyz}
\end{tcolorbox}

\begin{tcolorbox}[breakable]
\textbf{PBE/def2-SVP}
\scriptsize
\lstinputlisting{xyz_files/intercalated/geo_opts/PBE_def2-SVP.xyz}
\end{tcolorbox}

\begin{tcolorbox}[breakable]
\textbf{PBE0/def2-SVP}
\scriptsize
\lstinputlisting{xyz_files/intercalated/geo_opts/inter_PBE0_def2-SVP.xyz}
\end{tcolorbox}

\begin{tcolorbox}[breakable]
\textbf{GFN1-xTB}
\scriptsize
\lstinputlisting{xyz_files/intercalated/geo_opts/xTB1.xyz}
\end{tcolorbox}

\begin{tcolorbox}[breakable]
\textbf{GFN2-xTB}
\scriptsize
\lstinputlisting{xyz_files/intercalated/geo_opts/xTB2.xyz}
\end{tcolorbox}
\newpage
\newpage
\section*{Spectra Calculated with the Resolution of the Identity} 
\begin{figure}[H]
    \centering
    \begin{subfigure}[b]{0.45\textwidth}
        \centering
        \includegraphics[width=\textwidth]{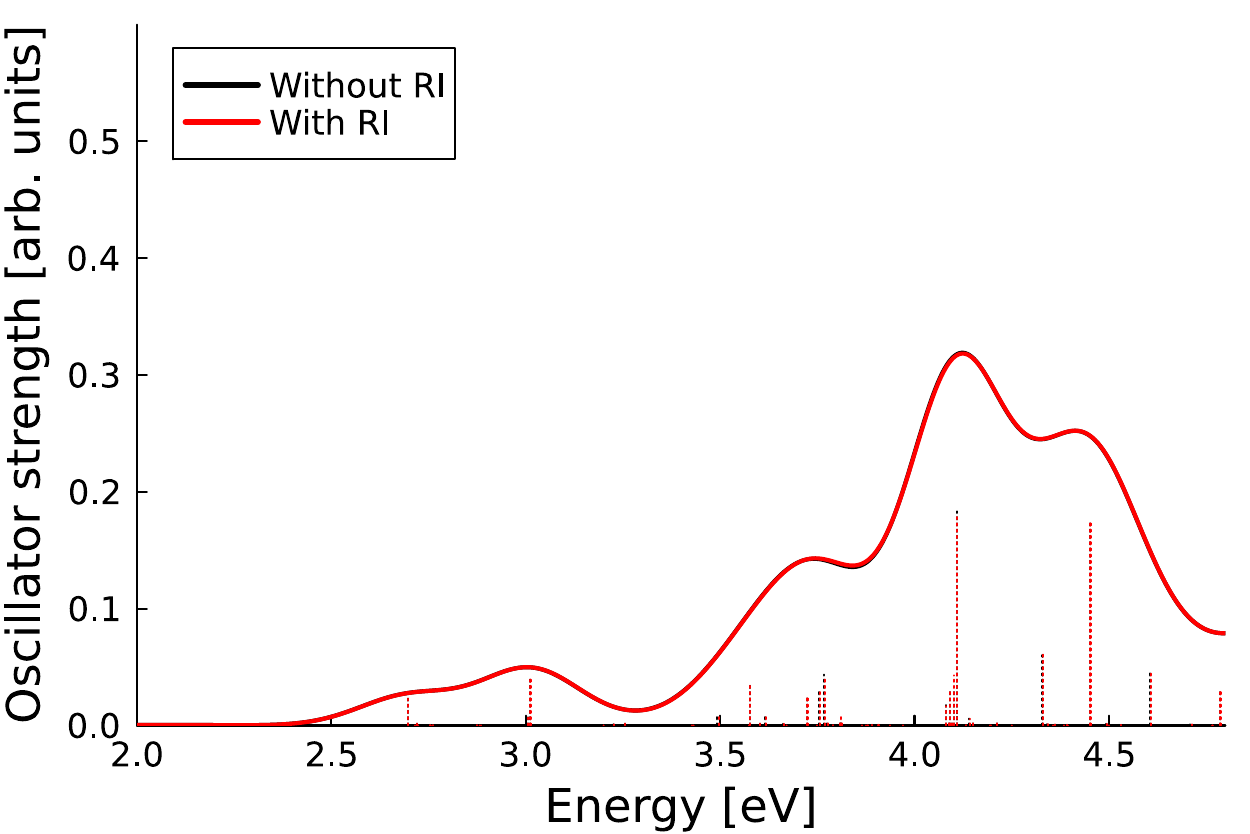}
        \caption{}
        \label{fig:enter-label-a}
    \end{subfigure}
    \hfill
    \begin{subfigure}[b]{0.45\textwidth}
        \centering
        \includegraphics[width=\textwidth]{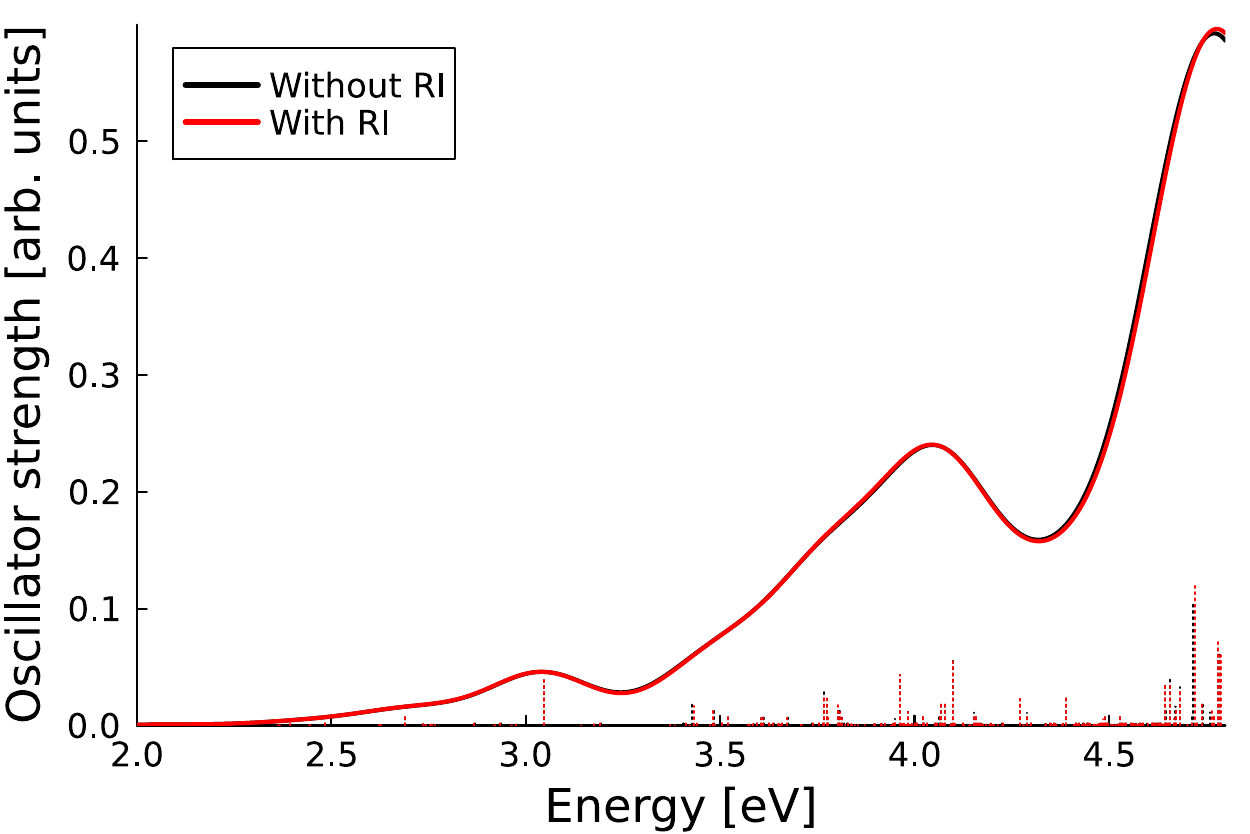}
        \caption{}
        \label{fig:enter-label-b}
    \end{subfigure}
    \caption{UV-Vis spectra calculated with PBE0/x2c-SVPall and \ac{SOC}, but without \ac{TDA}. Spectra calculated without (black) and with (red) RI for (a) the isolated complex and (b) the intercalated complex. The isolated and intercalated complex were optimized with PBE0/def2-SVP and PBE/def2-SVP, respectively. A FWHM of 0.3 eV was used to smear the spectra.}
    \label{RI}
\end{figure}

We also considered a third factor, the \ac{RI}, when we had to determine the necessary level of theory. The results here were obtained through PBE0/x2c-SVPall, and they are shown in Figure \ref{RI}. Akin to the previous factors (\ac{SOC} and \ac{TDA}), the \ac{RI} was the only differential factor in each pair of calculations.

There are no notable changes in the UV-Vis spectra, which seems to be especially true for the intercalated complex evident by their \ac{MAE} and MSE values. For the isolated complex, the values are 0.001 eV and 0.001 eV for the energies, respectively, and each corresponds to less than 1 kJ/mol.

When the same transition states are compared in terms of the intensity, the \ac{MAE} and \ac{MSE} values are 0.20 and –0.000, indicating little to no change in the oscillator strength. For the intercalated complex, both values are 0 in terms of energies and oscillator strength, which means there is no difference when \ac{RI} is used. Observing the intensities, both values are zero. Hence, there is no change when the \ac{RI} is used for the intercalated complex.

\section*{Basis Sets}
With a set of optimized structures, we wanted to determine the importance of the one-electron basis sets. However, this could only be investigated for the isolated complex since the larger basis sets became too computationally expensive for the intercalated complex. The calculations did not converge within 7 days.  

\begin{figure} [H]
    \centering
    \begin{subfigure}[b]{0.48\textwidth}
    \centering
    \includegraphics[width=\linewidth]{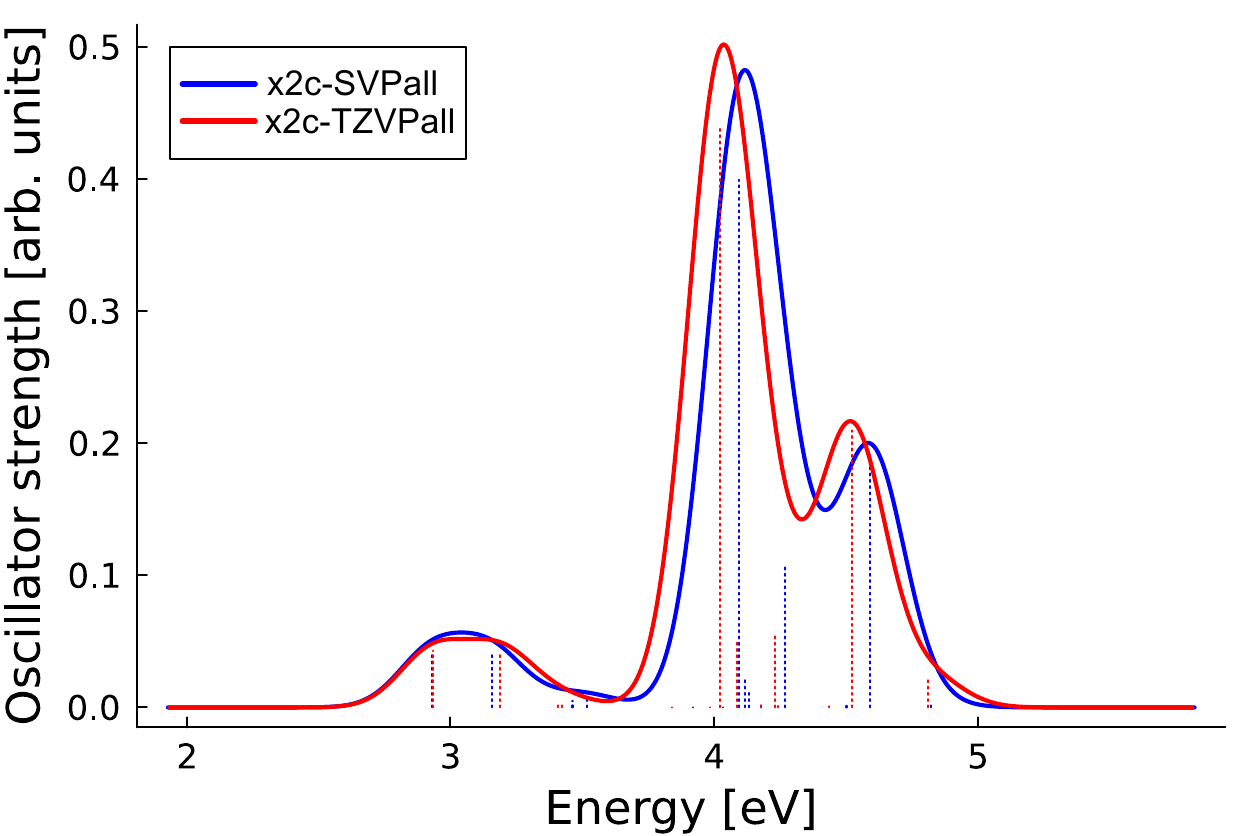}
        \caption{}
        \label{svp-tzvp}
     \end{subfigure}
     \begin{subfigure}[b]{0.48\textwidth}
    \centering
    \includegraphics[width=\linewidth]{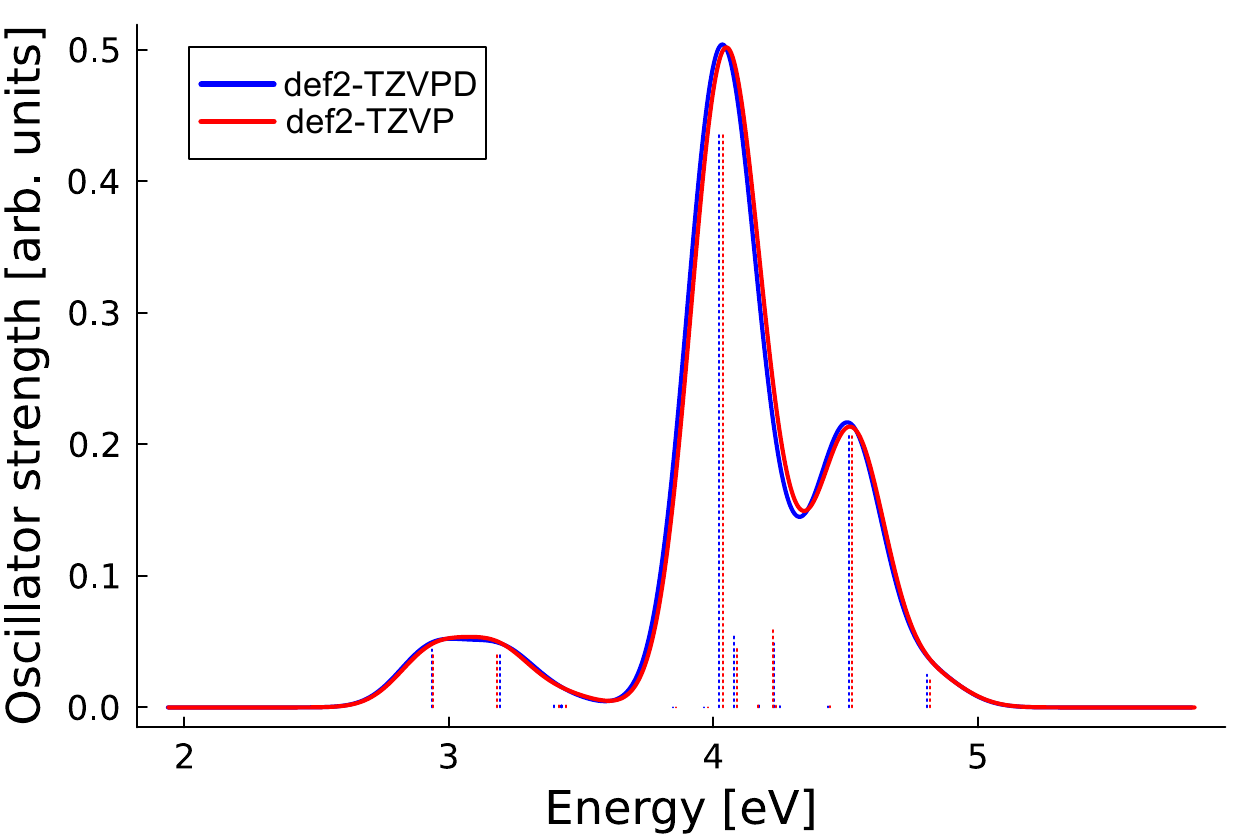}
        \caption{}
        \label{dif}
     \end{subfigure}
    \caption{UV-Vis spectra calculated for the isolated complex without \ac{SOC}. (a) The methods used to calculate the spectra were PBE0/x2c-SVPall (blue) and PBE0/x2c-TZVPall (red). (b) The methods used to calculate the spectra were PBE0/def2-TZVPD (blue) and PBE0/def2-TZVP (red). The complex was optimized with PBE0/def2-SVP.}
    \label{fig:enter-label}
\end{figure}

First, UV-Vis spectra were calculated with a double $\zeta$ and a triple $\zeta$ basis set (Figure \ref{svp-tzvp}). The shapes of the curves were similar, but the triple $\zeta$ basis set shifted the spectrum slightly to lower energies compared to the double $\zeta$ set. However, since the change was relatively modest and the calculation was significantly more time-consuming - already too time-consuming for the small intercalated complex model - a double $\zeta$ basis set represents a good compromise between cost and accuracy. 

The second test focused on diffuse functions. The two spectra, with and without diffuse functions, in Figure \ref{dif} aligned almost perfectly, which indicates that diffuse functions were not crucial. However,  this conclusion is based on the isolated complex, and it cannot be concluded that diffuse functions do not have an influence when relativistic effects or larger systems are considered. Additionally, there is no X2C relativistic Karlsruhe basis set with diffuse functions available in ORCA. The importance of diffuse functions for a relativistic approach can therefore not be evaluated here.

\newpage
\section*{Structure Optimizations with Different Functionals and Methods} 
\begin{figure}[H]
    \centering
    \includegraphics[width=1\linewidth]{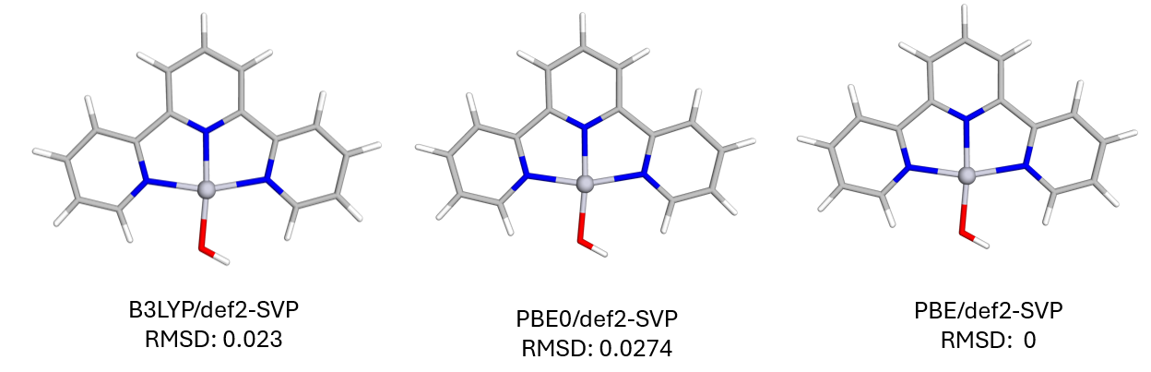}
    \caption{Three structures of the isolated complex optimized using B3LYP, PBE0, and PBE functionals with the def2-SVP basis set. The figure also shows the \ac{RMSD} values obtained when overlaying each structure with the one optimized with PBE/def2-SVP.}
    \label{non-rel_struct}
\end{figure}

The first set of results employs the hybrid functionals; B3LYP, PBE0, and PBE each paired with def2-SVP. The results are shown in Figure \ref{non-rel_struct}.

The geometry obtained with PBE0/def2-SVP has the largest deviation from the reference structure. However, there is no significant difference between its \ac{RMSD} value compared to the value obtained through B3LYP/def2-SVP as they are both less than 0.03 Å. The deviation from the reference structure is not visually discernible from the overlays.

\begin{figure}[H]
    \centering
    \includegraphics[width=1\linewidth]{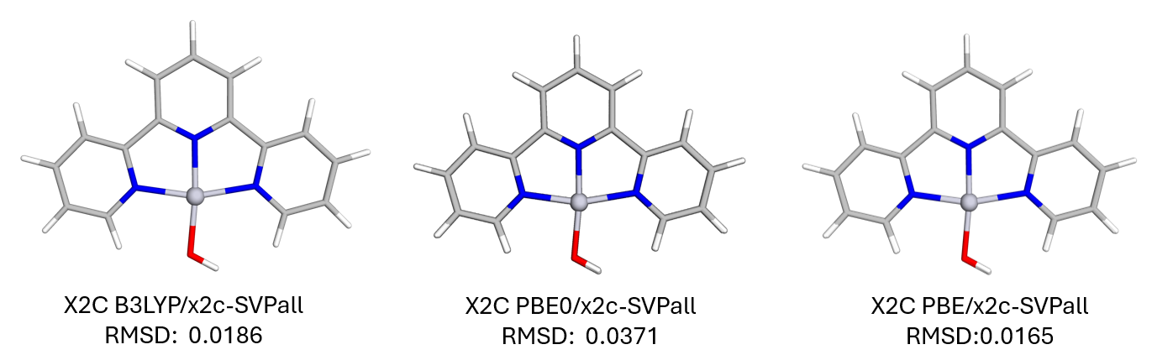}
    \caption{Three structures of the isolated complex, optimized using B3LYP, PBE0, and PBE functionals with the scalar-relativistic X2C Hamiltonian and x2c-SVPall basis set, but without \ac{SOC}. The figure also shows the \ac{RMSD} values obtained when overlaying each structure with the one optimized with PBE/def2-SVP.}
    \label{fig:rel_struct}
\end{figure}

We next discuss the results for the same set of functionals, but employed in a relativistic framework by utilizing the scalar-relativistic X2C Hamiltonian along with the x2c-SVPall basis set. The results are shown in Figure \ref{fig:rel_struct}.

When compared to each other in this set of calculations, B3LYP/x2c-SVPall and PBE/x2c-SVPall display the smallest deviation from the reference structure. Both of their \ac{RMSD} values are less than 0.03 Å. PBE0/x2c-SVPall has a value of close to 0.04 Å.

Hence, the results suggest that there is not a notable difference between the methods employing the hybrid functionals and their relativistic counterparts as the \ac{RMSD} values are less than 0.03 Å with the exception of PBE0/x2c-SVPall.

\begin{figure}[H]
    \centering
    \includegraphics[width=0.75\linewidth]{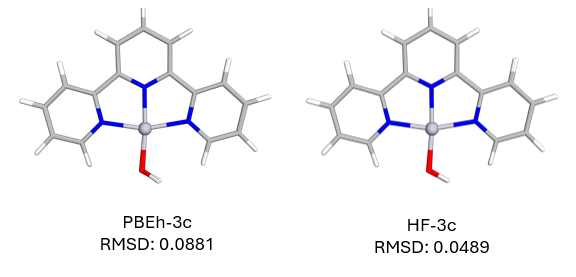}
    \caption{Two structures of the isolated complex optimized using the semi-empirical methods PBEh-3c and HF-3c. The figure also shows the \ac{RMSD} values obtained when overlaying each structure with the one optimized with PBE/def2-SVP.}
    \label{fig:semi-empirical_strucs}
\end{figure}
\newpage
\section*{Orbital Analysis of the First Band}

\begin{figure} [H]
    \centering
    \includegraphics[width=0.6\linewidth]{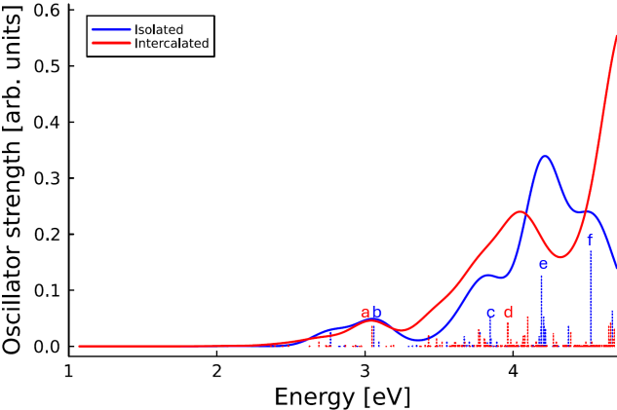}
    \caption{UV-Vis spectra calculated with PBE0/x2c-SVPall and \ac{SOC}, but without \ac{TDA} and \ac{RI} for the isolated complex (blue) and the intercalated complex (red). The isolated and intercalated complex were optimized with PBE0/def2-SVP and PBE/def2-SVP, respectively. A FWHM of 0.3 eV was used to smear the spectra.}
    \label{Iso_vs_inter}
\end{figure}

Since SOC was applied, it was necessary to start by analyzing which singlet and triplet states contributed to the peaks and which orbital transitions contributed to these states. 

\begin{figure}[H]
    \centering
    \includegraphics[width=0.95\linewidth]{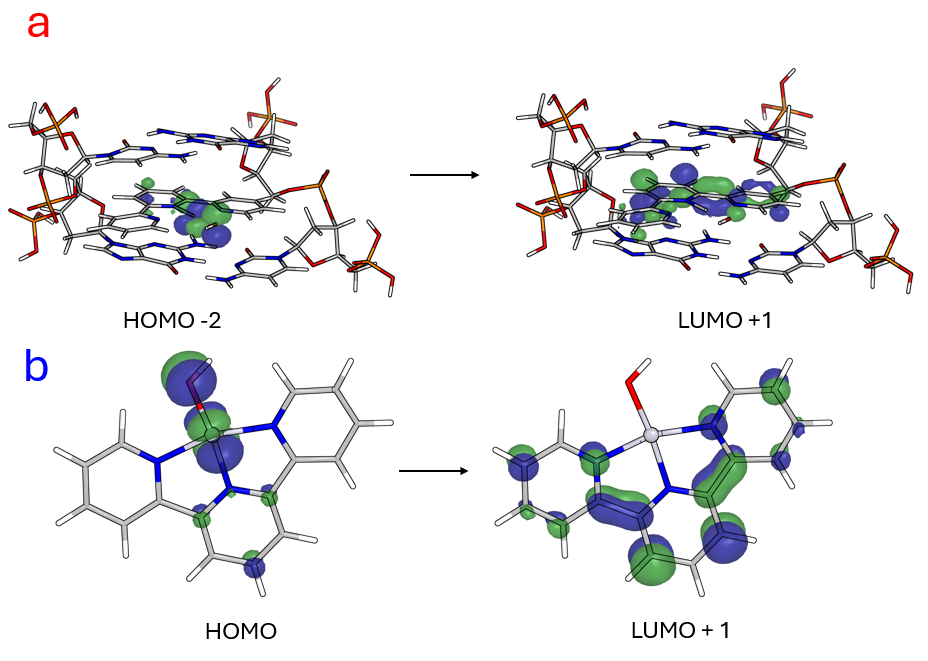}
    \caption{Main contributions to the orbital transitions for peaks \textbf{a} and \textbf{b} in Figure \ref{Iso_vs_inter}. }
    \label{orbital_a_b}
\end{figure}

Analyzing peaks \textbf{a} and \textbf{b} (3.05 eV and 3.06 eV, respectively) in Figure \ref{Iso_vs_inter}, the main contributing state was in both cases a singlet, contributing 85\% and 89\%, respectively. For each of these two singlet states, one main contribution was identified for the orbital transition (\textbf{a}: 98\%, \textbf{b}: 97\%). The orbitals are shown in Figure \ref{orbital_a_b}. For both the intercalated and isolated complex, the transition with the largest contribution was a metal to ligand transition. Peyratout et al.$^{17}$ state that the absorbance at these wavelengths corresponds to charge transfers, which do align with the orbital transitions that contribute the most to peaks \textbf{a} and \textbf{b}.

\newpage
\section*{$\omega$-Parameter Investigation}

\begin{figure}[H]
    \centering
    \begin{subfigure}[b]{0.48\textwidth}
        \centering
        \includegraphics[width=\textwidth]{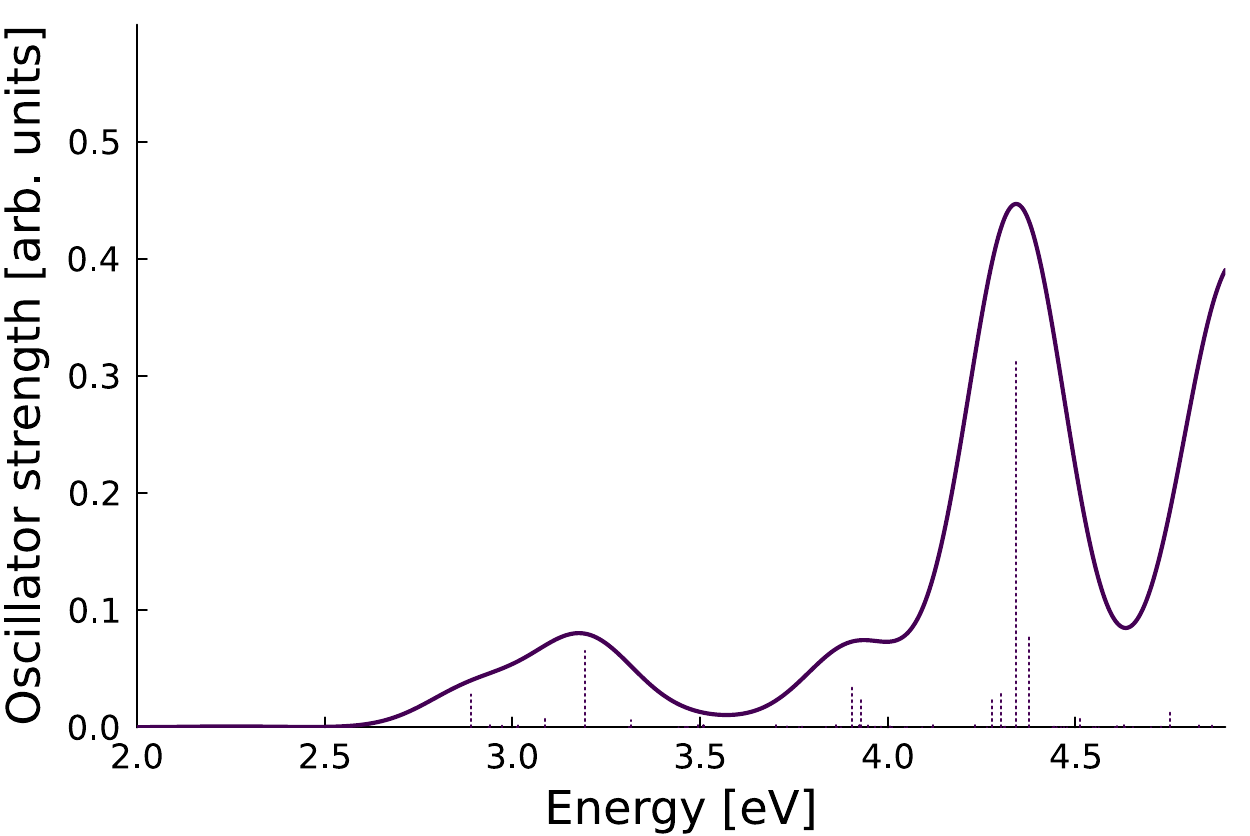}
        \caption{}
    \end{subfigure}
    \hfill
    \begin{subfigure}[b]{0.48\textwidth}
        \centering
        \includegraphics[width=\textwidth]{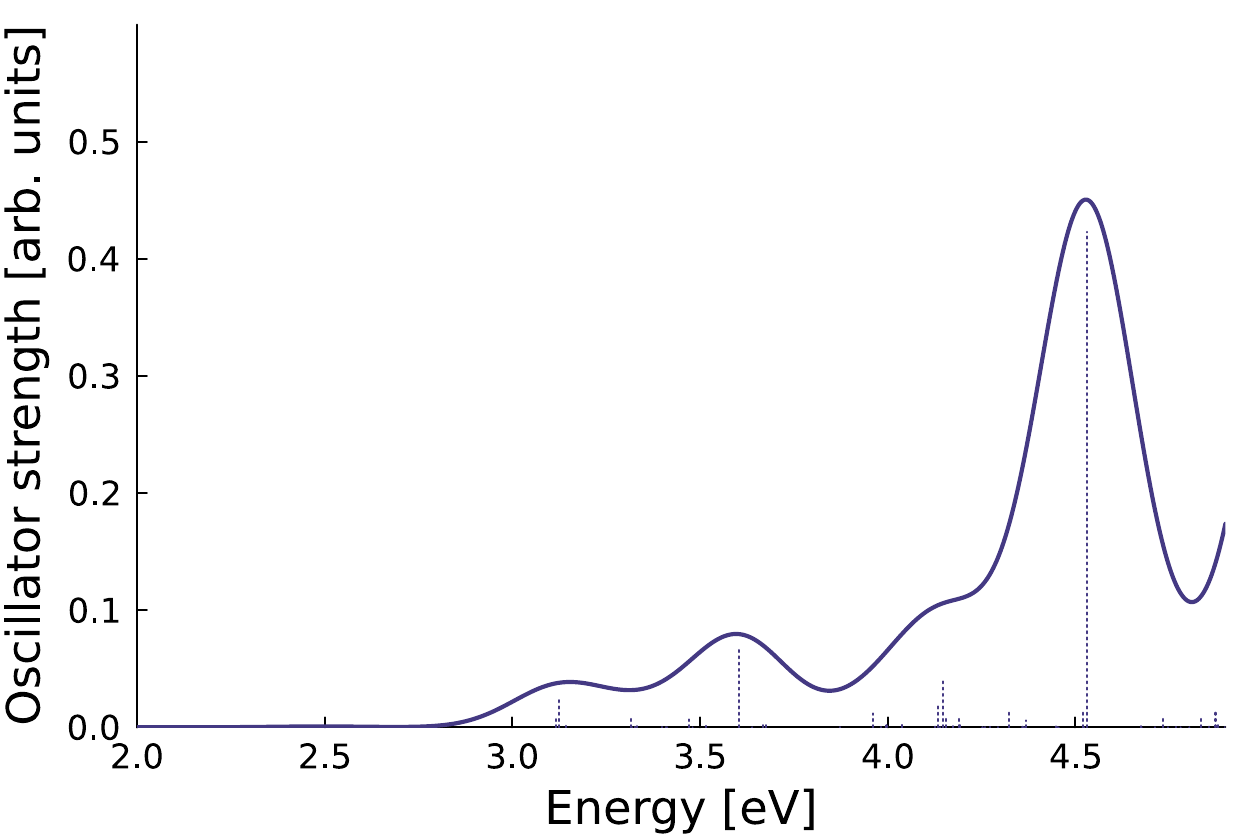}
        \caption{}
    \end{subfigure}
    \caption{UV-Vis spectra calculated for the isolated complex with LC-PBE/x2c-SVPall, \ac{SOC}, \ac{TDA}, and the \ac{RI} and with (a) $\omega$=0.20 and (b) $\omega$=0.25. A \ac{FWHM} of 0.3 eV was used to smear the spectra. The geometry was optimized with PBE/def2-SVP.}
    \label{SI:020mu_and_025mu}
\end{figure}

\begin{figure}[H]
    \centering
    \begin{subfigure}[b]{0.48\textwidth}
        \centering
        \includegraphics[width=\textwidth]{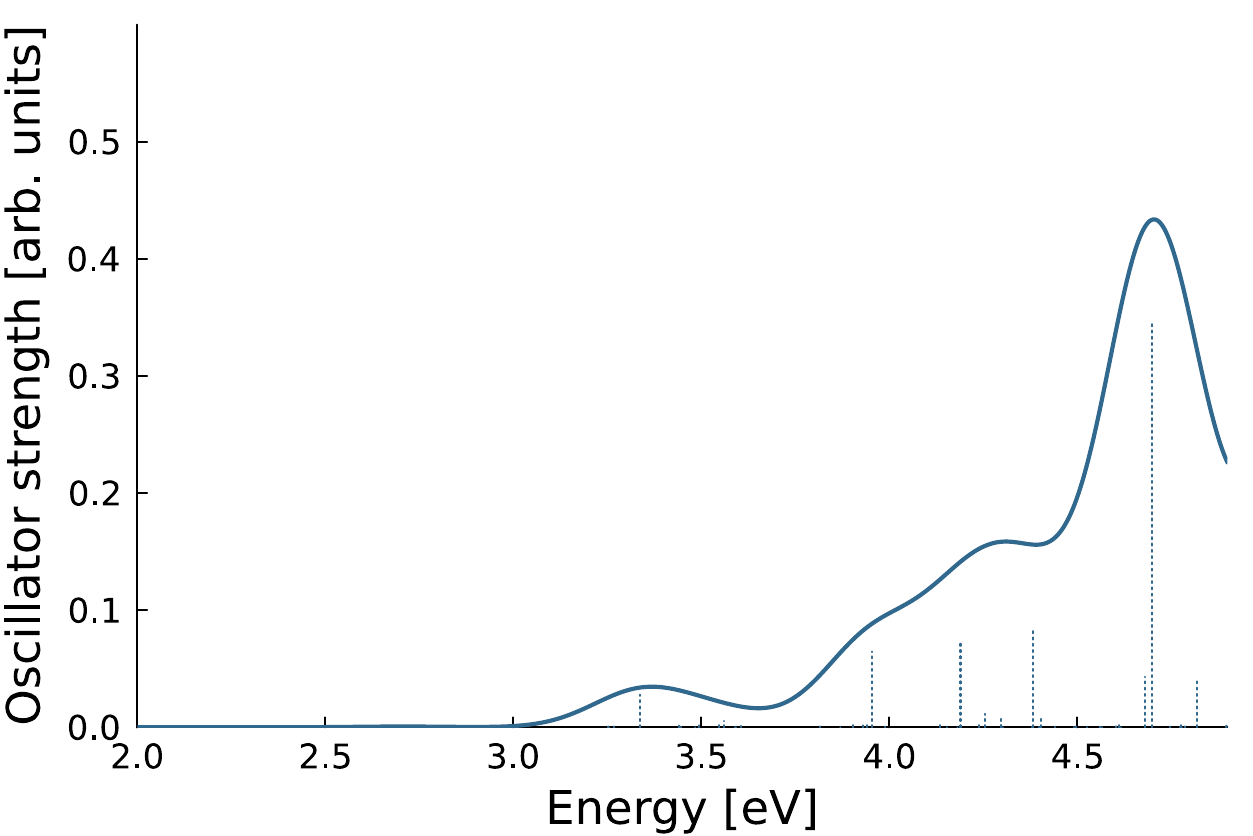}
        \caption{}
    \end{subfigure}
    \hfill
    \begin{subfigure}[b]{0.48\textwidth}
        \centering
        \includegraphics[width=\textwidth]{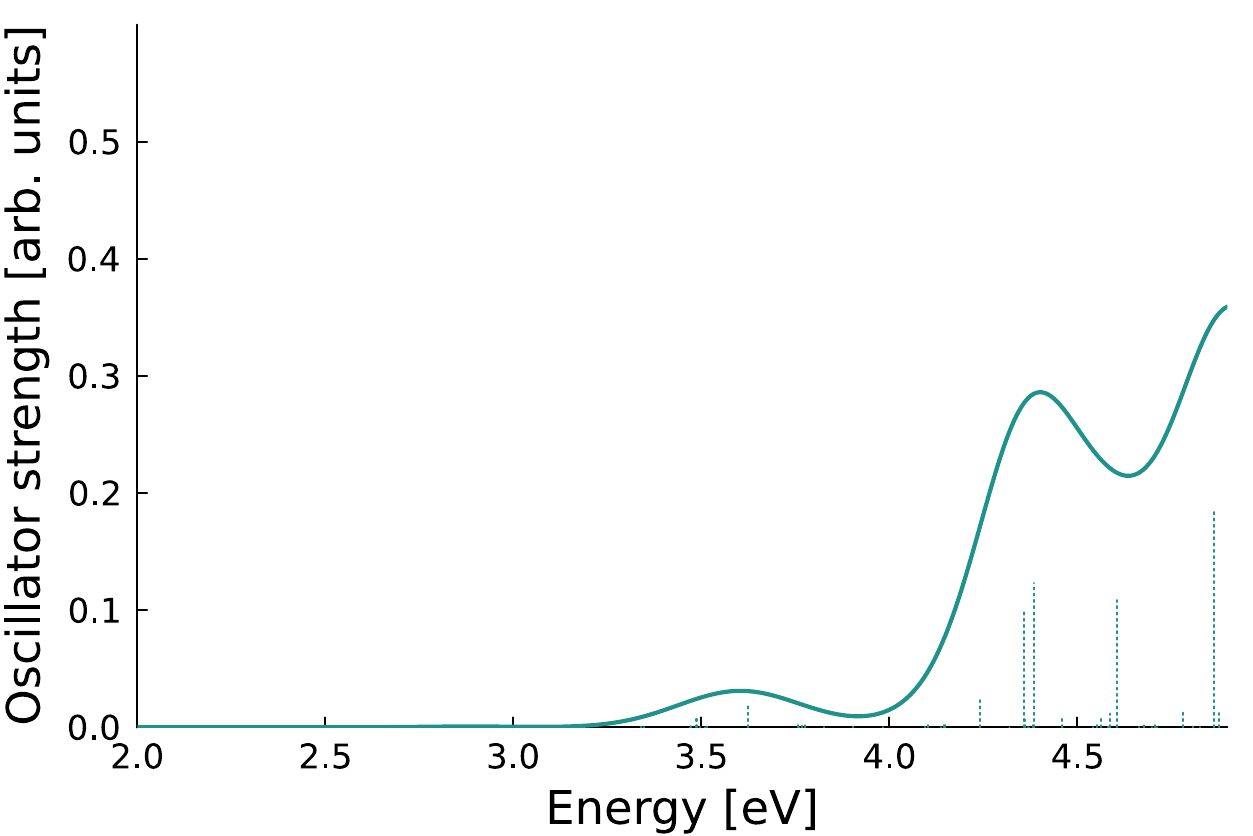}
        \caption{}
    \end{subfigure}
    \caption{UV-Vis spectra calculated for the isolated complex with LC-PBE/x2c-SVPall, \ac{SOC}, \ac{TDA}, and the \ac{RI} and with (a) $\omega$=0.30 and (b) $\omega$=0.35. A \ac{FWHM} of 0.3 eV was used to smear the spectra. The geometry was optimized with PBE/def2-SVP.}
    \label{SI:030mu_and_035mu}
\end{figure}

\begin{figure}[H]
    \centering
    \begin{subfigure}[b]{0.48\textwidth}
        \centering
        \includegraphics[width=\textwidth]{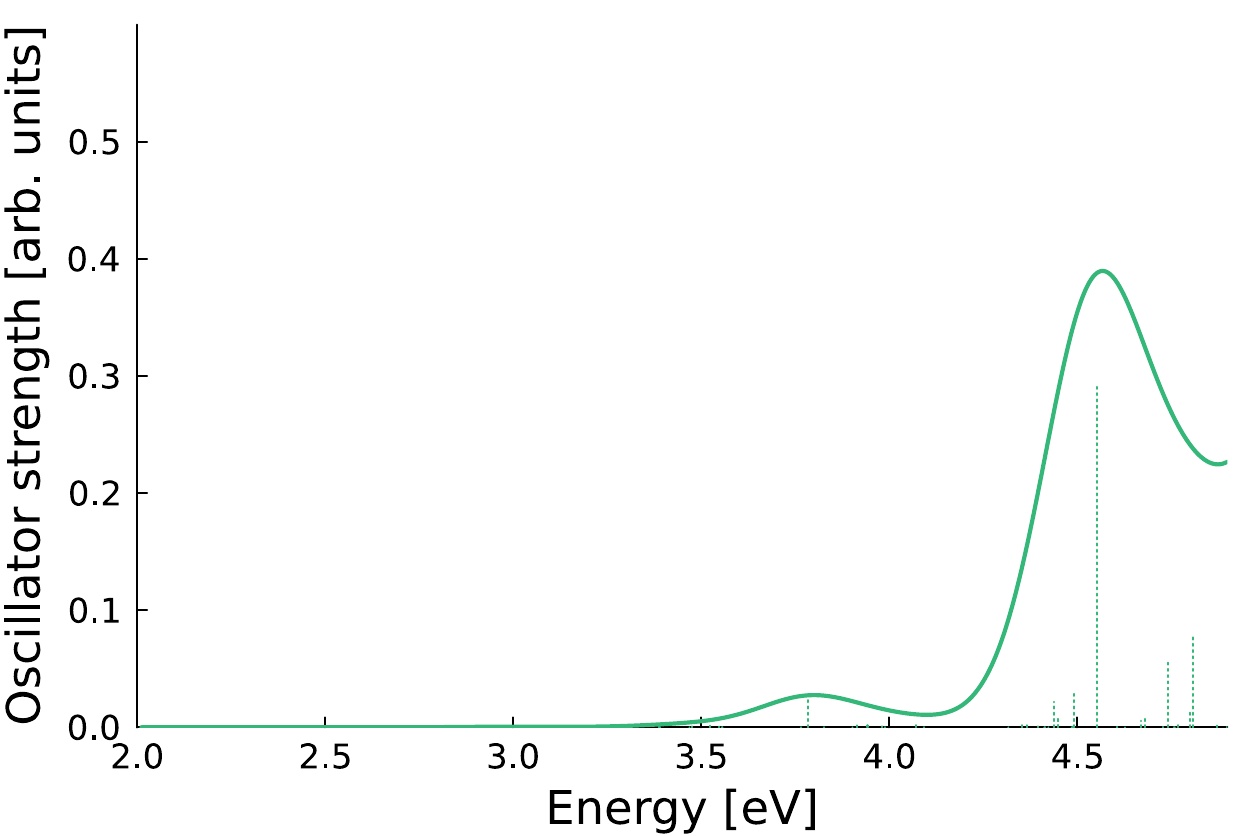}
        \caption{}
        \label{SI:040_mu}
    \end{subfigure}
    \hfill
    \begin{subfigure}[b]{0.48\textwidth}
        \centering
        \includegraphics[width=\textwidth]{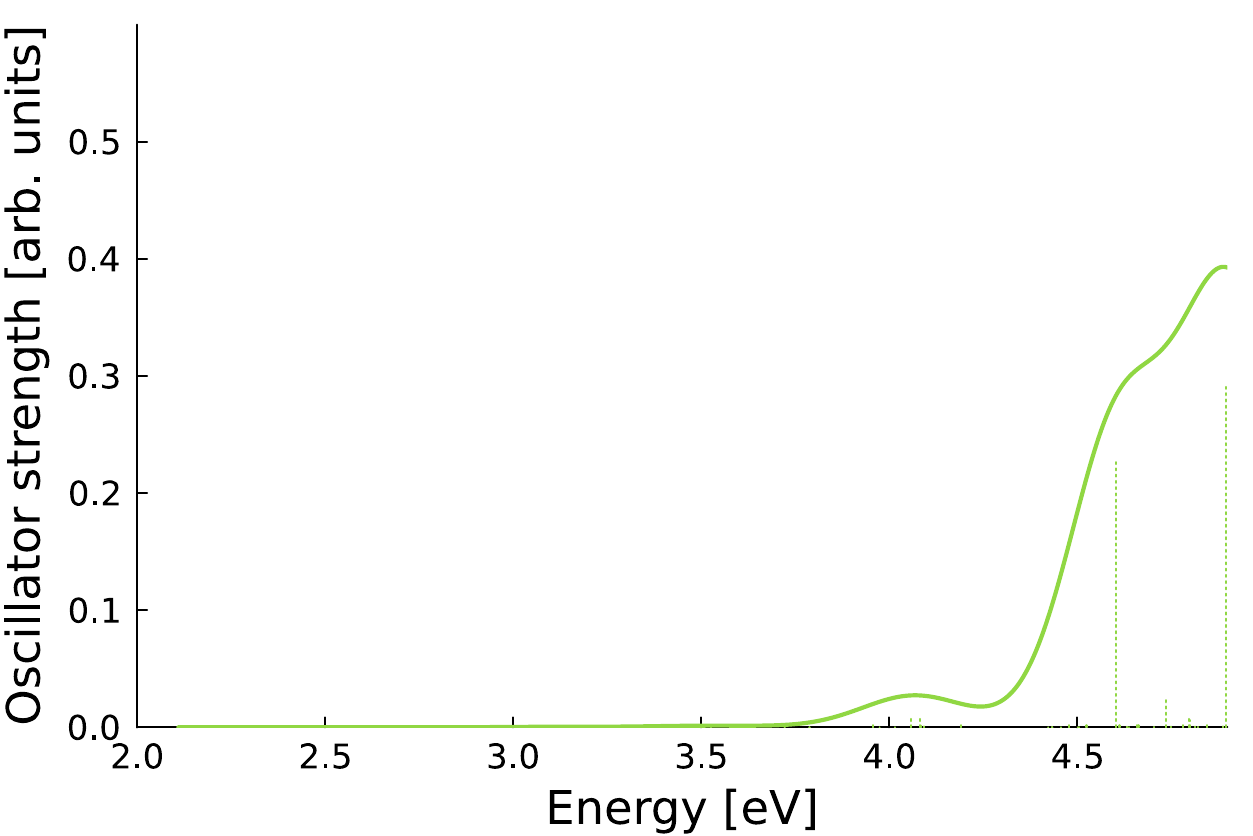}
        \caption{}
    \end{subfigure}
    \caption{UV-Vis spectra calculated for the isolated complex with LC-PBE/x2c-SVPall, \ac{SOC}, \ac{TDA}, and the \ac{RI} and with (a) $\omega$=0.40 and (b) $\omega$=0.47 (default value). A \ac{FWHM} of 0.3 eV was used to smear the spectra. The geometry was optimized with PBE/def2-SVP.}
    \label{SI:040mu_and_047mu}
\end{figure}
\end{document}